\documentstyle[11pt]{article}
\renewcommand{\theequation}{\arabic{section}.\arabic{equation}}

\begin{document}

\title{\huge \bf Gauge Conditions in the Canonical Hamiltonian  
Formulation of the Light-Front Quantum
Electrodynamics\thanks{This paper is a part of a qualifying
thesis for habilitation in the Faculty of Physics of the Warsaw
University.}} 

\date{May 25, 1998}

\author{\Large \bf Jerzy A. Przeszowski\\
\\
Institute of Fundamental Technological Research\\
Polish Academy of Sciences\\
ul. \'Swi\c{e}tokrzyska 11//21, \ 00-049 Warsaw, \ Poland\\
email: jprzeszo@ippt.gov.pl}

\maketitle

\normalsize

\vspace{2cm}

\begin{abstract}
We report here the status of different gauge conditions in the
canonical formulation of quantum electrodynamics on light-front
surfaces. We start with the massive vector fields as
pedagogical models  where all basic concepts and possible
problems manifestly appear. Several gauge choices are considered for
both the infinite and the finite volume formulation of
massless gauge field electrodynamics. We obtain the
perturbative Feynman rules in the first approach and the
quantum Hamiltonian for all sectors in the second approach.
Different space-time dimensions are discussed in all
models where they crucially change the physical meaning.
Generally, fermions are considered as the charged matter fields
but also one simple 1+1 dimensional model is discussed for
scalar fields. Finally the perspectives for further research
projects are discussed. 
\end{abstract}

\newpage

\tableofcontents

\newpage

\begin{flushright}
{\em Noninmensum gloriantes  in alienis laboribus}\\
2 Cor. 10.\raisebox{-3pt}{15.}
\end{flushright}

\part{Introduction}

Half a century ago Dirac \cite{Dirac1949} has proposed 3
different forms of relativistic dynamics depending on the 
types of
surfaces where independent modes were initiated. The first
possibility when a space-like surface is chosen (named by
Dirac {\it instant form}) has been used most frequently so far
and is usually called {\bf equal-time} quantization
(ET). The second choice is to take a surface of a single light wave
(called by Dirac {\it front form}). This kind of
quantization will be discussed in this work and is commonly 
referred to as {\bf light-front } formalism (LF). The third
possibility is to take a branch of hyperbolic surface $x^\mu
x_\mu = \kappa^2, \ x_0 > 0$ (named by Dirac {\it point form}).
Though a single attempt in this last approach can hardly be
found, we quote also this choice because very frequently
the LF formalism is erroneously called {\it light-cone
quantization} - a name which correctly should be connected with
a special case (when $\kappa^2 \rightarrow 0$) of the point form
of dynamics \cite{Dirac1949}. Then it took almost 20 years before
Dirac's idea of front form of dynamics was applied by
physicists.\\ 
At an ET surface, any two {\it different} points are spatially 
separated, therefore fields at these points are naturally
independent quantities. For a LF surface things are different
because for any point, there is a null-direction and all points
lying along this direction are light-like separated, thus
fields at these points may have nonvanishing commutators. For
all other directions on a LF surface, points are spatially
separated and fields at these points are independent. This
property of the LF formalism has been used to infer the
behaviour of light-cone commutators \cite{LCkomutatory1},
\cite{LCkomutatory2} from the respective LF commutators. \\ 
When Weinberg considered the scalar field theory in the ~{\bf
infinite momentum frame} (IMF) \cite{WeinbergIMF}, he has found
great simplifications of the "~old-fashioned" time-ordered
perturbation theory. Then the close relation between both
the IMF formalism \cite{SussFrye}, \cite{BarHal68},
\cite{ChaMa69}, \cite{DreLevYan69}, \cite{KogutSoper1970},
\cite{Yan1972}, \cite{Yan1973}, \cite{BroRosSua73}
and the LF formalism  \cite{LeuKlaStr70}, \cite{NevRoh71},
\cite{Soper71}, \cite{Tomboulis73}, \cite{Leutwyler74},
\cite{Casher76} was established, and these two names
were quite frequently interchangeably used. Since then, a lot
of successful attempts to phenomenology has been done 
\cite{BrodskyLepage}, most of them being based on
the LF perturbative calculations \cite{Namyslowski1984}.\\
Another possibility started over 10 years ago
when Brodsky and Pauli \cite{PauliBrodsky} have introduced
so-called {\bf discrete light-cone quantisation} (DLCQ), where 
periodic boundary conditions have been imposed on fields in a 
finite volume of the LF surface. This approach aimed at the
nonperturbative approach opened a broad scope of both the 
theoretical and numerical studies. Another important attempts
in the nonperturbative directions are the renormalization of
Hamiltonians \cite{WilsonGlazek} and the relativistic
bound-state problem \cite{Glazek1993}. Though, at the time,
they are beyond the scope of our paper, the next steps that one
can take starting from our analysis may be made precisely in these
directions. \\ 
Main topic of this paper is the canonical description of {\bf quantum
electrodynamics} (QED) within the LF approach. This model is
best known in the standard ET formulation but its LF
analysis has also been studied in many papers since the first
formulations in \cite{ChaMa69}, \cite{KogutSoper1970},
\cite{NevRoh71}. However, one 
important question of gauge fixing conditions, which can be
imposed of gauge field potentials, has been not solved so far. 
Nearly all papers use, seemingly the only possible, the gauge
fixing condition\footnote{Notation is given in
Appendix \ref{LFcoorappend}.} $A^{+} = A_{-}
= 0$ named the {\bf  light-cone gauge} (LC-gauge) or the IMF-gauge.
Evidently such a choice simplifies enormously many problems,
specially those which appear in the sectors of
charged matter fields. Also this gauge allows for the
propagation of only physical photons in the sector of gauge fields.   
However, there is a minor inconsistency in the LC-gauge, because
the Feynman perturbative rules found in the ET quantization
\cite{Bassettoetal1985} have the causal {\bf
Mandelstam-Leibbrandt} (ML) \cite{MandelLeibbr} prescription  
$$
\frac{1}{\left(n\cdot k + i \epsilon {\rm sgn} (n^* \cdot
k)\right)^n} $$
for the spurious poles of gauge field propagator, while the 
corresponding rules found in the LF quantization give the
{\bf Cauchy Principal Value} (CPV) prescription for these poles.
While for Abelian models both prescriptions are equally
correct then non-Abelian models need exclusively the
ML-prescription \cite{Wilsonloop}. This important caution
should be kept in mind during the present discussion where
Abelian interactions don't falsify the CPV prescription but
only prepare a future developments with non-Abelian interactions.\\
\noindent In the canonical LF approach one chooses one null-direction,
usually $x^{+}$, as a parameter of dynamical evolution and
treats all $\partial_{+}$ operators as 'time' derivatives.
Therefore the ET and the LF descriptions of the same
relativistic models formally may look quite different,
specially the structure of their canonical conjugate momenta. Here
we will not dogmatically follow  Dirac's general method of canonical
description for constraint system \cite{Dirac1962}. But to the
contrary, we will rather focus our attention on important
physical points. First, we will consider as {\em truly
constrained} variables only 
those fields which satisfy non-dynamical equations (i.e. 
with no $\partial_{+}$ terms). Such equations can be {\it formally}
solved for non-dynamical fields and therefore, non-dynamical and
dependent variables are understood as {\it synonyms} here.  
By definition, all other fields are considered as dynamical
and independent variables.  Second, canonical momenta
for dynamical
fields  are introduced and canonical
Hamiltonians are defined via the Legendre transformation. If
necessary, the terms which contain $\partial_{+}$ will be expressed
by means of corresponding canonical momenta and only in this way
the conjugate momenta can appear in the canonical LF 
Hamiltonian. For all variables which appear in the Hamiltonian 
we have dynamical equations of motion which are of the first order
with respect to $\partial_{+}$, and we demand that these
Hamilton equations follow from the canonical Hamiltonian
by means  of the classical Poisson-Dirac 
brackets. Next, having a consistent classical canonical
structure one can {\em define} a consistent quantum theory
where Poisson-Dirac brackets are replaced by quantum
commutation relations, and the Heisenberg equations
for quantum field operators have  the same form (up to some
ordering of noncommuting operators) as the corresponding classical
counterparts. Third, for the interacting system of gauge
and matter fields one can carry the above canonical
method in steps, starting with the gauge field sector where all
interactions  with matter are described by linear couplings to
arbitrary external currents. Having eliminated all gauge constraints
one can write an {\em equivalent} Lagrangian which describes
the gauge sector by means of fewer unconstrained
fields. Finally the canonical quantization of complete system
can be carried out quite easily having the constraints for gauge
and matter fields transparently separated and
solved.\footnote{This method of quantization is similar to that
proposed by Jackiw \cite{JackiwFaddeev}.}\\ 
\noindent Our paper is organized as follows.  In part II a
pedagogical model of massive vector electrodynamics with
fermion matter fields is analysed. Though this is not a
true gauge field model, it is discussed in three cases of no
gauge fixing, LF Weyl and Lorentz gauges. In part III the
QED with fermions is presented for several different gauges,
and perturbation Feynman rules for practical S-matrix
calculations are given in each case. Also QED
for charged scalar fields is presented in 1+1 dimensions
for the LF Weyl gauge. In part IV we analyse the DLCQ approach to
QED with periodic boundary conditions for gauge fields and antiperiodic
conditions for fermion fields imposed on the finite volume LF.
Also here several gauge conditions are  discussed and the canonical
analysis is carried out subsequently in different sectors for
various gauge field modes and fermion fields.  In Conclusions
all results are generally discussed 
special attention being paid to the further developments.
The Appendices contain all notations and definitions of different
functions used throughout the main text.\\ 
\noindent Most of these results have not been published so far, otherwise
due references to original papers are given in the paper.

\pagebreak

\part{Massive Electrodynamics}

\section{Theory without gauge fixing condition}\label{massbezgaugefix}

Our first model, which we analyse along the lines indicated in
the Introduction, will be the electrodynamics of massive vector
field $B_\mu$ with the mass term $m^2$. Its kinetic term has
the Maxwell form built with $B_{\mu \nu} =\partial_{\mu} B_{\nu}-
\partial_{\nu}B_{\mu}$ and it couples with fermion fields via
the electrodynamic current term $g \bar{\psi} \gamma^\mu \psi
B_{\mu}$. Though this model has been consistently quantized on
LF by Yan over 25 years ago \cite{Yan1973} via the Schwinger
action principle \cite{SchwingerActPrin}, we decided to derive
his results in our original method of quantization for
constrained systems. This model is even quite a pedagogical
example because it contains all obstacles and restrictions
which later will be encountered in more physical cases.

\subsection{Maxwell theory with a mass term}

We start our analysis with the following Lagrangian density: 
\begin{eqnarray}
{\cal L}_{mass} & = & - \frac 1 4 B_{\mu\nu} B^{\mu \mu} +
\frac{m^2}{2} B_\mu B^\mu  + \bar{\psi}(i
\partial_\mu \gamma^\mu - e B_\mu \gamma^\mu + M)\psi\nonumber\\
& = & (\partial_{+} B_i - \partial_i B_{+}) (\partial_{-}
B_i - \partial_i B_{-}) + \frac 1 2 (\partial_{+} B_{-} -
\partial_{+} B_{-})^2 - \frac 1 4 (\partial_{i} B_j - \partial_j
B_{i})^2 \nonumber\\
& + & \frac{m^2}{2} (2 B_{+} B_{-} - B_i^2) + \bar{\psi}(i
\partial_\mu \gamma^\mu - e B_\mu \gamma^\mu + M)\psi \ ,
 \end{eqnarray}
where the explicit LF coordinates are introduced for the vector
fields. It is known that only a half of the fermion degrees of
freedom are dynamical fields, therefore in order to separate the
vector field constraints from the fermion field ones, we have decided
to start with a subsystem where the fermion contribution is solely 
described by external currents $j^\mu$ coupled linearly with
$B_\mu$. Because in such a simplified model the dynamics of
fermions is omitted, one cannot argue in favour of the external
current conservation law $\partial_\mu j^\mu = 0$. Even more we
stress that all components of the current should be taken as
independent 
{\it classical} quantities which allow for taking functional
derivatives with respect to them\footnote{This property of
external currents is crucial in checking the consistency of
the simplified model with the one we have started with.}
\begin{equation}
\frac{\delta \ j^\mu(x)}{\delta \ j^\nu(y)} = \delta^\mu_\nu
\delta(x- y).
\end{equation}  
For such a simplified model one can easier find a set of
dynamical independent field variables which identically satisfy
the constraints and have correct dynamical properties. Later we can
write an {\it equivalent Lagrangian} with these independent
modes and external currents $j^\mu$, and then by reintroducing
fermion kinetic terms we end up with the {\it complete
equivalent Lagrangian} and finally, quantize a complete 
system without any effort. However this scenario is valid
only  if the condition of mutual independence of
fermion and vector fields is satisfied. Otherwise some
obstacles will appear and (fortunately or not) this situation
will appear in our first model.\\

\subsubsection{Vector fields with external currents}\label{massQEDpwzp}

Our first task will be to analyse a subsystem of vector fields,
so we start with the Lagrangian density
\begin{eqnarray}
{\cal L}^{mass}_B & = & (\partial_{+} B_i - \partial_i B_{+}) (\partial_{-}
B_i - \partial_i B_{-}) + \frac 1 2 (\partial_{+} B_{-} -
\partial_{-} B_{+})^2 - \frac 1 4 (\partial_{i} B_j - \partial_j
B_{i})^2 \nonumber\\
& + & \frac{m^2}{2} (2 B_{+} B_{-} - B_i^2) + B_\mu j^\mu \ ,
\label{massQEDLagrmassB} 
\end{eqnarray}
and taking the variations of vector fields $B_\mu$ we generate 
the Euler-Lagrange equations
\begin{eqnarray}
\partial_{+} (\partial_{+} B_{-} - \partial_{-} B_{+} -
\partial_i B_i) & = & (m^2 - \Delta_\perp ) B_{+} +
j^{-}\ ,\label{massQEDELeqmassBm} \\
- \partial_{-} (\partial_{+} B_{-} - \partial_{-} B_{+} +
\partial_i B_i) & = & (m^2 - \Delta_\perp ) B_{-} +
j^{+}\ ,\label{massQEDELeqmassBp}\\ 
(2 \partial_{+} \partial_{-} - \Delta_\perp + m^2)B_i & = &
\partial_i (\partial_{+} B_{-} + \partial_{-} B_{+} -
\partial_j B_j) + j^i. \label{massQEDELeqmassBi}
\end{eqnarray}
As the consistency condition for these equations we have
\begin{equation}
\partial_{+} B_{-} + \partial_{-} B_{+} - \partial_j B_j = -
\frac{1}{m^2} \left(\partial_{+} j^{+} + \partial_{-} j^{-}
+ \partial_{i} j^{i}\right)\label{massQEDmasswarzgodn}
\end{equation}
so, together with (\ref{massQEDELeqmassBm}), we end up with 
two independent equations with $\partial_{+}B_{-}$ - but the excess
of dynamical equations indicates the presence of constraints. This
observation will become transparent after introducing the canonical
momentum $\Pi^{-}$. Generally at LF, one introduces canonical
momenta $\Pi^\mu$ as the functional derivatives of Lagrangian
by $\partial_{+}B_\mu$, respectively. In our present
case this gives 
\begin{eqnarray}
\Pi^{-} & = & \partial_{+} B_{-} - \partial_{-} B_{+}\ , \\
\Pi^{i} & = & \partial_{-} B_{i} - \partial_{i} B_{-}\ ,
\label{massQEDdefPiimass}\\ 
\Pi^{+} & = & 0\ , \label{massQEDdefPipmass}
\end{eqnarray}
which according to Dirac's nomenclature indicates the presence
of {\em primary constraints} (\ref{massQEDdefPiimass}) and 
(\ref{massQEDdefPipmass}) because they do not determine
$\partial_{+}B_{i}$ and $\partial_{+} B_{+}$. Constraint
$\Pi^{+} \approx 0$ is characteristic also for the gauge fields 
and here the component $B_{+}$ is a dependent variable
which can be determined from Eq.(\ref{massQEDmasswarzgodn})
as\footnote{Integral operators and their convolutions which
appear hereafter are introduced in Appendix \ref{InvLaploperdod}.}
\begin{equation}
B_{+} = - \frac{1}{2\partial_{-}} * \left[ \Pi^{-} - \partial_i
B_i + \frac{1}{m^2} \left(\partial_{+} j^{+} + \partial_{-} j^{-}
+\partial_{i} j^{i}\right)\right].\label{massQEDmassvectBp}
\end{equation}
Other constraints $\Pi^{i} - \partial_{-} B_{i} + \partial_{i}
B_{-} \approx 0$ regularly appear on LF for covariant
relativistic fields and they can be ignored only after the
canonical Hamiltonian is calculated
\begin{eqnarray}
{\cal H}_{B}^{can} & = & \Pi^{-} \partial_{+} B_{-} + \Pi^{i}
\partial_{+} B_i - {\cal L}_{B} \nonumber\\
& = & \frac 1 2 (\Pi^{-})^2 + \frac 1 4 (\partial_i B_j -
\partial_j B_i)^2 + \frac 1 2 m^2 B_i^2 - B_{-}j^{-} - B_i j^i
\\
& + & B_{+} \left[ - \partial_{-}(\Pi^{-} + \partial_i B_i) +
(\Delta_\perp  -  m^2) B_{-} - j^{+} \right]\nonumber.
\end{eqnarray}
We notice that the expression in the square bracket 
identically vanishes due to Eq.(\ref{massQEDELeqmassBp})
which determines the field component $B_{-}$ 
\begin{equation}
B_{-} = \frac{1}{\Delta_\perp - m^2}* \left[ \partial_{-}(\Pi^{-} + 
\partial_i B_i) + j^{+}\right].
\end{equation}
For other dynamical field components we have the equations of motion
\begin{eqnarray}
\left( 2 \partial_{+} \partial_{-} - \Delta_\perp + m^2\right) \Pi^{-} &
= & - \partial_{+}j^{+} + \partial_{-} j^{-}\ ,\\
\left( 2 \partial_{+} \partial_{-} - \Delta_\perp + m^2\right) B_{i} &
= & - \frac{\partial_i}{m^2} \left( \partial_{+} j^{+} +
\partial_{-} j^{-} + \partial_i j^i\right) + j^i\ ,
\end{eqnarray}
which contain the troublesome expression $\partial_{+} j^{+}$. Here, contrary
to the complete theory with fermion dynamics, we cannot impose 
the external current conservation for curing this problem but
rather introduce new field variables
\begin{eqnarray}
{\Pi} & = & \Pi^{-} + \frac{1}{2\partial_{-}}*j^{+}\ ,\\
\widetilde{B}_i & = & B_i + \frac{\partial_i}{m^2}
\frac{1}{2\partial_{-}}*j^{+}\ ,
\end{eqnarray}
which satisfy the dynamical equations 
\begin{eqnarray}
\left( 2 \partial_{+} \partial_{-} - \Delta_\perp + m^2\right)
\Pi & = & \partial_{-} j^{-} +
(\Delta_\perp-m^2)\frac{1}{2\partial_{-}}*j^{+}\ , 
 \label{massQEDmassPieq}\\
\left( 2 \partial_{+} \partial_{-} - \Delta_\perp +m^2 \right)
\widetilde{B}_{i} & 
= & - \frac{\partial_i}{m^2} \left( \partial_{-} j^{-} +
\partial_k j^k + (\Delta_\perp- m^2) 
\frac{1}{2\partial_{-}}*j^{+}\right) + j^i\label{massQEDmasstildeBieq}.
\end{eqnarray}
Now the independent dynamical equations have the form of 
Hamilton equations and the Hamiltonian density can be
rewritten in terms of these new fields
\begin{eqnarray}
{\cal H}_{can} & = & \frac 1 2 (\Pi -
\frac{1}{2\partial_{-}}*j^{+})^2 + \frac 1 4 
(\partial_i \widetilde{B}_j - \partial_j \widetilde{B}_i)^2 + \frac 1 2
m^2 \left( \widetilde{B}_i -
\frac{\partial_i}{m^2}\frac{1}{2\partial_{-}}*j^{+} \right)^2 \nonumber\\
& - & j^{-} \left[ \frac{1}{\Delta_\perp - m^2} * \partial_{-}(
{\Pi} + \partial_i \widetilde{B}_i) - \frac{j^{+}}{2m^2} \right] -
j^i \left( \widetilde{B}_i - 
\frac{\partial_i}{m^2} \frac{1}{2\partial_{-}}*j^{+}
\right).\label{massQEDHameffmassB} 
\end{eqnarray}
Next, one can demand that Eqs.(\ref{massQEDmassPieq} and
\ref{massQEDmasstildeBieq}) follow canonically from this
Hamiltonian and this produces Dirac brackets\footnote{Though in
our analysis we didn't follow Dirac's procedure, we have
decided to call these brackets Dirac ones in order to stress
their consistency with all constraints present in the primary
description of the model.} at LF
\begin{eqnarray}
2 \partial^x_{-} \left\{\widetilde{\Pi}(x^{+}, \vec{x}),
\widetilde{\Pi}(x^{+}, \vec{y})\right\}_{DB} & = & - (\Delta_\perp
- m^2) \delta^{3}(\vec{x} - \vec{y})\ ,\\ 
2 \partial^x_{-} \left\{\widetilde{\Pi}(x^{+}, \vec{x}),
\widetilde{B}_i(x^{+}, \vec{y})\right\}_{DB} & = & 0 \ , \\
2 \partial^x_{-} \left\{\widetilde{B}_i(x^{+}, \vec{x}),
\widetilde{B}_j(x^{+}, \vec{y})\right\}_{DB} & = & - \left( \delta_{ij} -
\frac{\partial_i \partial_j}{m^2}\right) \delta^{3}(\vec{x} -
\vec{y}).
\end{eqnarray}
In this way we have found a canonical structure with
Hamiltonian and brackets which is almost ready for further
canonical quantization, however, before we will make this next
step, we should check its consistence with the primary
Lagrangian (\ref{massQEDLagrmassB}). When we functionally
differentiate the primary Lagrangian with respect to the external
currents $j^\mu$, we obtain primary fields $B_\mu$.
Similarly, differentiation of the consistent canonical Hamiltonian 
should produce $- B_\mu$ but in this case
Eq.(\ref{massQEDHameffmassB}) do not correctly reproduce the
expression  for $B_{+}$ (\ref{massQEDmassvectBp}). Physically this
means that influence of the fermion fields on the massive vector fields
cannot be reduced here to the presence of charged matter currents, but 
contrary, these two kinds of fields have non-vanishing mixed 
brackets at LF. Keeping this limitation in mind one can yet 
introduce {\em the effective Lagrangian} for vector field sector.
First we parameterize the primary vector fields as 
\begin{eqnarray}
B_{+} & = & \partial_{+} \frac{\phi}{m} +
\frac{1}{\partial_{-}} * \left[ m \phi + \partial_i C_i +
\frac{1}{\partial_{-}}* j^{+}\right]\ ,\\   
B_{-} & = & \partial_{-} \frac{\phi}{m}\ ,\\
B_{i} & = & \partial_{i} \frac{\phi}{m} + C_i\ ,
\end{eqnarray}
where new fields $\phi$ and $C_i$ satisfy equations of motion 
\begin{eqnarray}
\left( 2 \partial_{+} \partial_{-} - \Delta_\perp + m^2\right) \phi &
= & - \frac{1}{m} \left(\partial_{+} j_{-} + \partial_{-} j_{+} -
\partial_i j_i \right) - m \frac{1}{\partial_{-}} * j^{+}\ ,
\label{massQEDmassQEDeqphi} \\  
\left( 2 \partial_{+} \partial_{-} - \Delta_\perp + m^2\right) C_i &
= & j^i + \partial_i \frac{1}{\partial_{-}} *
j^{+}.\label{massQEDmassQEDeqCi} 
\end{eqnarray}
When this parameterization is introduced into the primary
Lagrangian (\ref{massQEDLagrmassB}), we end up with the
expression 
\begin{eqnarray}
{\cal L}^{eff}_{B} & = & \partial_{+} C_i \partial_{-} C_i - \frac 1 2
\left( \partial_i C_j\right)^2 - \frac{m^2}{2} C_i^2\nonumber\\
& + & \partial_{+} \phi \partial_{-} \phi - \frac 1 2
\left( \partial_i \phi\right)^2 - \frac{m^2}{2} \phi^2 - \frac
1 2 \left( \frac{1}{\partial_{-} }*j^{+}\right)^2 +
\frac{\partial_{-} \phi}{m} j^{-} + \frac{\partial_{+} \phi}{m}
j^{+} + \frac{\partial_{i} \phi}{m} j^{i}\nonumber\\
& + & C_i j^i - (m \phi + \partial_i C_i)\frac{1}{\partial_{-}}
* j^{+}\ ,
\end{eqnarray}
where evidently fields $C_i$ and $\phi$ are independent
modes. We stress that now there are no constraints in this
effective Lagrangian and one can reintroduce the fermion kinetic
terms and substitute external currents by the fermion currents.
Next the canonical quantization leads to the same
results as in \cite{Yan1973}, especially the mixed commutators
for the $\phi$ field and the fermion  fields are nonzero.

\subsection{Gross-Treiman model}

More than 25 years ago, Gross and Treiman have proposed a
modified model for describing the interaction of massive vector
fields with fermions \cite{GrossTreiman1971}. They have been specially
interested in the commutators with a smoother behaviour near
the light-cone, and their model can be described by the following
modified Lagrangian:
\begin{eqnarray}
{\cal L}_{GT} & = & (\partial_{+} V_i - \partial_i V_{+}) (\partial_{-}
V_i - \partial_i V_{-}) + \frac 1 2 (\partial_{+} V_{-} -
\partial_{+} V_{-})^2 - \frac 1 4 (\partial_{i} V_j - \partial_j
V_{i})^2 \\
& + & \frac{m^2}{2} \left(V_{\mu} - \frac{\partial_\mu
\phi}{m}\right)\left(V^{\mu} - \frac{\partial^\mu
\phi}{m}\right)  + \bar{\psi}(i
\partial_\mu \gamma^\mu - e V_\mu \gamma^\mu + M)\psi - \frac 1
2 \partial_\mu \phi \partial^\mu \phi + \frac{m^2}{2}
\phi^2\nonumber\ ,
\end{eqnarray}
where another scalar field $\phi$ has been added with
apparently incorrect signs in quadratic terms. Now we can take
the sector of vector and scalar 
fields by omitting the fermion dynamics
\begin{eqnarray}
{\cal L}_{GT}^{V\phi} & = & (\partial_{+} V_i - \partial_i V_{+})
(\partial_{-} V_i - \partial_i V_{-}) + \frac 1 2 (\partial_{+}
V_{-} - \partial_{+} V_{-})^2 - \frac 1 4 (\partial_{i} V_j -
\partial_j V_{i})^2 \nonumber\\
& + & \frac{m^2}{2} \left(V_{\mu} - \frac{\partial_\mu \phi}{m}
\right)\left(V^{\mu} - \frac{\partial^\mu \phi}{m}\right) +
V_\mu j^\mu - \frac 1 2 \partial_\mu \phi \partial^\mu \phi +
\frac{m^2}{2} \phi^2\ , \label{massQEDLagr'GrossTreimann}
\end{eqnarray}
and then  we change variables
\begin{equation}
V_{\mu} = B_\mu + \frac{\partial_\mu \phi}{m}.
\end{equation}
We notice that the Lagrangian (\ref{massQEDLagr'GrossTreimann})
splits into two parts ${\cal L}^{V\phi}_{GT} = {\cal L}_B +
{\cal L}_\phi$, where ${\cal L}_B$ is the previously analysed
Lagrangian (\ref{massQEDLagrmassB}) and ${\cal L}_\phi$
describes solely negative metric scalar field
\begin{equation}
{\cal L}_\phi = - \frac 1 2 \partial_\mu \phi \partial^\mu \phi
+ \frac{m^2}{2} \phi^2 + \frac{\partial_\mu \phi}{m} j^\mu.
\end{equation} 
Therefore we need to analyze only this last contribution of $\phi$
fields here and then adopt previous results for $B_\mu$ fields.
The Euler-Lagrange equations 
\begin{equation}
(2 \partial_{+} \partial_{-} - \Delta_\perp + m^2) \phi = 
\frac{1}{m} (\partial_{+} j^{+}+ \partial_{-} j^{-} +
\partial_{i} j^{i}), \label{massQEDELeqGrTr}
\end{equation}
suggest the change of field variables 
\begin{equation}
\phi = \widetilde{\phi} + \frac{1}{2m}
\frac{1}{\partial_{-}}*j^{+}\ ,
\end{equation}
which removes $\partial_{+} j^{+}$ term from the equation of motion
\begin{equation}
(2 \partial_{+} \partial_{-} - \Delta_\perp + m^2) \widetilde{\phi} = 
\frac{1}{m} \left((\Delta_\perp - m^2)\frac{1}{2\partial_{-}}* j^{+}+
\partial_{-} j^{-} + \partial_{i} j^{i}\right) \label{massQEDELeqmodGrTr}.
\end{equation}
If we introduce the canonical momentum for the $\phi$ field 
\begin{equation}
\pi_{\phi} = - \partial_{-} \phi + \frac{j^{+}}{m}\ ,
\end{equation}
then the canonical Hamiltonian density is 
\begin{eqnarray}
{\cal H}_{\phi}^{can} & = & \pi_{\phi}\partial_{+} \phi - {\cal
L}_\phi  =  - \frac 1 2 (\partial_i \phi)^2 - \frac{m^2}{2} \phi^2 +
\frac{\phi}{m} (\partial_{-}j^{-} + \partial_{i}j^{i})\nonumber\\
& =&  - \frac 1 2 \left(\partial_i \widetilde\phi +
\frac{\partial_i}{m} \frac{1}{2\partial_{-}} * j^{+}\right)^2 -
\frac{m^2}{2} \left( \widetilde{\phi} + \frac{1}{m}
\frac{1}{2\partial_{-}}*j^{+} \right)^2 \nonumber
\\& + &  \frac{1}{m}\left(\widetilde{\phi} + \frac{1}{m}
\frac{1}{2\partial_{-}}*j^{+}\right) (\partial_{-}j^{-} +
\partial_{i}j^{i}) \label{massQEDmasstildephieq} \ ,
\end{eqnarray}
and we find the Dirac bracket at LF
\begin{equation}
2 \partial_{-}^x \left\{ \widetilde{\phi}(x^{+}, \vec{x}),
\widetilde{\phi}(x^{+}\vec{y})\right\}_{DB} = \delta^3(\vec{x} - \vec{y}).
\end{equation}
At last we may add the above results to  those found
previously and we write down the density of effective
Hamiltonian as
\begin{eqnarray}
{\cal H}^{eff}_{GT} & = & {\cal H}^{eff}_{B}+ {\cal
H}^{can}_{\phi} = - \frac 1 2 \left(\partial_i \widetilde\phi
\right)^2 - \frac{m^2}{2}  \widetilde{\phi}^2 
+  \frac 1 2 \Pi^2 +  \frac 1 4 
\left(\partial_i \widetilde{B}_j - \partial_j \widetilde{B}_i\right)^2 +
\frac{m^2}{2}  \widetilde{B}_i^2
\nonumber\\ 
& - & 
 j^{-} \partial_{-}\left[ \frac{1}{\Delta_\perp - m^2}*
(\Pi + \partial_i \widetilde{B}_i) +
\frac{ \widetilde{\phi}}{m} \right] \nonumber\\ 
& - &  j^i \left( \widetilde{B}_i +
\partial_{i} \frac{1}{m}\widetilde{\phi} \right) 
- j^{+} \frac{1}{2\partial_{-}} * \left[ \frac{\Delta_\perp -
m^2}{m} \widetilde{\phi} - \Pi + \partial_i \widetilde{B}_i
\right] \ ,
\end{eqnarray}
and nonzero Dirac brackets at LF for all independent field variables
\begin{eqnarray}
2 \partial^x_{-} \left\{\Pi(x^{+}, \vec{x}),
\Pi(x^{+}, \vec{y})\right\}_{DB} & = & (\Delta_\perp -
m^2)\delta^{3}(\vec{x} - \vec{y})\label{massQEDDBmassPiPi} \ ,\\
2 \partial^x_{-} \left\{\widetilde{B}_i(x^{+}, \vec{x}),
\widetilde{B}_j(x^{+}, \vec{y})\right\}_{DB} & = & - \left( \delta_{ij} -
\frac{\partial_i \partial_j}{m^2}\right) \delta^{3}(\vec{x} -
\vec{y}) \label{massQEDDBmasstildeBitildeBi}\ ,\\
2 \partial_{-}^x \left\{ \widetilde{\phi}(\vec{x}),
\widetilde{\phi}(\vec{y})\right\}_{DB} & = & \delta(\vec{x} - \vec{y}).
\label{massQEDDBmasstildephitildephi}
\end{eqnarray}
Next, we can easily check the consistency of this effective
Hamiltonian with the starting point in
(\ref{massQEDLagr'GrossTreimann}) - in both cases currents
couple linearly to $V_\mu$ fields\footnote{ In the case of
$V_{+}$ one needs to use the equation of motion
(\ref{massQEDELeqmodGrTr}) in order to have the term
$\partial_{+} \widetilde{\phi}$.} 
\begin{eqnarray}
V_i & = & {B}_i + \partial_i \frac{{\phi}}{m} =
 \widetilde{B}_i + \partial_i \frac{\widetilde{\phi}}{m} \ ,\\
V_{-} & =& {B}_{-} + \partial_{-}
\frac{{\phi}}{m} =   
\partial_{-}\left[ \frac{1}{\Delta_\perp - m^2}*
(\Pi + \partial_i \widetilde{B}_i) +
\frac{ \widetilde{\phi}}{m} \right]\ , \\
V_{+}  & = & {B}_{+} + \partial_{+}
\frac{{\phi}}{m} =  \frac{1}{2\partial_{-}} * \left[ \frac{\Delta_\perp -
m^2}{m} \widetilde{\phi} - \Pi + \partial_i \widetilde{B}_i
\right].
\end{eqnarray}
From these results we expect that the above structure of brackets will
survive also in the interacting theory when complete dynamics
of  fermions will be reintroduced. In order to achieve
this aim most easily, we build here the equivalent Lagrangian  
\begin{eqnarray}
\widetilde{\cal L}^{V \phi}_{GT} & = & \partial_{+} \Pi
\frac{1}{\Delta_\perp - 
m^2} * \partial_{-} \Pi + \partial_{+} \widetilde{B}_i \left(
\delta_{ij} - \partial_i \partial_j \frac{1}{\Delta_\perp -
m^2} * \right)\partial_{-} \widetilde{B}_i \nonumber\\
& - & \partial_{+} \widetilde{\phi} \partial_{-} \widetilde{\phi} -
{\cal H}^{eff}_{GT} 
\end{eqnarray}
which is nonlocal at LF but directly leads to 
Dirac brackets (\ref{massQEDDBmassPiPi},
\ref{massQEDDBmasstildeBitildeBi}, 
\ref{massQEDDBmasstildephitildephi}) and equations of motion
(\ref{massQEDmassPieq}, \ref{massQEDmasstildeBieq},
\ref{massQEDmasstildephieq}) are generated as the Euler-Lagrange
equations. 

\subsubsection{Interaction with fermion fields}

Having found the effective equivalent prescription of vector and
scalar fields, we can add fermion kinetic terms 
and study the complete  Gross-Treiman model 
\begin{eqnarray}
{\cal L}^{eff}_{GT} & = & \partial_{+} \Pi \frac{1}{\Delta_\perp -
m^2} * \partial_{-} \Pi + \partial_{+} \widetilde{B}_i \left(
\delta_{ij} - \partial_i \partial_j \frac{1}{\Delta_\perp -
m^2} * \right)\partial_{-} \widetilde{B}_i - \partial_{+}
\widetilde{\phi} \partial_{-} \widetilde{\phi} \nonumber\\
& + &  \frac 1 2 \left(\partial_i \widetilde\phi \right)^2 +
\frac{m^2}{2} \widetilde{\phi}^2 - \frac 1 2 \Pi^2 - \frac 1 4 
\left(\partial_i \widetilde{B}_j - \partial_j \widetilde{B}_i\right)^2 -
\frac{m^2}{2} \widetilde{B}_i^2 + i \sqrt{2}\psi^{\dag}_{+}
\partial_{+} \psi_{+} \nonumber\\ 
& + &  i\sqrt{2}\psi^{\dag}_{-} \partial_{-} \psi_{-} - M
\left(\psi_{-}^{\dag} \gamma^0 \psi_{+}+ \psi_{+}^{\dag}
\gamma^0 \psi_{-}\right) - e \sqrt{2} \psi_{-}^{\dag} \psi_{-}
\partial_{-}\left[ \frac{1}{\Delta_\perp - m^2}* 
(\Pi + \partial_i \widetilde{B}_i) +
\frac{ \widetilde{\phi}}{m} \right] \nonumber\\ 
& - & e \left(\psi_{-}^{\dag} \alpha^i \psi_{+}+
\psi_{+}^{\dag} \alpha^i \psi_{-}\right) \left( \widetilde{B}_i
+ \partial_{i} \frac{1}{m}\widetilde{\phi} \right) - e \sqrt{2}
\psi_{+}^{\dag} \psi_{+} \frac{1}{2\partial_{-}} * \left[
\frac{\Delta_\perp - m^2}{m} \widetilde{\phi} - \Pi +
\partial_i \widetilde{B}_i \right] \nonumber\\ 
&+ & \psi_{-}^{\dag} \alpha^i \partial_i  \psi_{+}+
\psi_{+}^{\dag} \alpha^i \partial_i \psi_{-}\ ,
\end{eqnarray}
where we have introduced two components  of
fermion fields $\psi_{\pm} = \Lambda_{\pm}
\psi$\footnote{See Appendix \ref{Diracmatappend} for the complete
LF notation of the Dirac matrices and the projection
operators $\Lambda_{\pm}$.}. The
components $\psi_{-}$ and  $\psi^{\dag}_{-}$ of fermion fields
satisfy the non-dynamical Euler-Lagrange equations
\begin{eqnarray}
\sqrt{2} \left[i \partial_{-} - e (\partial_{-}\Phi)\right]
\psi_{-} & = &  \xi \equiv  \left[ M \gamma^0 - i \partial_i
\alpha^i + e \alpha^i \left(B_i + \frac{\partial_i}{m}
\widetilde{\phi}\right) \right]\psi_{+}\ ,\\
\sqrt{2} \left[- i \partial_{-} - e (\partial_{-}\Phi)
\right] \psi_{-}^{\dag} & = &  \xi^{\dag} \equiv  \psi_{+}
\left[ M \gamma^0 + i \stackrel{\leftarrow}{\partial_i} \alpha^i 
+ e \alpha^i \left(B_i + \frac{\partial_i}{m}
\widetilde{\phi}\right) \right]\ ,\\
\Phi & = &   \left[ \frac{1}{\Delta_\perp - m^2}* (\Pi +
\partial_i \widetilde{B}_i) + \frac{\widetilde{\phi}}{m}\right]\ ,
\end{eqnarray}
where additional notations $\xi$, $\xi^{\dag}$ and $\Pi$
are introduced for convenience and clarity. Next we can
write the formal solutions for these non-dynamical fermion
components, 
\begin{eqnarray}   
\psi_{-} & = &  \frac{1}{\sqrt{2}} e^{-ie \Phi}\frac{1}{i
\partial_{-}}*\left(e^{ie \Phi}\xi\right)\ ,\\  
\psi_{-}^{\dag} & = &  \frac{1}{\sqrt{2}} 
\left(\xi^{\dag} e^{-ie \Phi}\right) *\frac{1}{i
\partial_{-}} e^{ie \Phi}.
\end{eqnarray}
In this way there are only dynamical fields left and their
quantization is straightforward. The Hamiltonian density has
the following form: 
\begin{eqnarray}
{\cal H}^{eff}_{GT} & = & -\frac 1 2 \left(\partial_i
\widetilde\phi \right)^2 - \frac{m^2}{2} \widetilde{\phi}^2 +
\frac 1 2 \Pi^2 + \frac 1 4 \left(\partial_i \widetilde{B}_j -
\partial_j \widetilde{B}_i\right)^2 + \frac{m^2}{2}
\widetilde{B}_i^2 \nonumber\\ 
& + & \frac{1}{\sqrt{2}} \xi^{\dag} e^{- ie \Phi}
\frac{1}{i \partial_{-}}*\left(e^{ie \Phi}\xi\right) 
+ e \sqrt{2} \psi_{+}^{\dag} \psi_{+} \frac{1}{2\partial_{-}} *
\left[ \frac{\Delta_\perp - m^2}{m} \widetilde{\phi} - \Pi +
\partial_i \widetilde{B}_i \right],
\end{eqnarray}
while the non-vanishing (anti)commutators at LF have the expected
form 
\begin{eqnarray}
2 \partial^x_{-} \left[\Pi(x^{+}, \vec{x}),
\Pi(x^{+}, \vec{y})\right] & = & i(\Delta_\perp -
m^2)\delta^{3}(\vec{x} - \vec{y})\label{massQEDcommassPiPi}\ ,\\
2 \partial^x_{-} \left[\widetilde{B}_i(x^{+}, \vec{x}),
\widetilde{B}_j(x^{+}, \vec{y})\right] & = & - i \left(
\delta_{ij} - \frac{\partial_i \partial_j}{m^2}\right)
\delta^{3}(\vec{x} - \vec{y})
\label{massQEDcommasstildeBitildeBi}\ ,\\ 
2 \partial_{-}^x \left[ \widetilde{\phi}(x^{+}, \vec{x}),
\widetilde{\phi}(x^{+}, \vec{y})\right ] & = & i\delta(\vec{x}
- \vec{y}) \label{massQEDcommasstildephitildephi}\ ,\\
\left\{ \psi_{+}^{\dag}(x^{+}, \vec{x}), \psi_{+}(x^{+},
\vec{y})\right \} & = & \frac{1}{\sqrt{2}} \Lambda_{+}
\delta(\vec{x} - \vec{y}).
\end{eqnarray}
The quantum theory defined above is formulated in a natural way
in the Heisenberg representation, where the whole dynamics of
the system is connected with the quantum field operators. For
perturbative calculations the interaction picture is a more
convenient choice, but here we will not proceed in this
direction, leaving this important issue to a more physically
relevant models which will be discussed later. Rather we end up
our discussion of the present model with pointing out its two
interesting properties. The first one is connected with the
scalar field $\widetilde{\phi}$ which evidently interacts with
fermion fields 
\begin{eqnarray}
\hspace{-25pt}
(2 \partial_{+} \partial_{-} - \Delta_\perp + m^2) \widetilde{\phi} &=& 
- \frac{e}{m} (\Delta_\perp - m^2)\frac{1}{2\partial_{-}}*
\left(\sqrt{2} \psi_{+}^{\dag} \psi_{+} \right) \nonumber\\
&-&  \frac{ie\sqrt{2}}{2m} \left[\xi^{\dag}
e^{-ie \Phi}  \frac{1}{i \partial_{-}}*\left(e^{ie \Phi}\xi
\right) - \left(\xi^{\dag}e^{-ie \Phi}\right)* \frac{1}{i
\partial_{-}} e^{ie \Phi} \xi \right]\nonumber\\
& - & \frac{e\sqrt{2}}{2m} \partial_i\left[\psi^{\dag}_{+}
e^{-ie \Phi} \alpha^i \frac{1}{i \partial_{-}}*\left(e^{ie \Phi}\xi
\right) + \left(\xi^{\dag}e^{-ie \Phi}\right)* \frac{1}{i
\partial_{-}} e^{ie \Phi}\alpha^i \psi_{+} \right]
\label{massQEDphitildeeq}. 
\end{eqnarray}
However, one can define another field 
\begin{equation}
\phi = \widetilde{\phi} - \frac{e}{m} \sqrt{2} \frac{1}{2\partial_{-}} *
\left( \psi_{+}^{\dag} \psi_{+} \right)
\end{equation}
which, due to equations of the fermion fields
\begin{eqnarray}
i \partial_{+} \psi_{+} & = &  
 e   \psi_{+} \frac{1}{2\partial_{-}} *
\left[ \frac{\Delta_\perp - 
m^2}{m} \widetilde{\phi} - \Pi + \partial_i \widetilde{B}_i
\right]\nonumber\\
& + & \frac 1 2 \left[ M \gamma^0 - i \partial_i \alpha^i + e \alpha^i
\left(B_i + \frac{\partial_i}{m} \widetilde{\phi}\right) \right]
e^{-ie \Phi} \frac{1}{i \partial_{-}}*\left(e^{ie \Phi}\xi
\right), \\ 
-i \partial_{+} \psi_{+}^{\dag} & = &  e \psi_{+}^{\dag}
\frac{1}{2\partial_{-}} * \left[ \frac{\Delta_\perp - m^2}{m}
\widetilde{\phi} - \Pi + \partial_i \widetilde{B}_i \right]\nonumber\\
& + &\frac 1 2 \left(\xi^{\dag}e^{-ie \Phi} \right)* \frac{1}{i
\partial_{-}} e^{ie \Phi} \left[ M \gamma^0 + i
\stackrel{\leftarrow}{\partial_i} \alpha^i + e \alpha^i
\left(B_i + \frac{\partial_i}{m} \widetilde{\phi}\right) \right],
\end{eqnarray}
will satisfy the free field equation of motion
\begin{equation}
(2 \partial_{+} \partial_{-} - \Delta_\perp + m^2) {\phi} = 0.
\end{equation}
However, the price for such a free evolution has to be paid at
the level of LF commutators where now the non-vanishing relations
with $\phi$ are
\begin{eqnarray}
2 \partial^x_{-} \left[\phi(x^{+}, \vec{x}),
\phi(x^{+}, \vec{y})\right] & = & i\delta^{3}(\vec{x} -
\vec{y})\label{massQEDcommassphiphi}\ ,\\ 
2 \partial^x_{-} \left[{\phi}(x^{+}, \vec{x}),
\psi_{+}(x^{+}, \vec{y})\right] & = & \frac{e}{m}
\psi_{+}(x^{+}, \vec{x})\delta^{3} (\vec{x} - \vec{y})
\label{massQEDcommassphipsi}\ ,\\ 
2 \partial^x_{-} \left[{\phi}(x^{+}, \vec{x}),
\psi_{+}^{\dag}(x^{+}, \vec{y})\right] & = & - \frac{e}{m}
\psi_{+}^{\dag}(x^{+}, \vec{x}) \delta^{3}
(\vec{x} - \vec{y}) \label{massQEDcommassphipsidag}.
\end{eqnarray}
The second property is connected with the components of 
the vector $V_\mu$: 
\begin{eqnarray}
V_i & = & \widetilde{B}_i + \partial_i \frac{\widetilde{\phi}}{m}\ , \\
V_{-} & =& \partial_{-}\left[ \frac{1}{\Delta_\perp - m^2}*
(\Pi + \partial_i \widetilde{B}_i) +
\frac{ \widetilde{\phi}}{m} \right]\ ,\\
V_{+} & = & \frac{1}{2\partial_{-}} * \left[ \frac{\Delta_\perp
- m^2}{m} \widetilde{\phi} - \Pi + \partial_i \widetilde{B}_i \right]\ ,
\end{eqnarray}
which satisfy equations of motion 
\begin{eqnarray}
(2 \partial_{+} \partial_{-} - \Delta_\perp + m^2) V_{+} & = & 
e \sqrt{2} \psi_{+}^{\dag}\psi_{+}\ ,\\
(2 \partial_{+} \partial_{-} - \Delta_\perp + m^2) V_{-} & = & 
\frac{e}{\sqrt{2}}
\left(\xi^{\dag}e^{-ie \Phi} \right)* \frac{1}{i \partial_{-}}
\frac{1}{i \partial_{-}}*\left(e^{ie \Phi} \xi \right),\\ 
(2 \partial_{+} \partial_{-} - \Delta_\perp + m^2) V_i & = & 
\frac{e}{\sqrt{2}}\left[ \psi^{\dag}_{+} \alpha^i e^{-ie \Phi}
\frac{1}{i \partial_{-}}*\left(e^{ie \Phi}\xi\right) + \left
(e^{-ie \Phi}\xi^{\dag} \right)*\frac{1}{i \partial_{-}} e^{ie
\Phi}\alpha^i \psi_{+}\right]\nonumber ,\\ 
&&
\end{eqnarray}
and have the only non-vanishing commutation relations 
\begin{equation}
2 \partial^x_{-} \left[V_\mu(x^{+}, \vec{x}), V_\nu(x^{+},
\vec{y})\right] = - i g_{\mu \nu}\delta^{3}(\vec{x} - 
\vec{y})\label{massQEDcommmassCmuCnu}.
\end{equation}
Therefore $V_\mu$ can be interpreted as the massive vector
field in the so-called Feynman gauge. We see that modification of
the massive theory, which solves the problems connected with
singular behaviour of fields on LF, can be interpreted  in terms
of gauge condition, though for non-vanishing mass $m^2 \neq 0$,
this model is {\em not a true gauge} theory. Therefore we
suggest that also other modifications by means of truly gauge
fixing terms may be worth to be studied in the canonical LF
formulation of massive electrodynamics.

\section{Lorentz covariant gauge}\label{rozdzialcechLor}

\setcounter{equation}{0}

In the previous section we have learned that addition of the
extra scalar degrees of freedom can change LF commutators to
the form proper for the Feynman gauge, which is a special case
of the Lorentz covariant gauges. General covariant gauge is
implemented into 
Lagrangian by means of the scalar Lagrange multiplier field
$\Lambda$ and it takes the form $\partial_\mu B^\mu = - \alpha
\Lambda $. For $\alpha = 0$ it becomes the Lorentz gauge and
for $\alpha = 1$ it becomes the Feynman gauge. \\ 
As previously, we start with the classical fields theory 
defined by the Lagrangian density
\begin{eqnarray}
{\cal L}_{mass}^{cov} & = & \left(\partial_{+} {B}_i -
\partial_i {B}_{+}\right)\left(\partial_{-} {B}_i - \partial_i
{B}_{-}\right) + \frac 1 2 \left(\partial_{+} {B}_{-} -
\partial_{-}{B}_{+}\right)^2 - \frac 1 4 \left( \partial_i
{B}_j - \partial_j {B}_i\right)^2 \nonumber\\
& + & m^2 \left( B_{-} B_{+} - \frac 1 2 B_i^2 \right)
 +  \bar{\psi} \left( i \gamma^\mu \partial_\mu - e \gamma^\mu
{B}_{\mu} - M\right) \psi \nonumber\\
& + & {\Lambda}\left( \partial_{+} B_{-} + \partial_{-} B_{+} -
\partial_i B_i\right) + \frac{\alpha}{2} \Lambda^2. 
\label{procaQEDmassLagrcov}
\end{eqnarray}
We expect that the canonical structure of vector fields $B_\mu$
is independent of the fermion fields, therefore we will study the
sector of vector fields with external arbitrary currents
$j^\mu$ first.

\subsection{Vector field sector}

Omitting the fermion kinetic terms in (\ref{procaQEDmassLagrcov})
and inserting $j^\mu$ in the place of $e \bar{\psi}\gamma
\psi$ we obtain 
\begin{eqnarray}
{\cal L}_{jmass}^{cov} & = &
\left(\partial_{+} {B}_i - \partial_i
{B}_{+}\right)\left(\partial_{-} {B}_i - \partial_i
{B}_{-}\right)  +  \frac 1 2 \left(\partial_{+} {B}_{-} -
\partial_{-}{B}_{+}\right)^2 - \frac 1 4 \left( \partial_i {B}_j
- \partial_j {B}_i\right)^2 \nonumber\\
& + & m^2 \left( B_{-} B_{+} - \frac 1
2 B_i^2 \right) +  {B}_{\mu} {j}^{\mu}  +
\Lambda \left( \partial_{+} B_{-} + \partial_{-} B_{+} -
\partial_i B_i\right) + \frac{\alpha}{2} \Lambda^2\ , 
\end{eqnarray}
and next we generate the Euler-Lagrange equations for all field
variables 
\begin{eqnarray}
 \left( 2 \partial_{+} \partial_{-} - \Delta_\perp + m^2 \right)
{B}_{-} & = & (1 - \alpha)\partial_{-} \Lambda - j^{+} \ ,
\label{procaQEDmassELeqAm}\\ 
 \left( 2 \partial_{+} \partial_{-} - \Delta_\perp  + m^2\right)
{B}_{+} & = & (1 - \alpha)\partial_{+} \Lambda - j^{-}\ ,
\label{procaQEDmassELeqAp1}\\ 
 \left( 2 \partial_{+} \partial_{-} - \Delta_\perp + m^2 \right)
{B}_i & = & (1 - \alpha)\partial_i \Lambda + j^i\ ,
\label{procaQEDmassELAi} \\ 
\partial_{+} {B}_{-} & = & - \partial_{-} B_{+} + \partial_{i}
B_i - \alpha \Lambda \label{procaQEDcovgaugecon}.
\end{eqnarray}
The consistency condition for these equations has the form of
a dynamical equation for 
$\Lambda$
\begin{equation}
\left( 2 \partial_{+} \partial_{-} - \Delta_\perp + \alpha m^2
\right) \Lambda = \partial_{+} j^{+}+ \partial_{-} j^{-} +
\partial_{i} j^{i}\label{procaQEDeqLambda}\ ,
\end{equation}
and we see that in order to have no imaginary mass tachyon, we
must keep the parameter $\alpha \geq 0$. Just like in the
previous section, we have two different equations with
$\partial_{+} B_{-}$ (\ref{procaQEDmassELeqAm}) and
(\ref{procaQEDcovgaugecon}) so there is a constraint 
\begin{equation}
2\partial_{-} \left( \partial_j B_j - \alpha \Lambda -
\partial_{-} B_{+} \right) = (\Delta_\perp - m^2) B_{-} + ( 1 -
\alpha) \partial_{-} \Lambda - j^{+}.\label{procaQEDmass2consist} 
\end{equation}
If we define the canonical momentum conjugated to the field
$B_{-}$ 
\begin{equation}
{\Pi}^{-} = \partial_{+}{B}_{-} - \partial_{-} {B}_{+} +
\Lambda \label{procaQEDmassdefPiBm}
\end{equation}
then,  using  the gauge condition (\ref{procaQEDcovgaugecon})
we obtain another constraint 
\begin{equation}
2 \partial_{-} B_{+}  = \partial_i B_i - \Pi^{-} + (1 -
\alpha) \Lambda\label{procaQEDconstrAp}.
\end{equation}
In the dynamical equation for $\Lambda$ (\ref{procaQEDeqLambda})
there is the $\partial_{+} j^{+}$ term and in order to remove it,
we redefine the Lagrange multiplier field 
\begin{equation}
\lambda = \Lambda - \frac{1}{2\partial_{-}}* j^{+}.
\end{equation}
In this manner we have selected the set of fields $(\Pi^{-}$, $B_i$,
$\lambda$) which satisfy dynamical equations of motion
\begin{eqnarray}
(2 \partial_{+} \partial_{-} - \Delta_\perp + \alpha
m^2){\lambda} & = & (\Delta_\perp - \alpha m^2)
\frac{1}{2\partial_{-}}* {j}^{+} + \partial_{-} 
j^{-} + \partial_i {j}^{i}\label{procaQEDHameqlambda}\ ,\\
(2 \partial_{+} \partial_{-} - \Delta_\perp + m^2){\Pi}^{-} & =
&  m^2 (1 - \alpha) \left(\lambda + \frac{1}{2\partial_{-}}
\ast j^{+}\right) + 2 \partial_{-} {j}^{-} + \partial_i
{j}^{i}\label{procaQEDHameqPim}\ ,\\ 
(2 \partial_{+} \partial_{-} - \Delta_\perp + m^2) {B}_{i} & = &
(1 - \alpha)\partial_i \lambda + {j}^i + \frac{1- \alpha}{2}
\frac{1}{\partial_{-}}*\partial_i j^{+}\ ,
\label{procaQEDHameqAi}
\end{eqnarray} 
so they are independent canonical variables. Also we have
dependent fields $B_{-}$ and $B_{+}$: 
 \begin{eqnarray}
B_{-} & = & \partial_{-} \frac{1}{\Delta_\perp - m^2} \ast
\left( \Pi^{-} + \partial_i B_i - 2 \lambda\right) 
\label{procaQEDmodconstrAm}\ ,\\ 
 B_{+} & = & \frac{1}{2 \partial_{-}} \ast \left[
\partial_i B_i - \Pi^{-} + (1 - \alpha) \lambda + \frac{1 -
\alpha}{2} \frac{1}{\partial_{-}}* j^{+} \right]
\label{procaQEDmodconstrAp}.
\end{eqnarray}
Though the other canonical momenta 
apparently represent primary constraints (according to Dirac's
nomenclature) 
\begin{eqnarray} 
&&{\Pi}^{i}(x) - \partial_{-} {A}_{i}(x) + \partial_{i}
{A}_{-}(x) \approx 0 \ ,\\ 
&&{\Pi}^{+}(x) \approx 0\ ,\\
&&{\Pi}_\lambda(x) \approx 0\ ,
\end{eqnarray}
we notice that they are absent form the canonical Hamiltonian
density\footnote{We have used relations
(\ref{procaQEDmodconstrAm}) and (\ref{procaQEDmodconstrAp}) to
remove dependent variables from the Hamiltonian.}
\begin{eqnarray}
{\cal H}_{can} & = & (\partial_{+} {B}_{-}) {\Pi}^{-} +
(\partial_{+} {B}_{i}) {\Pi}^i - {\cal L} = \frac 1 2
\left({\Pi}^{-} - \lambda - \frac{1}{2\partial_{-}}* j^{+}\right
)^2 + \frac 1 4 \left( \partial_i {B}_j - \partial_j
{B}_i\right)^2 + \frac{m^2}{2} B_i^2 \nonumber \\ 
& + & \left(\lambda + \frac{1}{2\partial_{-}}* j^{+} \right)
\partial_i B_i - {j}^{-} \frac{1}{  \Delta_\perp - m^2} * 
\partial_{-}\left( \Pi^{-} + \partial_i B_i - 2
\lambda\right)  - {B}_i {j}^i \\
& - & \frac{\alpha}{2} \left(\lambda + \frac{1}{2\partial_{-}}*
j^{+} \right)^2\nonumber.
\end{eqnarray}
Next we construct Eqs. (\ref{procaQEDHameqlambda}),
(\ref{procaQEDHameqPim}), (\ref{procaQEDHameqAi}) as
Hamilton equations by imposing the following Dirac
brackets\footnote{Though we have 
not used Dirac's procedure of quantization, these brackets are
evidently consistent with all constraints, therefore we call
them {\em Dirac brackets}.} on independent variables at LF: 
\begin{eqnarray}
2\partial^x_{-}\left \{{B}_{i}(\vec{x}), {\Pi}^{-}(\vec{y})
\right \}_{DB} & = & \partial_i^x \delta^3(\vec{x} - \vec{y})\ ,\\
2\partial_{-}^x\left \{{\Pi}^{-}(\vec{x}), {\Pi}^{-}(\vec{y})
\right \}_{DB} & = & \Delta_\perp \delta^3(\vec{x} - \vec{y})\ ,\\
2 \partial_{-}^x \left \{{B}_{i}(\vec{x}), {B}_j (\vec{y})
\right \}_{DB} & = & - \delta_{ij} \delta^3(\vec{x} - \vec{y})\
,\\
2 \partial_{-}^x \left \{ \lambda(\vec{x}), {B}_j (\vec{y})
\right \}_{DB} & = & - \partial_{i}^x \delta^3(\vec{x} -
\vec{y})\ ,\\ 
2 \partial_{-}^x \left \{ \lambda(\vec{x}), {\lambda} (\vec{y})
\right \}_{DB} & = & m^2 \delta^3(\vec{x} -
\vec{y})\ ,\\
2 \partial_{-}^x \left \{ \lambda(\vec{x}), \Pi^{-} (\vec{y})
\right \}_{DB} & = & m^2 \delta^3(\vec{x} -
\vec{y})\ ,
\end{eqnarray}
while all other brackets vanish. The structure of these
brackets evidently indicates that our independent canonical
variables are not independent modes yet. Therefore as
independent modes we propose to take the following linear but
nonlocal combinations of fields: 
\begin{eqnarray}
Q & = & \frac{\lambda}{m} \ ,\\
\varphi & = & \frac{m}{\Delta_\perp-m^2} * \left( \Pi^{-} - 2
\lambda + \partial_i B_i\right) - \frac{1}{m} \lambda \ ,\\
C_i & = & B_i - \partial_i \frac{1}{\Delta_\perp-m^2} *
\left(\Pi^{-} - 2 \lambda + \partial_j B_j\right) .
\end{eqnarray}
They satisfy the equations of motion
\begin{eqnarray}
(2 \partial_{+} \partial_{-} - \Delta_\perp + m^2){\varphi} & = & 
- \frac{\Delta_\perp + m^2}{m}  \frac{1}{2 \partial_{-}}*j^{+}
- \frac{1}{m} \left(\partial_{-}j^{-} + \partial_i
{j}^{i}\right) \label{procaQEDHameqphi}\ ,\\ 
(2 \partial_{+} \partial_{-} - \Delta_\perp + \alpha m^2)Q & = &
\frac{\Delta_\perp - \alpha m^2}{m} \frac{1}{2\partial_{-}}*
j^{+} + \frac{1}{m} \left(\partial_{-} j^{-} + \partial_i
{j}^{i} \right)\label{procaQEDHameqlambda2}\ ,\\ 
(2 \partial_{+} \partial_{-} - \Delta_\perp + m^2) {C}_{i} & = &
{j}^i + \partial_i \frac{1}{\partial_{-}}* j^{+}\ ,
 \label{procaQEDHameqCi} 
\end{eqnarray} 
and their  Dirac brackets have a diagonal form 
\begin{eqnarray}
2\partial^x_{-}\left \{ C_i (\vec{x}), C_j(\vec{y})\right
\}_{DB} & = & - \delta_{ij}\delta^3(\vec{x} - \vec{y}) \ ,
\label{procaQEDDBCiCj}\\
2\partial^x_{-}\left \{ \varphi (\vec{x}), \varphi(\vec{y})\right
\}_{DB} & = & -\delta^3(\vec{x} - \vec{y})\ ,
\label{procaQEDDBvarphivarphi}\\
2\partial^x_{-}\left \{ Q (\vec{x}), Q(\vec{y})\right
\}_{DB} & = & \delta^3(\vec{x} - \vec{y}).
\label{procaQEDDBlambdalambda} 
\end{eqnarray}
The Hamiltonian density, when expressed in terms of the
independent modes, looks like
\begin{eqnarray}
{\cal H}^{eff}_{jcov} & = & \frac{1 - \alpha}{2} \left(m Q
+ \frac{1}{2\partial_{-}}* j^{+}\right) ^2 + \frac 1 2
\left(\partial_i C_i + \frac{1}{\partial_{-}}* j^{+}\right) ^2
+ \frac 1 4 \left( \partial_i{C}_j - \partial_j {C}_i\right)^2
+ \frac{m^2}{2} C_i^2 \nonumber \\ 
&- & \frac{1}{2}(\partial_i Q)^2 - \frac{m^2}{2} Q^2
+ \frac 1 2 (\partial_i\varphi)^2  + \frac{m^2}{2}\varphi^2
+ \frac{\varphi + Q}{m} \left(
\partial_{-} j^{-} + \partial_{i} j^{i} \right) - j^i C_i
\nonumber\\ 
&+& \left[\frac{\Delta_\perp - m^2}{m}Q +
\frac{\Delta_\perp + m^2}{m} \varphi - \frac{1}{\partial_{-}}* 
j^{+}\right] \frac{1}{2\partial_{-}}* j^{+}  
 \label{procaQEDeffcanHam}
\end{eqnarray}
and, in spite of the presence of terms quadratic in currents
$j^{+}$, it describes the same physical system as the primary
Lagrangian  with constraints. This allows us to believe that 
the canonical structure for independent modes of vector fields,
that we have just formulated, will survive in the complete
theory with fermion fields. In order to incorporate our
hitherto obtained results into the full interacting theory,
we construct the effective Lagrangian density by means of 
another Legendre transformation 
\begin{eqnarray}
{\cal L}_{eff}^{jmass} & = & \partial_{-} C_i \partial_{+} C_i 
-  \partial_{+} Q \partial_{-} Q
+ \partial_{-} \varphi \partial_{+} \varphi - {\cal
H}^{eff}_{jcov} = \nonumber\\ 
& = & \partial_{-} C_i \partial_{+} C_i - \partial_{+} Q
\partial_{-} Q + \partial_{-} \varphi \partial_{+} \varphi -
\frac{1 - \alpha}{2} \left(m Q + \frac{1}{2\partial_{-}}*
j^{+}\right) ^2 - \frac 1 2 \left(\partial_i C_i +
\frac{1}{\partial_{-}}* j^{+}\right)^2 \nonumber\\ 
& - & \frac 1 4 \left( \partial_i{C}_j - \partial_j {C}_i
\right)^2 - \frac{m^2}{2} C_i^2 + \frac{1}{2}(\partial_i
Q)^2 + \frac{m^2}{2} Q^2 - \frac 1 2 (\partial_i\varphi)^2 -
\frac{m^2}{2}\varphi^2 + j^i C_i \nonumber\\ 
&- & \left[\frac{\Delta_\perp - m^2}{m}Q + \frac{\Delta_\perp +
m^2}{m} \varphi - \frac{1}{\partial_{-}}* j^{+}\right]
\frac{1}{2\partial_{-}}* j^{+} - \frac{\varphi + Q}{m} \left(
\partial_{-} j^{-} + \partial_{i} j^{i} \right). 
\label{procaQEDmasscoveffLagr}
\end{eqnarray}
One can easily check that the brackets
(\ref{procaQEDDBCiCj}-\ref{procaQEDDBlambdalambda}) and the
equations of motion
(\ref{procaQEDHameqphi}-\ref{procaQEDHameqCi}) will follow 
directly from this effective Lagrangian within the
canonical procedure without constraints\footnote{Strictly
speaking there would be trivial primary constraints which
usually appear for covariant relativistic fields at LF.} 

\subsection{Interactions with fermions}

Our next step, from the effective Lagrangian for the gauge
sector to the full interacting theory with fermions, can be done
easily by inserting fermion currents instead of $j^\mu$ and
addition of the fermion kinetic terms
\begin{eqnarray}
{\cal L}_{eff}^{QED} & = & \partial_{-} C_i \partial_{+} C_i -
\partial_{+} Q \partial_{-} Q + \partial_{-} \varphi
\partial_{+} \varphi + i \sqrt{2} \psi^{\dag}_{+} \partial_{+} 
\psi_{+} + \sqrt{2} \psi^{\dag}_{-} \left[ i \partial_{-} - e
\partial_{-} \left(\frac{\varphi + Q}{m}\right) \right]
\psi_{-} \nonumber\\ 
& - & \frac{1 - \alpha}{2} \left(m Q - \frac{1}{2\partial_{-}}*
J^{+}\right) ^2 - \frac 1 2 \left(\partial_i C_i -
\frac{1}{\partial_{-}}* J^{+}\right)^2 \nonumber\\ 
& - & \frac 1 4 \left( \partial_i{C}_j - \partial_j {C}_i
\right)^2 - \frac{m^2}{2} C_i^2 + \frac{1}{2}(\partial_i
Q)^2 + \frac{m^2}{2} Q^2 - \frac 1 2 (\partial_i\varphi)^2 -
\frac{m^2}{2}\varphi^2  \nonumber\\ 
&+ & \left[\frac{\Delta_\perp - m^2}{m}Q + 
\frac{\Delta_\perp + m^2}{m} \varphi + \frac{1}{\partial_{-}}* 
J^{+}\right] \frac{1}{2\partial_{-}}* J^{+} - \xi^{\dag}
\psi_{-} - \psi_{-}^{\dag}\xi \label{procaQEDLagreffQED} \ ,
\end{eqnarray}
where 
\begin{eqnarray}
\xi & = & \left(- i \partial_i \alpha^i + M \beta \right)\psi_{+}
+ e \left [C_i + \partial_i \left ( \frac{\varphi +
Q}{m}\right ) \right] \alpha^i\psi_{+}\ ,\\ 
\xi^{\dag} & = & \left ( i \partial_i\psi^{\dag}_{+} \alpha^i +
M \psi_{+}^{\dag}\beta \right) + e \psi_{+}^{\dag} \alpha^i
\left[C_i + \partial_i \left( \frac{\varphi +Q}{m}\right)
\right] \ ,\\
J^{+} & = & e\sqrt{2} \psi_{+}^{\dag} \psi_{+}.
\end{eqnarray}
As usually in the LF formulation, $\psi_{-}$ and
$\psi^{\dag}_{-}$ fermions are non-dynamical and can be expressed
in terms of dynamical fermions $\psi_{+}$ and $\psi^{\dag}_{+}$
\begin{eqnarray}
\psi_{-} & = & \frac{1}{\sqrt{2}}\ \frac{1}{i\partial_{-} - e
\partial_{-} \left(\frac{\varphi +Q}{m} \right)}*\xi\ ,\\ 
\psi_{-}^{\dag} & = & \frac{1}{\sqrt{2}} \ \xi^{\dag} *
\frac{1}{i\partial_{-} - e \partial_{-} \left( \frac{\varphi
+Q}{m} \right)}\ ,
\end{eqnarray}
and one obtains the nonlocal Hamiltonian density expressed
solely in terms of dynamical independent modes
\begin{eqnarray}
{\cal H}_{total} & = & \frac{1 - \alpha}{2} \left(m Q
- \frac{1}{2\partial_{-}}* J^{+}\right) ^2 + \frac 1 2
\left(\partial_i C_i - \frac{1}{\partial_{-}}* J^{+}\right) ^2
+ \frac 1 4 \left( \partial_i{C}_j - \partial_j {C}_i\right)^2
+ \frac{m^2}{2} C_i^2 \nonumber \\ 
&- & \frac{1}{2}(\partial_i Q)^2 - \frac{m^2}{2} Q^2
+ \frac 1 2 (\partial_i\varphi)^2  + \frac{m^2}{2}\varphi^2
+ \frac{1}{\sqrt{2}} \xi^{\dag} \frac{1}{i \partial_{-} - e
\partial_{-} \left( \frac{\varphi + Q}{m} \right) }* \xi
\nonumber\\ 
&- & \left[\frac{\Delta_\perp - m^2}{m}Q + \frac{\Delta_\perp +
m^2}{m} \varphi + \frac{1}{\partial_{-}}* J^{+}\right]
\frac{1}{2\partial_{-}}* J^{+} .
\label{procaQEDtotalHamilt} 
\end{eqnarray}
As we have expected, the non-vanishing equal $x^{+}$
(anti)commutators have the forms compatible with the former vector
field brackets 
\begin{eqnarray}
2\partial^x_{-}\left [ C_i (\vec{x}), C_j(\vec{y})\right ] & = &
- i \delta_{ij}\delta^3(\vec{x} - \vec{y})
\label{procaQEDCiCjcommut} \ ,\\
2\partial^x_{-}\left [ \varphi (\vec{x}), \varphi(\vec{y})\right ]
& = & - i \delta^3(\vec{x} - \vec{y})
\label{procaQEDphiphicommut}\ ,\\ 
2\partial^x_{-}\left [ Q (\vec{x}), Q(\vec{y})\right ]
& = & i \delta^3(\vec{x} - \vec{y})
\label{procaQEDQQcommut}\ ,\\
\left \{ \psi^{\dag}_{+}(\vec{x}), \psi_{+}(\vec{y}) \right \} & =
& \frac{1}{\sqrt{2}} \Lambda_{+} \delta^3(\vec{x} - \vec{y})\ ,
\label{procaQEDpsipsidagantcom} 
\end{eqnarray}
and, just like in the ET approach to the covariant gauge
condition, there are negative metric excitations
here.\footnote{Opposite signs in (\ref{procaQEDphiphicommut})
and (\ref{procaQEDQQcommut}) allow the combination of
fields $\varphi + Q$ to commute with itself
on LF and this simplifies considerably the consistent
definition of the integral operator $[i \partial_{-} - e
\partial_{-} \left(\frac{\varphi + Q}{m}\right)]^{-1}$.} 
From these relations one
can derive the dynamical equations for the interacting
system\footnote{Here we do not discuss explicitly the ordering
problems and the proper definition of singular
products for noncommuting operators. These very important
aspects of the complete definition of Quantum Field Theory    
remain out of the scope of this work where we are ultimately
interested in the Feynman rules for perturbative calculations.}
\begin{eqnarray}
(2 \partial_{+} \partial_{-} - \Delta_\perp+ m^2){\varphi} & = & 
+ \frac{\sqrt{2} e}{2 m} \partial_{-} \left[ \xi^{\dag} \ast
\frac{1}{ i\partial_{-} - e \left( \frac{\varphi + Q}{m}
\right)}  \frac{1}{ i\partial_{-} - e \left( \frac{\varphi + Q}{m}
\right)} \ast \xi \right]\nonumber\\
& + & \frac{\sqrt{2} e}{2 m} \partial_{i} \left[
\psi_{+}^{\dag} \alpha^i \frac{1}{ i\partial_{-} - e \left(
\frac{\varphi + Q}{m}\right)} \ast \xi + \xi^{\dag}
\ast  \frac{1}{ i\partial_{-} - e \left( \frac{\varphi + Q}{m}
\right)} \alpha^i \psi_{+} \right]\nonumber\\
& + & \frac{\Delta_\perp + m^2}{m} \frac{1}{2\partial_{-}} \ast J^{+}
 \ ,\label{procaQEDeffeqphi}\\ 
(2 \partial_{+} \partial_{-} - \Delta_\perp + \alpha m^2){Q} & = &
- \frac{\sqrt{2} e}{2 m} \partial_{-} \left[ \xi^{\dag} \ast
\frac{1}{ i\partial_{-} - e \left( \frac{\varphi + Q}{m}
\right)}  \frac{1}{ i\partial_{-} - e \left( \frac{\varphi + Q}{m}
\right)} \ast \xi \right]\nonumber\\
& - & \frac{\sqrt{2} e}{2 m} \partial_{i} \left[
\psi_{+}^{\dag} \alpha^i \frac{1}{ i\partial_{-} - e \left(
\frac{\varphi + Q}{m}\right)} \ast \xi + \xi^{\dag}
\ast  \frac{1}{ i\partial_{-} - e \left( \frac{\varphi + Q}{m}
\right)} \alpha^i \psi_{+} \right]\nonumber\\
& - & \frac{\Delta_\perp - \alpha m^2}{m}
\frac{1}{2\partial_{-}} \ast J^{+} 
\ ,\label{procaQEDeffeqQ} \\ 
(2 \partial_{+} \partial_{-} - \Delta_\perp + m^2){C_i} & = &
- \frac{\sqrt{2} e}{2 m} \left[ \psi_{+}^{\dag} \alpha^i
\frac{1}{ i\partial_{-} - e \left( \frac{\varphi +
Q}{m}\right)} \ast \xi + \xi^{\dag} \ast \frac{1}{
i\partial_{-} - e \left( \frac{\varphi + Q}{m} 
\right)} \alpha^i \psi_{+} \right] \nonumber\\
& - & \partial_i  \frac{1}{\partial_{-}} \ast J^{+}
\ ,\label{procaQEDeffeqCi} \\ 
i \sqrt{2} \partial_{+} \psi & = & \frac{1}{\sqrt{2}} \left[- i
\partial_i \alpha^i + M \beta  + e \alpha^i C_i +
\alpha^i \partial_i \left ( \frac{\varphi + Q}{m}\right )
\right] \frac{1}{i \partial_{-} - e \partial_{-} \left (
\frac{\varphi + Q}{m}\right ) }* \xi \nonumber\\ 
& + & \frac{e\sqrt{2}}{2} \frac{1}{\partial_{-}} \ast
\left[\frac{\Delta_\perp - \alpha m^2}{m}Q + \frac{\Delta_\perp
+ m^2}{m} \varphi + \partial_i C_i - (1-\alpha)
\frac{1}{\partial_{-}}* J^{+}\right] \psi_{+} 
\ ,\nonumber\\
\\ 
- i \sqrt{2} \partial_{+} \psi^{\dag} & = & \frac{1}{\sqrt{2}} 
\xi^{\dag} *\frac{1}{i \partial_{-} - e \partial_{-} \left (
\frac{\varphi + Q}{m}\right )} 
 \left[ i \stackrel{\leftarrow}{\partial}_i
\alpha^i + M \beta + e \alpha^i C_i + 
\alpha^i \partial_i \left ( \frac{\varphi + Q}{m}\right )
\right]\nonumber\\
& + & \frac{e\sqrt{2}}{2} \psi_{+}^{\dag}\frac{1}{\partial_{-}} \ast
\left[\frac{\Delta_\perp - \alpha m^2}{m}Q + \frac{\Delta_\perp
+ m^2}{m} \varphi + \partial_i C_i - (1-\alpha)
\frac{1}{\partial_{-}}* J^{+}\right]\ .\nonumber\\   
\end{eqnarray} 
These equations show that one can introduce another quantum
field $\Lambda$ 
\begin{equation}
\Lambda = m Q + e \frac{1}{2\partial_{-}}* J^{+}\ ,
\end{equation}
which satisfies the free field equation 
\begin{equation}
(2 \partial_{+} \partial_{-} - \Delta_\perp + \alpha
m^2){\Lambda} = 0 \ , \label{procaQEDeffeqLambda}
\end{equation}
but at the price of having extra non-vanishing commutators on LF
\begin{eqnarray}
\left [ \psi_{+} (\vec{x}), \Lambda(\vec{y})\right ]
& = & e \delta^3(\vec{x} - \vec{y}) \psi_{+}(\vec{x})\ ,
\label{procaQEDpsiLambdacommut}\\ 
\left [ \psi_{+}^{\dag} (\vec{x}), \Lambda(\vec{y})\right ]
& = & - e \delta^3(\vec{x} - \vec{y}) \psi_{+}^{\dag} (\vec{x})
\label{procaQEDdagpsiLambdacommut}. 
\end{eqnarray}
This alternative, of having either the non-free field $Q$ which
commutes with fermions or the free field $\Lambda$ which
satisfies (\ref{procaQEDpsiLambdacommut}) and
(\ref{procaQEDdagpsiLambdacommut}), is a specific feature of the
LF formulation - contrary to the ET case where $\Lambda$ is the
only opportunity.\\  
The mere presence of q-number commutators would destroy a
simple picture of perturbative calculations based on Feynman
diagrams and Wick contractions, therefore in the next
subsection we will build the perturbation theory taking $Q$ as
an independent field and going to the interaction picture, where
only c-number terms will appear as contractions for all
fields.

\subsection{Perturbation theory}

The perturbative calculations of the S-matrix elements are most
easily performed in the interaction picture where all 
quantum field operators have free dynamics while interactions
appear in the evolution of quantum states. Below we will work
in this representation; however for clarity we will omit
subscripts {\scriptsize I} for the  field operators. Here the
interaction representation is 
defined by taking the free Hamiltonian $H_0$ as 
the limit of the total Hamiltonian (\ref{procaQEDtotalHamilt}) 
\begin{eqnarray}
{\cal H}_{0}  =  \lim_{e\rightarrow 0} {\cal H}_{total} & = & 
\frac 1 2 
\left(\partial_i C_i \right) ^2
+ \frac 1 4 \left( \partial_i{C}_j - \partial_j {C}_i\right)^2
+ \frac{m^2}{2} C_i^2 -  \frac{1}{2}(\partial_i Q)^2 - \alpha 
\frac{m^2}{2} Q^2\nonumber\\
& +& \frac 1 2 (\partial_i\varphi)^2  + \frac{m^2}{2}\varphi^2
+ \frac{1}{\sqrt{2}} \xi^{\dag}_0
\frac{1}{i \partial_{-} }* \xi_0 \label{procaQEDfreeHamilt}\ ,
\end{eqnarray}
where
\begin{eqnarray}
\xi_0 & = & \left(- i \partial_i \alpha^i + M \beta
\right)\psi_{+} \ ,\\
\xi^{\dag}_0 & = & \left (i \partial_i\psi^{\dag}_{+} \alpha^i
+ M \psi_{+}^{\dag}\beta \right).
\end{eqnarray}
Next the interaction Hamiltonian, which is defined as the difference
of these two Hamiltonians, can be written, for the later
convenience, as a sum of two contributions
\begin{equation}
{\cal H}_{int}  =  {\cal H}_{total} - {\cal H}_{int} = {\cal
H}^1_{int} + {\cal H}^2_{int}.
\end{equation}
The first part describes the interaction of fermion and LF
spatial components of the vector field 
\begin{eqnarray}
{\cal H}^1_{int} & = & \frac{1}{\sqrt{2}} \xi^{\dag}_0 \left(
\frac{1}{ i \partial_{-} - e \partial_{-} \phi} - \frac{1}{i
\partial_{-}} \right) \ast \xi_0 \nonumber\\
& + & \frac{e}{\sqrt{2}} \psi^{\dag}_{+} \alpha^i (C_i +
\partial_{i} \phi)\frac{1}{ i \partial_{-} - e \partial_{-}
\phi} \ast \xi_0 + \frac{e}{\sqrt{2}} \xi^{\dag}_{0} 
\frac{1}{ i \partial_{-} - e \partial_{-} \phi} \ast  \alpha^i (C_i +
\partial_{i} \phi) \psi_{+}\nonumber\\
& + & \frac{e^2}{\sqrt{2}} \psi^{\dag}_{+} \alpha^j (C_j +
\partial_{j} \phi)\frac{1}{ i \partial_{-} - e \partial_{-}
\phi} \ast \alpha^i (C_i + \partial_{i} \phi) \psi_{+}\ ,
\label{procaQEDintHam1} 
\end{eqnarray}
while the second one is connected with the current $J^{+}$ 
\begin{eqnarray}
{\cal H}^2_{int} & = & - \left( 2 \partial_i C_i + m \varphi
- \alpha m Q + \Delta_\perp \phi \right )\frac{1}{2\partial_{-}}*
J^{+} - \frac{1 - \alpha}{8} J^{+}
\frac{1}{\partial_{-}^2}\ast  J^{+},\label{procaQEDintHam2}
\end{eqnarray}
where we have introduced another notation 
\begin{eqnarray}
\phi = \frac{\varphi + Q}{m}.
\end{eqnarray}

\subsubsection{Field operators in the interaction representation}

From the free Hamiltonian (\ref{procaQEDfreeHamilt}) and the
equal-$x^{+}$ (anti)commutation relations one derives the free
field equations
\begin{eqnarray}
(2 \partial_{+} \partial_{-} - \Delta_\perp + m^2){C_i} & = &
0\ ,\\
(2 \partial_{+} \partial_{-} - \Delta_\perp + m^2){\varphi} & =
& 0 \ ,\\
(2 \partial_{+} \partial_{-} - \Delta_\perp + \alpha m^2) {Q} &
= & 0\ ,\\
(2 \partial_{+} \partial_{-} - \Delta_\perp + M^2)\psi_{+} & =
& 0 \ ,\\
(2 \partial_{+} \partial_{-} - \Delta_\perp + M^2) \psi_{+}^{\dag} & =
& 0 .
\end{eqnarray} 
Such free fields have their Fourier representations for all $x^{+}$
and for vector field modes\footnote{The case of free fermion
fields was given by Yan \cite{Yan1972} therefore we will omit
them here and only quote all the needed results.} we write
\begin{eqnarray}
{C_{i}}(x) & = & \int_{-\infty}^{\infty} \frac{d^2k_\perp}{(2
\pi)^2} \int_{0}^{\infty} \frac{dk_{-}}{4\pi k_{-}}
 \ \left[e^{-i {k}\cdot{x}} c_i(\vec{k}) + e^{+i
{k}\cdot {x}} c^{\dagger}_i(\vec{k})\right]_{k_{+} =
\frac{k_\perp^2+m^2}{2k_{-}}}\ ,\\
{Q}({x}) & = &\int_{-\infty}^{\infty} \frac{d^2k_\perp}{(2
\pi)^2} \int_{0}^{\infty} \frac{dk_{-}}{4\pi k_{-}}
 \ \left[e^{-i {k}\cdot{x}} q(\vec{k}) + e^{+i
{k}\cdot {x}} q^{\dagger}(\vec{k})\right]_{k_{+} =
\frac{k_\perp^2+\alpha m^2}{2k_{-}}}\ ,\\
{\varphi}(x) & = & \int_{-\infty}^{\infty} \frac{d^2k_\perp}{(2
\pi)^2} \int_{0}^{\infty} \frac{dk_{-}}{4\pi k_{-}}
 \ \left[e^{-i {k}\cdot{x}} p(\vec{k}) + e^{+i
{k}\cdot {x}} p^{\dagger}(\vec{k})\right]_{k_{+} =
\frac{k_\perp^2+m^2}{2k_{-}}}\ ,
\end{eqnarray}
where the creation and annihilation operators have non-vanishing
commutators
\begin{eqnarray}
\left[ q(\vec{k}), q^{\dagger}(\vec{k'})\right] & = & 
- \left[ p(\vec{k}), p^{\dag}(\vec{k'})\right] =
(2\pi)^3 \ 2 k_{-} \ \delta^3(\vec{k} -
\vec{k'})\label{procaQEDfourcomcc}\ ,\\ 
\left[ c_{i}(\vec{k}), c^{\dagger}_j(\vec{k'})\right] & =
&(2\pi)^3 \ 2 k_{-} \ \delta_{ij}\delta^3(\vec{k} -
\vec{k'}).\label{procaQEDfourcomcicj} 
\end{eqnarray}
Now it is an easy exercise to calculate the chronological (in
$x^{+}$) products of these free field operators
\begin{eqnarray}
\left< 0 \left| T^{+} C_i(x) C_j(y)\right| 0 \right> & = &  
\delta_{ij} \Delta_F(x-y, m^2)\ ,\\
\left< 0 \left| T^{+} \varphi(x) \varphi(y)\right| 0 \right> &
= &   \Delta_F(x-y, m^2)\ ,\\
\left< 0 \left| T^{+} Q(x) Q(y)\right| 0 \right> & = &
-  \Delta_F(x-y, \alpha m^2)\ ,
\end{eqnarray}
where the covariant Feynman massive propagator function $\Delta_F(x,
\mu^2)$ is defined in Appendix \ref{dodfenGrefun}. In the
interaction Hamiltonian, the linear combinations of independent
vector modes 
\begin{eqnarray}
\bar{B}_{+} & = & \frac{1}{2\partial_{-}}* \left[ 2 \partial_i
C_i + m\varphi  - \alpha m Q + \Delta_\perp \phi \right ]\ ,
\label{procaQEDdefbarBp}\\
\bar{B}_{-} & = & \partial_{-} \phi \label{procaQEDdefbarBm} \
, \\
\bar{B}_i & = & C_i(x) + \partial_{i} \phi
\label{procaQEDdefbarBi} \ ,
\end{eqnarray}
are coupled with the fermion currents, therefore in the Dyson-Wick
perturbation procedure we effectively encounter contractions
given by the chronological products of $\bar{B}_\mu$. Now after
some algebra we find
\begin{eqnarray}
\left< 0 \left| T \bar{B}_\mu(x) \bar{B}_\nu(y)\right| 0
\right> & = & \left[- g_{\mu \nu} {\Delta}_F(x-y,m^2) +
\partial_\mu^x \partial_\nu^y \frac{\Delta_F(x-y,m^2) -
\Delta_F(x-y,\alpha m^2)}{m^2} \right] \nonumber\\
& + & i\ g_{\mu -} g_{\nu -}\frac{1 - \alpha}{4}
\frac{1}{(\partial_{-})^2} * \delta(x - y)\ ,
\label{procaQEDBmuBnuvev} 
\end{eqnarray}
where the last non-covariant contribution arises when one uses
the equation 
\begin{equation}
(2 \partial_{+} \partial_{-} - \Delta_\perp + \mu^2)
\Delta_F(x,\mu^2) = - i \delta^4(x)
\end{equation}
in transforming the result into the form with second
derivatives $\partial_{+}$ of the propagator functions.\\
In later discussion it will be very convenient to reintroduce 
the dependent fermion fields 
\begin{eqnarray}
\psi_{-} & = & \frac{1}{\sqrt{2}} \frac{1}{i \partial_{-}}*
\left(-i \partial_i \alpha^i + M \beta \right)\psi_{+} \ ,\\
\psi^{\dag}_{-} & = & \frac{1}{\sqrt{2}} \left(i \partial_i
\psi^{\dag}_{+} \alpha^i + M \psi^{\dag}_{+} \beta \right)
*\frac{1}{i \partial_{-}}\ ,
\end{eqnarray}
and consider the complete spinor fields $\psi = \psi_{+} +
\psi_{-}$ and $\psi^{\dag} = \psi^{\dag}_{+}+ \psi^{\dag}_{-}$.
Taking the well known 
chronological product for independent fermions  \cite{Yan1972} 
\begin{eqnarray}
\left< 0 \left| T \psi_{+}(x) \psi_{+}^{\dag}(y)\right| 0
\right> & = & i \sqrt{2} \Lambda_{+} \partial_{-}^x
\Delta_F(x-y, M^2)\ ,
\end{eqnarray}
we derive the chronological products for dependent fermion fields 
\begin{eqnarray}
\hspace{-15pt}&&\left< 0 \left| T^{+} \psi_{-}(x) \psi_{+}^{\dag}(y)\right| 0
\right> + \left< 0 \left| T \psi_{+}(x)
\psi_{-}^{\dag}(y)\right| 0 \right> = \left( i \partial_i^x
\gamma^i +  M\right) \gamma^0 \Delta_F(x-y, M^2),\\
&&\left< 0 \left| T^{+} \psi_{-}(x) \psi_{-}^{\dag}(y)\right| 0
\right>  =  i \partial_{+}^x \gamma^{+} \gamma^0 \Delta_F(x-y,
M^2) - \gamma^{+} \gamma^0 \frac{1}{2 \partial_{-}}*
\delta(x-y) \ ,
\end{eqnarray}
and finally for the complete fermion fields
\begin{eqnarray}
\left< 0 \left| T^{+} \psi(x) \psi^{\dag}(y)\right| 0
\right> =  \left(i \partial_{\mu }^x \gamma^{\mu} +  M\right)
\gamma^0 \Delta_F(x-y, M^2) - \gamma^{+} \gamma^0 \frac{1}{2
\partial_{-}}* \delta(x-y) \ ,
\end{eqnarray}
or 
\begin{eqnarray}
\left< 0 \left| T \psi(x) \bar{\psi}(y)\right| 0
\right>  =   \left(i\partial_{\mu }^x \gamma^{\mu} +  M \right)
\Delta_F(x-y, M^2) - \gamma^{+} \frac{1}{2 \partial_{-}}*
\delta(x-y). \label{procaQEDpsibarpsiprop}
\end{eqnarray}
Now we may re-express the interaction Hamiltonian in terms of
$\bar{B}_\mu$, $\psi$ and $\psi^{\dag}$. First we take ${\cal
H}^1_{int}$ and,  using the identity 
\begin{equation}
\frac{1}{i \partial_{-} - e \bar{B}_{-}} - \frac{1}{i
\partial_{-}} = \left(\frac{1}{i \partial_{-} - e \bar{B}_{-}}  e
\bar{B}_{-}\right) * \frac{1}{i \partial} = \frac{1}{i
\partial} \ast \left( e \bar{B}_{-} \frac{1}{i \partial_{-} - e
\bar{B}_{-}} \right)\,
\end{equation}
we write it as 
\begin{eqnarray}
{H}_{int}^{1} & = & \frac{1}{\sqrt{2}} \xi^{\dag}_0 *\frac{1}{i
\partial_{-}} e \bar{B}_{-} \ast \left(1 + \frac{1}{i
\partial_{-} - e \bar{B}_{-}}  e\bar{B}_{-}\right) \ast 
\frac{1}{i\partial_{-}}* \xi_0  \nonumber\\
& + & \frac{e}{\sqrt{2}} \psi_{+}^{\dag} \alpha^i \bar{B}_i
\ast \left(1 + \frac{1}{i \partial_{-} - e \bar{B}_{-}}
e\bar{B}_{-}\right) \ast \frac{1}{i\partial_{-}}* \xi_0 \nonumber \\
& + & \frac{e}{\sqrt{2}} \xi_{0}^{\dag} *\frac{1}{i
\partial_{-}} \ast \left(1 + \frac{1}{i
\partial_{-} - e \bar{B}_{-}}  e\bar{B}_{-}\right) \ast 
\alpha^i \bar{B}_i \psi_{+} \\
& + & \frac{e}{\sqrt{2}} \psi_{+}^{\dag} \alpha^i \bar{B}_i
\ast \frac{1}{i \partial_{-} - e \bar{B}_{-}} * \alpha^j
\bar{B}_j \psi_{+} \nonumber. 
\end{eqnarray}
Then $\xi_0$ and $\xi^{\dag}$ can be expressed in terms of
$\psi_{-}$ and $\psi^{\dag}_{-}$ fields and finally, we arrive
at the factorized form 
\begin{equation}
{H}_{int}^{1} =  \bar{\psi} e \left(\gamma^{-} \bar{B}_{-} +
\gamma^{i} \bar{B}_{i}\right) \ast \left[1 + \frac{e}{2} \gamma^{+} \frac{1}{i
\partial_{-} - e \bar{B}_{-}}  \left( \gamma^{-} \bar{B}_{-} + \gamma^{j}
\bar{B}_{j}\right)\right]\ast \psi.  \label{procaQEDmodintHam1}
\end{equation}
The second part of the interaction Hamiltonian is much simpler
\begin{equation}
{\cal H}_{int}^{2} = \bar{\psi} e \gamma^{+}
\bar{B}_{+}\psi  - \frac{1 - \alpha}{8} J^{+}
\frac{1}{\partial_{-}^2} \ast J^{+},
\end{equation}
where for simplicity, in the last term we have left the
notation $J^{+}$ for fermion current.\\
These formulas show that here the noncovariant terms
appear both in the propagators and the interaction
Hamiltonians, contrary to the equal-time results where all
the corresponding expressions are explicitly covariant. However,
following the analysis by Yan \cite{Yan1972}, one can hope that
also here these non-covariant terms will cancel in pairs
(\ref{procaQEDintHam2}) with (\ref{procaQEDBmuBnuvev}) and
(\ref{procaQEDpsibarpsiprop}) with (\ref{procaQEDmodintHam1}).

\subsubsection{LF perturbative calculations of the S-matrix
elements}\label{ProcaSmatrix} 

Here we will check the above conjuncture by studying 
the formal structure of perturbative calculations.
The functional techniques \cite{SchwingerFunct},
\cite{Yan1972} are
quite useful for this purpose 
because they allow us to analyse contractions of gauge and
fermion fields separately. First we check 
the contractions of vector  field $\bar{B}_{+}$ and treat  all 
other components of vector fields and the fermion
fields as classical objects for time being. The Wick theorem
for transformation of chronological products into the normal
products says 
\begin{equation}
T^{+} \exp\left\{ -i\int j^{+} \bar{B}_{+} \right\} \exp\left\{i
\frac{1-\alpha}{8} \int j^{+}  \frac{1}{\partial_{-}^2} \ast
j^{+} \right \} = \exp\left \{\frac{i}{2} \int j^{+} {{\cal
D}}_{+ +}  j^{+}\right \} \ :\ 
\exp\left\{ -i \int j^{+} \bar{B}_{+} \right\}:
\end{equation}
where 
\begin{equation}
{\cal D}_{+ +}(x)  =  i \left< 0 \left| T \bar{B}_+(x)
\bar{B}_{+}(y)\right| 0 \right> + \frac{1 -
\alpha}{4} \frac{1}{\partial_{-}^2}*\delta(x) = i
\partial_{+}^x \partial_{+}^y \frac{\Delta_F(x-y,m^2) -
\Delta_F(x-y,\alpha m^2)}{m^2},
\end{equation}
and as usually, colons denote the normal product. Therefore we
can simultaneously omit the instantaneous current 
interaction in ${\cal H}_{int}^2$ and  take the covariant 
propagator for contractions of all vector field components
\begin{eqnarray}
&&\hspace{-10pt}{{\cal D}_F}_{\mu \nu}(x) =  - g_{\mu \nu}
{\Delta}_F(x-y,m^2) + \partial_\mu^x \partial_\nu^y
\frac{\Delta_F(x-y,m^2) - \Delta_F(x-y,\alpha m^2)}{m^2} =
 \nonumber\\ 
&& \hspace{30pt} =   i \int \frac{d^4k}{(2
\pi)^4} \frac{e^{- ik\cdot (x)}}{2k_{+}k_{-} - k^2_\perp - m^2+ i
\epsilon} \left[ - g_{\mu \nu} + (1 - \alpha)\frac{k_\mu
k_\nu}{2k_{+}k_{-} - k^2_\perp - \alpha m^2 + i
\epsilon}\right]. \label{procavectpropagator} 
\end{eqnarray}
Thus effectively the complete interaction Hamiltonian is
bilinear in fermion fields 
\begin{equation}
{ H}_{int} =  \bar{\psi} * e \gamma^\mu \bar{B}_\mu 
\left[1 + \frac{e}{2} \gamma^{+} \frac{1}{i \partial_{-} - e
\bar{B}_{-}}  e \gamma^{\nu} \bar{B}_{\nu} \right] \ast \psi =
\bar{\psi} \ast {\cal V}[\bar{B}_\mu] \ast \psi. 
\label{procaQEDeffintHam} 
\end{equation}
Next when all the vector fields are kept as c-numbers $b_\mu$,
one can easily study the contractions of fermion fields
\cite{SchwingerFunct}, \cite{Yan1972} 
\begin{eqnarray}
T^{+} \exp - i \bar{\psi} \ast {\cal V}[b_\mu] \ast \psi =
{\exp}\left[{\rm Tr} \ln 
\left(1 - \bar{S}_{F} \ast {\cal V}\right)\right] : \ {\rm exp}
\left[ - i \bar{\psi} \ast {\cal V} \ast (1 - \bar{S}_{F}\ast
{\cal V} )^{-1} \ast \psi\right]:\label{procaQEDfermcontr} 
\end{eqnarray}
where now
\begin{eqnarray}
i \bar{S}_F(x,y) & = & \left< 0 \left| T \psi(x)
\bar{\psi}(y)\right| 0 \right> = i S_F(x-y) - \gamma^{+}
\frac{1}{2\partial_{-}}* \delta(x-y) \label{procaQEDnoncovfermprop}\\
iS_F(x) & = &  \left( i\gamma^\mu \partial_\mu^{x} +
M\right) \Delta_F(x,M^2) .
\end{eqnarray}
One can check that the following factorization property holds 
\begin{eqnarray}
1 - \bar{S}_{F}\ast {\cal V} & = & 1 - \left(S_F - \frac{\gamma^{+}}{2 i
\partial_{-}} \right)\ast  e \gamma^\mu  b_\mu \left[ 1 +
\frac{\gamma^{+}}{2} \frac{1}{i \partial_{-} - e 
b_{-}}* e \gamma^\nu b_\nu\right] \nonumber\\
& = & \left( 1 -  S_F e \gamma^\mu b_\mu \right) \ast \left[ 1 +
\frac{\gamma^{+}}{2} \frac{1}{i \partial_{-} - e b_{-}}*
e \gamma^\nu b_\nu \right]
\nonumber\\ 
& = & \left(1 - S_F \ast e \gamma^\mu b_\mu \right) \ast {\cal
V}_2\ , 
\end{eqnarray}
and then it is easy to notice that the non-covariant factor
$H_2$ disappears from the normal product part of
Eq.(\ref{procaQEDfermcontr}). It still formally remains in the closed loop
contribution 
$$
{\exp} {\rm Tr} \ln \left( 1 +
\frac{\gamma^{+}}{2} \frac{1}{i \partial_{-} - e 
b_{-}}* e \gamma^{\mu} b_{\mu} \right) \approx {\exp} {\rm
Tr} \ln \left( 1 + \frac{1}{i \partial_{-} - e b_{-}}* e b_{-}
\right) .
$$
However, the last expression can be shown to be independent of
$b_{-}$.\footnote{One can proved perturbatively showing that
the closed loop diagrams disappear, but in the present model,
from $b_{-} = \partial_{-}\phi$, one also gets
$$\left[i\partial_{-} - e(\partial_{-}\phi)\right]^{-1}(x,y) =
\exp ie\phi(x) (i\partial_{-})^{-1}(x,y) \exp -ie\phi(y)$$ and
its functional determinant is evidently independent of $\phi$.}
In this manner we have 
shown that the formal perturbative series based on the
noncovariant interaction Hamiltonian (\ref{procaQEDfermcontr})
and the canonical noncovariant fermion propagator
(\ref{procaQEDnoncovfermprop}) will give the same result as the
calculation based on the covariant interaction Hamiltonian
$\bar{\psi} * e \gamma^\mu {b}_\mu * \psi$ and the
covariant fermion propagator  $S_F$
\begin{eqnarray}
T^{+} \exp - i \bar{\psi} \ast {\cal V}[b_\mu] \ast \psi =
{\exp}\left[{\rm Tr} \ln \left(1 - {S}_{F} \ast e \gamma^\mu
{b}_\mu \right)\right] :\ {\rm exp}  \left[ - i
\bar{\psi} \ast  (1 - {S}_{F}\ast e \gamma^\nu {b}_\nu)^{-1} \ast
\psi\right]:\label{procaQEDfermequiv} 
\end{eqnarray}
provided all divergent expressions are properly regularized.

\section{LF Weyl gauge}\label{massWeylrozdzial}

\setcounter{equation}{0}

Another gauge condition which we have chosen for the massive 
vector fields is the LF Weyl gauge $B_{+} = B^{-} = 0$. This
gauge condition is particularly suitable for the canonical
procedure because it explicitly removes this component of
vector field  whose canonically conjugated momentum is zero.
Contrary to the previous covariant gauge, the LF Weyl gauge can
be strongly implemented in the Lagrangian density thus reducing
the number of field variables.

\subsection{Vector fields sector}

First we take the model of massive vector fields 
coupled to the external currents and impose 
explicitly the LF-Weyl gauge condition. We see that the
covariant mass term takes the form $ - \frac 1 2 m^2 (B_i)^2$
\begin{equation}
{\cal L}^{mass}_{jWeyl}  =  \partial_{+} {B}_i (
\partial_{-} {B}_i - \partial_i {B}_{-}) + \frac 1 2
\left(\partial_{+} {B}_{-} \right)^2 - \frac 1 4 \left(
\partial_i {B}_j- \partial_j {B}_i\right)^2 
- \frac{ m^2}{2} B_i^2 + {B}_{-} {j}^{-} + {B}_i j^i
\label{massWeylmAbelLagr} 
\end{equation}
and all Euler-Lagrange equations are dynamical
\begin{eqnarray}
\left( 2\partial_{+} \partial_{-} - \Delta_\perp + m^2
\right){B}_i & = & \partial_i ( \partial_{+} B_{-} - \partial_i
B_i ) + j^i \ ,\label{massWeylmassWeELeqBi} \\
\partial_{+} ( \partial_{+} B_{-} - \partial_i B_i ) & = &
j^{-}. \label{massWeylmassWeELeqBm}
\end{eqnarray}
Thus if we rewrite these equations in the first-order form 
\begin{eqnarray}
\left( 2\partial_{+} \partial_{-} - \Delta_\perp \right){B}_i
& = & \partial_i \Pi + j^i\ , \label{massWeylmassHameqBi}\\
\partial_{+} \Pi & = & j^{-}\ ,\\ 
\partial_{+} B_{-} & = & \Pi + \partial_i B_i \ ,\label{massWeylmassHameqBm}
\end{eqnarray}
and find the canonical Hamiltonian\footnote{The canonical
momenta are simple here
\begin{eqnarray*}
\Pi^{-} & = & \partial_{+} B_{-} - \partial_i B_i\ , \\
\Pi^{i} & = & \partial_{-} B_{i} - \partial_i B_{-}.
\end{eqnarray*}
}
\begin{equation}
{\cal H}_{can}^{mass} =  \Pi^{-}\partial_{+}B_{-} + \Pi^{i}
\partial_{+}B_i - {\cal L}^{mass}_{jWeyl} = \frac 1 2 (\Pi)^2 
+ \frac 1 2 (\partial_i B_j)^2 + \Pi \partial_i B_i + \frac{
m^2}{2} B_i^2- B_i j^i - B_{-} j^{-}\label{massWeylmassWeylHamden1},
\end{equation}
then the Poisson brackets will follow immediately
\begin{eqnarray}
\left\{2\partial_{-} B_i(x^{+}, \vec{x}), B_j(x^{+}, \vec{y})
\right\}_{PB} & = & - \delta_{ij} \delta^{3}(\vec{x} -
\vec{y})\ ,\\
\left\{B_{-}(x^{+}, \vec{x}), \Pi(x^{+}, \vec{y}) \right\}_{PB} & = & 
 \delta^{3}(\vec{x} - \vec{y}).
\end{eqnarray}
Just as in the previous case, our independent canonical
variables $(B_i, B_{-})$ are not independent modes. Rather 
we should take  their linear combinations
\begin{eqnarray}
C_{i} & = & B_i + {\partial_i}\frac{1}{\Delta_\perp-m^2} *
\Pi \ , \label{massWeylmassfromCitoAim}\\ 
C_{-} & = & B_{-} -  \partial_{-}\frac{1}{\Delta_\perp-m^2} *
\left[ 2 \partial_j B_j + \Delta_\perp \frac{1}{\Delta_\perp-m^2}
 * \Pi\right] \ ,\label{massWeylmassfromCmtoAmm}
\end{eqnarray}
which satisfy separated equations of motion\footnote{The field
$C_{-}$ is a {\em noncovariant multipole field} which is
characteristic for noncovariant gauge conditions.}
\begin{eqnarray}
\left( 2\partial_{+} \partial_{-} - \Delta_\perp + m^2\right){C}_i
& = & j^i - 2\partial_i \frac{1}{m^2 - \Delta_\perp} *
\partial_{-}j^{-}\ ,\\ 
\partial_{+} \Pi & = & j^{-}\ ,\\
\partial_{+} C_{-} & = & - m^2 \frac{1}{\Delta_\perp-m^2} * \Pi -
\frac{1}{\Delta_\perp-m^2} * \partial_i j^i \nonumber\\
& - & \Delta_\perp \frac{1}{\Delta_\perp-m^2} * \left(
\frac{1}{\Delta_\perp-m^2} *  \partial_{-} j^{-}\right)\ ,
\end{eqnarray}
and still have diagonal Poisson brackets
\begin{eqnarray}
\left\{2\partial_{-} C_i(x^{+}, \vec{x}), C_j(x^{+}, \vec{y})
\right\}_{PB} & = & - \delta_{ij} \delta^{3}(\vec{x} -
\vec{y})\ ,\\
\left\{C_{-}(x^{+}, \vec{x}), \Pi(x^{+}, \vec{y}) \right\}_{PB} & = & 
 \delta^{3}(\vec{x} - \vec{y}).
\end{eqnarray}
At last we can express the Hamiltonian density 
(\ref{massWeylmassWeylHamden1}) in terms of independent modes
\begin{eqnarray}
{\cal H}^{mass}_{total} & = & \frac 1 2 (\partial_i C_j)^2 +
\frac{m^2}{2} C_i^2 - \frac{m^2}{2} \Pi  \frac{1}{\Delta_\perp-m^2}
* \Pi - C_{-}j^{-}\nonumber\\
& - & \Pi \frac{1}{\Delta_\perp-m^2}* \left[\partial_i j^{i} +
\Delta_\perp \frac{1}{\Delta_\perp-m^2}* \partial_{-} j^{-}
\right] - C_i \left[j^i - 2\partial_i \frac{1}{\Delta_\perp-m^2}
* \partial_{-} j^{-}\right]\label{massWeyleffhammassweyl}
\end{eqnarray}
and notice that no instantaneous interaction of currents occurs
in this model. We see that the fermion field contribution is
quite similar to that discussed in the previous case of
covariant gauge and no modification of the perturbative vector
field propagators should appear here. Therefore all we need to know
here are  the free vector field propagators.

\subsection{Free quantum fields}

We restrict our discussion to the free field case because in the
interaction representation  field operators have free dynamics. From
the canonical analysis we take the form of canonical
commutators at LF
\begin{eqnarray}
\left[2\partial_{-} C_i(\vec{x}), C_j(\vec{y}) \right] & = & 
- i \delta_{ij} \delta^{3}(\vec{x} - \vec{y}) \ ,
\label{massWeylkommassweylCiCj}\\ 
\left[C_{-}(\vec{x}), \Pi(\vec{y}) \right] & = & 
i\delta^{3}(\vec{x} - \vec{y})\ ,\label{massWeylkommassweylPiCm}
\end{eqnarray}
and the free Hamiltonian density
\begin{equation}
{\cal H}^{mass}_{0} = \frac 1 2 (\partial_i C_j)^2 +
\frac{m^2}{2} C_i^2 - \frac{m^2}{ 2} \Pi \frac{1}{\Delta_\perp
-m^2} \ast \Pi .
\end{equation}
They generate free field equations
\begin{eqnarray}
\left( 2\partial_{+} \partial_{-} - \Delta_\perp + m^2\right){C}_i
& = & 0\ ,\\
\partial_{+} \Pi & = & 0\ ,\\
\partial_{+} C_{-} & = & - m^2 \frac{1}{\Delta_\perp-m^2} * \Pi. 
\end{eqnarray}
All fields, the covariant $C_i$ and the noncovariant 
($\Pi$,  $C_{-}$) have their Fourier representations for all
$x^{+}$ 
\begin{eqnarray}
{C_{i}}(x) & = & \int_{-\infty}^{\infty} \frac{d^2k_\perp}{(2
\pi)^2} \int_{0}^{\infty} \frac{dk_{-}}{4\pi k_{-}}
 \ \left[e^{-i {k}\cdot{x}} c_i(\vec{k}) + e^{+i
{k}\cdot {x}} c^{\dagger}_i(\vec{k})\right]_{k_{+} =
\frac{k_\perp^2+m^2}{2k_{-}}}\ ,\\
{\Pi}(\vec{x}) & = &\int_{-\infty}^{\infty} \frac{d^2k_\perp}{(2
\pi)^2} \int_{0}^{\infty} \frac{dk_{-}}{2\pi}
\ \left[e^{-i \vec{k}\cdot
\vec{x}} p(\vec{k}) + e^{+i \vec{k}\cdot \vec{x}}
p^{\dagger}(\vec{k})\right]\ ,\\
{C_{-}}(x^{+},\vec{x}) & = & - m^2
x^{+}\frac{1}{\Delta_\perp-m^2} * \Pi (\vec{x}) \nonumber\\
&+ &\int_{-\infty}^{\infty} \frac{d^2k_\perp}{(2
\pi)^2} \int_{0}^{\infty} \frac{dk_{-}}{2\pi}
 \ \left[e^{-i \vec{k}\cdot
\vec{x}} c_{-}(\vec{k}) + e^{+i \vec{k}\cdot \vec{x}}
c^{\dagger}_{-}(\vec{k})\right]\ ,
\end{eqnarray}
and  the commutator relations 
for the creation and annihilation  operators are
\begin{eqnarray}
\left[ c_{-}(\vec{k}), p^{\dagger}(\vec{k'})\right] & = & 
\left[ c^{\dagger}_{-}(\vec{k}), p(\vec{k'})\right] =
i (2\pi)^3 \delta^3(\vec{k} - \vec{k'})\label{massWeylmfourcomcp}\\
\left[ c_{i}(\vec{k}), c^{\dagger}_j(\vec{k'})\right] & =
&(2\pi)^3 \ 2 k_{-} \ \delta_{ij}\delta^3(\vec{k} -
\vec{k'}).\label{massWeylmfourcomcicj} 
\end{eqnarray}
Now a quite straightforward calculation gives the chronological
products for independent modes
\begin{eqnarray}
<0| T\ C_i(x) C_j(y) | 0> & = & \delta_{ij}
\Delta_F(x-y)\ ,\\  
<0| T\ C_{-}(x) \Pi(y) | 0> & = & E_F^1(x_L - y_L)
\delta^{2}(x_\perp-y_\perp)\ ,\label{massWeylpropmassWeylCmPi}\\ 
<0| T\ C_{-}(x) C_{-}(y) | 0> & = & - m^2 E_F^2(x_L - y_L)
\frac{1}{\Delta_\perp-m^2} * \delta^2(x_\perp-y_\perp),
\label{massWeylpropmassWeylCmCm}
\end{eqnarray}
where all other propagators vanish\footnote{Definitions and properties of 
the noncovariant propagator functions 
$E_F^{1,2}(x)$ are given in Appendix \ref{dodfenGrefun}.}.
We notice that the causal propagation of noncovariant
fields ($\Pi$, $C_{-}$) takes place only in the LF longitudinal
directions $x_L = (x_{+}, x_{-})$ and in the transverse
directions $x_\perp = (x_2, x_3)$ these propagators are either
local (\ref{massWeylpropmassWeylCmPi}) or are given by the
modified inverse Laplace operator (\ref{massWeylpropmassWeylCmCm}). 
This means that their causal properties i.e. the ML-prescription
which appears in them, do not depend on the space-time
dimensionality, and  thus have no direct infrared singularities
\footnote{Evidently the infrared singularity would appear in
the integral operator $(\Delta_\perp - m^2)^{-1}$ when the 
limit $m^2 \rightarrow 0$ is taken in two transverse dimensions.}.\\
Finally, in order to find the propagators for primary vector
fields, we use the relations
\begin{eqnarray}
B_{i} & = & C_i - {\partial_i} \frac{1}{\Delta_\perp-m^2}
*\Pi\ , \label{massWeylfromBitoCi}\\ 
B_{-} & = &  C_{-} +  \partial_{-}\frac{1}{\Delta_\perp-m^2} *
\left[ 2 \partial_j C_j -  \Delta_\perp \frac{1}{\Delta_\perp-m^2} *
\Pi\right] \ ,\label{massWeylfromBmtoCm}
\end{eqnarray}
and then after some algebra we obtain the expressions 
\begin{eqnarray}
\hspace{-30pt}<0| T\ B_i(x) B_j(x) | 0> &  = & \delta_{ij}
\Delta_F(x-y, m^2)\ ,\\
\hspace{-30pt} <0| T\ B_{-}(x) B_i(y) | 0> & = &  \partial^x_i
\frac{1}{\Delta_\perp-m^2} *\left[ 2 \partial_{-}^x \Delta_F(x-y)
+ E_F^1(x_L - y_L) \delta^2(x_\perp - y_\perp)\right]\nonumber\\
& = & \partial_i^x \int_0^{x^+ -y^{-}} d\xi \Delta_F(\xi,
\vec{x} - \vec{y})\ ,\\
\hspace{-30pt}<0| T\ B_{-}(x) B_{-}(y) | 0> & = & 2 \partial^x_{-}
\Delta_\perp \frac{1}{({\Delta_\perp-m^2})^2} *\left[ 2 \partial_{-}^x
\Delta_F(x-y) 
+ E_F^1(x_L - y_L) \delta^2(x_\perp -
y_\perp)\right]\nonumber\\ 
 & - & m^2 E_F^2(x_L - y_L) \frac{1}{\Delta_\perp-m^2}*
\delta^2(x_\perp-y_\perp) \nonumber\\ 
& = & 2 \partial_{-}^x \int_0^{x^+ -y^{-}} d\xi \Delta_F(\xi,
\vec{x} - \vec{y})  +  m^2 \int_0^{x^+ -y^{-}} d\xi
\int_{0}^{\xi} d\eta \Delta_F(\eta, \vec{x} - \vec{y})\ ,
\nonumber\\
\end{eqnarray}
with the successively increasing number of independent modes
contributions\footnote{In the second and the third case we
have used the property (\ref{wlasDeltaFEF}) of the propagator
functions $\Delta_F(x)$ i $E_F^1(x)$.}. These all components
have a concise form in their Fourier representation 
\begin{eqnarray}
<0| T\ B_{\mu}(x) B_{\nu}(y) | 0> & =& i \int
\frac{d^4k}{(2\pi)^4} \frac{e^{i k\cdot (x-y)}}{k^2 - m^2 + i
\epsilon} \left( - g_{\mu \nu} + \frac{k_\nu N_\mu + k^\mu
N_\nu}{k_{+} + i \epsilon'{\rm sgn}(k_{-}) } \right.\nonumber\\ 
&& + m^2 \left. \frac{N_\mu N_\nu}{[k_{+} + i \epsilon'{\rm
sgn}(k_{-})]^2} \right),\label{31massWeylpropag}
\end{eqnarray}
where we have introduced the LF Weyl gauge vector $N_\mu=
(N_{-}= 1, N_{+} = 0, N_\perp = 0) $ which chooses the 
LF-Weyl gauge condition $N^\mu B_\mu = B_{+} = 0$. In this way 
we have obtained the causal ML-prescription for the spurious
poles in the vector field propagator via the canonical
quantization procedure on a single LF surface of quantization.
We see that the noncovariant modes $\Pi$ and $C_{-}$ are
inevitable for this encouraging result. Thus we can speculate 
that the canonical procedure at a single LF cannot lead to the
ML-prescription for the LC-gauge, where all noncovariant
nonphysical modes are excluded.

\newpage

\part{Light Front QED }

\section{Class of LF Weyl gauges}\label{chapivWeyl}

\setcounter{equation}{0}

Now we begin the discussion of the Abelian gauge field
models where the gauge transformation is a local symmetry of
the theory. Therefore now it is not a problem of having smoother
behaviour of canonical LF commutators but the fundamental
property of a gauge field prescription which makes us choose
some gauge fixing condition. The first choice that we 
discuss here is the class of LF Weyl gauge conditions which is
a generalization of the exact LF Weyl gauge $A_{+} = 0$. It is
usually introduced into the Lagrangian density by the so-called
{\em gauge fixing term} 
\begin{equation}
\frac{1}{2\alpha} \left( N^\mu A_\mu\right)^2\ ,
\label{guafixconklWeyla}
\end{equation}
where in our case the LF Weyl gauge vector is $N^\mu=(N^{+}= 1,
N^{-} = 0, N^\perp = 0)$. This term is very convenient in the
path-integral approach to the quantization of gauge field
theories \cite{AbersLee1973} because it constitutes a
non-singular quadratic part of gauge field Lagrangian thus
allowing for immediate inversion. However the path-integral
procedure gives no prescription for spurious poles in the gauge
field propagator which should have the form
\begin{eqnarray}
D_{\mu \nu}(x) & = & i \left < 0 \left | T A_{\mu } (x) A_\nu(0)
\right| 0 \right > \nonumber\\
& = &\int \frac{d^4k}{(2\pi)^4} {e^{- i k \cdot x}}\left\{
\frac{1}{2 k_{+} k_{-} - k_\perp^{2} + i \epsilon} \left[  
g_{\mu \nu} - \frac{N_\mu k_\nu + N_\nu k_\mu}{[ k_{+} + i
\epsilon '{\rm sgn}(k_{-})]}\right] - \alpha \frac{k_\mu
k_\nu}{[ k_{+} + i \epsilon '{\rm sgn}(k_{-})]^2}\right\}
\nonumber\\
&&\label{naiveAAprop} 
\end{eqnarray}
with the causal ML-prescription $ [k_{-} + i \epsilon '
{\rm sgn}(k_{+})]^{-1,2}$. We see that while the above
propagator is finite for $\alpha \rightarrow 0$ where it
describes the exact LF Weyl gauge, the expression
(\ref{guafixconklWeyla}) is ill-defined for $\alpha = 0$.
Therefore we choose another form of the gauge fixing
condition via the Lagrange multiplier field $\Lambda$
\begin{equation}
\Lambda  A_{+} - \frac{\alpha}{2} \Lambda^2,
\end{equation}
which is equivalent to the previous choice as long as $\alpha
\neq 0$. Evidently it is also regular for $\alpha =
0$ when it describes the exact LF-Weyl gauge $A_{+} = 0$. In
the present case $\Lambda$ is not a dynamical field, contrary to
the case of covariant gauges. Below we will  consecutively
discuss the sector of gauge fields in two distinct cases of 1+1 and D+1
dimensions.\\ 

\subsection{Gauge field sector in 1+1 dimensions}\label{11classWeylsect}

We want to discuss first the case of 1+1 dimensions where the
infrared singularities in the transverse momenta for
$k_\perp^2 = 0$ are excluded from the considerations. Thus we
can focus our attention on singularities in the longitudinal
momentum $k_{-}$. Just as we did for previous models,  also
here we start from  the sector of gauge fields coupled linearly
to external currents $j^\mu$
\begin{eqnarray}
{\cal L}^{1+1}_{\alpha Weyl} & = & \frac 1 2 \left(\partial_{+}
{A}_{-} - \partial_{-} A_{+} \right)^2 + {A}_{-} {j}^{-}
+ A_{+} j^{+} + \Lambda A_{+} - \frac{\alpha}{2} \Lambda^2,
\end{eqnarray} 
and the Euler-Lagrange equations are
\begin{eqnarray}
\partial_{+}\left( \partial_{+} {A}_{-} - \partial_{-} A_{+}
\right)  & = &  j^{-} \ , \label{11alWeELAm}\\
\partial_{-}\left( \partial_{+} {A}_{-} - \partial_{-} A_{+}
\right) &  = &  - j^{+} - \Lambda \ , \label{11alWeELLam}\\
A_{+} &  = & \alpha \Lambda.\label{11alWeELAp}
\end{eqnarray}
Without going into details, we give the canonical Hamiltonian
structure which follows from the above equations after removing
non-dynamical field variables $(A_{+}, \Lambda)$. The canonical
Hamiltonian density  
\begin{eqnarray}
{\cal H}^{can,1+1}_{\alpha Weyl} & = & = \frac 1 2
\Pi^{-}\left( 1 - \alpha 
\partial_{-}^2 \right) \Pi^{-} - A_{-} j^{-} +\alpha j^{+}
\partial_{-} \Pi^{-} + \frac{\alpha}{2} (j^{+})^2,
\label{11alWeyeffHam} 
\end{eqnarray}
and non-vanishing commutator at LF
\begin{equation}
\left[ \Pi (x^{+}, x^{-}), A_{-}(x^{+},y^{-}) \right]  =  - i \ 
\delta(x^{-} - y^{-})\label{11alpWeycomPiAm}.
\end{equation}
generate effective dynamical  equations
\begin{eqnarray}
\partial_{+}  \Pi^{-}&  =& j^{-}\label{1+1klasWeylhamPi}\\
\partial_{+} A_{-} & = & (1 - \alpha \partial_{-}^2 )\Pi^{-} -
\alpha \partial_{-} j^{+}.\label{1+1klasWeylhamAm}
\end{eqnarray} 
We stress that this quantum theory describes a larger system
than the physical excitations. This is due to the lack of 
the Gauss law as an equation of motion
\begin{equation}
G = \partial_{-} (\partial_{+} A_{-} - \partial_{-} A_{+})
 + j^{+} = \partial_{-} \Pi^{-} + j^{+} = 0.
\end{equation}
Evidently one cannot {\em strongly} impose the condition $G = 0$ 
because this would be in a conflict with the commutator
(\ref{11alpWeycomPiAm}); thus one has to implement it {\em
weakly} as a condition on states \cite{GaussLawlit}
\begin{equation}
\langle phys' | G(x) |phys \rangle = 0.\label{11alphaWeyGauss}
\end{equation}
The interaction part of the Hamiltonian (\ref{11alWeyeffHam}) 
indicates that external currents are coupled with the linear
combinations of fields
\begin{eqnarray}
\bar{A}_{-} & = & A_{-}\ ,\\
\bar{A}_{+} & = & - \alpha \partial_{-} \Pi^{-}\ ,
\end{eqnarray}
and also here there is an instantaneous interaction term
$\frac{\alpha}{2} j^{+}j^{+}$ which, during the Wick contraction
procedure, would modify the perturbative propagator from the
canonical chronological product to the following expression: 
\begin{eqnarray}
D_{\mu \nu}(x-y) = i \langle 0 | T^{+} \ \bar{A}_{\mu}(x)
\bar{A}_{\nu}(y) |0\rangle - \alpha g_{-\mu} g_{-\nu}\delta^2(x-y).
\end{eqnarray}
First we calculate the chronological product for independent
modes in the free field case when they can be represented by
\begin{eqnarray}
\Pi^{-}(x) & = & \pi(x^{-})\ ,\\
A_{-}(x) & = & x^{+} (1 - \alpha \partial_{-}^2 )\pi(x^{-}) +
a_{-}(x^{-}) \ ,
\end{eqnarray}   
with the Fourier representation
\begin{eqnarray}
a_{-}(x^{-}) & = & \int_0^\infty \frac{dk_{-}}{2 \pi}
 \ \left[e^{-i k_{-}x^{-}} a(k_{-}) + e^{+i k_{-} x^{-}}
a^{\dagger}(k_{-})\right]\label{Fouram11alpWey} \ ,\\
\pi(x^{-}) & = &\int_0^\infty \frac{dk_{-}}{2 \pi}
 \ \left[e^{-i k_{-}x^{-}} p(k_{-}) + e^{+i k_{-} x^{-}}
p^{\dagger}(k_{-})\right]\label{Fourpi11alpWey},
\end{eqnarray}
and the commutation relations for the creation and annihilation
operators\footnote{The weak Gauss law condition
(\ref{11alphaWeyGauss}) shows that for free fields, all
physical states are created by the $p(k)^{\dag}$ operators and
hence they have zero norm. This means that no physical photons
are present in 1+1 dimensions.}  
\begin{equation}
\left[ a_{-}(k_{-}), p^{\dagger}(k'_{-})\right]  =  
\left[ a_{-}^{\dag}(k_{-}), p(k'_{-})\right] =
i \ 2 \pi \ \delta(k_{-} - k'_{-}).\label{11alpWeycomampi}
\end{equation}
Now the free propagators are 
\begin{eqnarray}
\langle 0 | T^{+} \ \Pi^{-}(x) \Pi^{-}(y)|0\rangle & = & 0 \ ,\\
\langle 0 | T^{+} \ A_{-}(x) \Pi^{-}(y)|0\rangle & = &
E^1_F(x-y) \ ,\\
\langle 0 | T^{+} \ A_{-}(x) A_{-}(y) |0\rangle & = &
(1 - \alpha \partial_{-}^2 )E^2_F(x-y),
\end{eqnarray}
where noncovariant functions $E^1_F(x)$ and $E^2_F(x)$ are
defined in Appendix \ref{dodfenGrefun}. The perturbative gauge
field propagator has the causal form 
\begin{equation}
D_{\mu \nu}(x) = \int \frac{d^2k}{(2\pi)^2} {e^{- i k \cdot
x}}\left\{ \frac{1}{2 k_{+} k_{-} + i \epsilon} \left[g_{\mu
\nu} - \frac{N_\mu k_\nu + N_\nu k_\mu}{[ k_{+} + i \epsilon
'{\rm sgn}(k_{-})]}\right] - \alpha \frac{k_\mu k_\nu}{[ k_{+}
+ i \epsilon '{\rm sgn}(k_{-})]^2}\right\}, 
\end{equation}
and simultaneously the current interaction term is omitted in
the interaction Hamiltonian. Then one could incorporate the
fermion interaction, but we will not discuss it here because
this would be a mere repetition of the relevant considerations from
the subsection (\ref{ProcaSmatrix}) with a rather evident
neglect of transverse coordinates and components. \\
We conclude that in 1+1 dimensions, the class of the LF Weyl
gauges effectively leads to the causal  form of
the perturbative Feynman rules though at the canonical level it
contains non-covariant terms which ultimately cancel in pairs.\\

\subsection{Gauge field sector in D+1 dimensions}

As a physically relevant model we take the higher-dimensional
case where excitations of physical photons are possible. However 
if we would take 3+1 dimensions then we would encounter
infrared singularities connected with the inverse Laplace
operator in 2 transverse directions. Therefore we have decided
to use here the dimensional regularization of this singularity
by working with $d= D - 1> 2$ transverse coordinates $x_\perp =
(x_2, x_3, \ldots , x_{D})$. Thus we would like to start with
the Lagrangian density 
\begin{eqnarray}
{\cal L}_{\alpha Weyl} & = & (\partial_{+} {A}_i - \partial_i
A_{+}) ( \partial_{-} {A}_i - \partial_i {A}_{-}) + \frac 1 2
\left( \partial_{+} {A}_{-} - \partial_{-} {A}_{+} \right)^2 -
\frac 1 4 \left(\partial_i {A}_j- \partial_j {A}_i\right)^2
\nonumber\\
&+ & A_{+} \Lambda - \frac{\alpha}{2} \Lambda^2 + {A}_{+}
{j}^{+}+ {A}_{-} {j}^{-} + {A}_i j^i \label{D+1alphWeylLagr}, 
\end{eqnarray}   
which generates the Euler-Lagrange equations
\begin{eqnarray}
\partial_{+}\left( \partial_{+}{A}_{-} - \partial_{-} A_{+} -
\partial_j A_j \right) & = & - \Delta^d_\perp A_{+}+ j^{-}\ ,
\label{eqAm}\\ 
- \partial_{-}\left( \partial_{+}{A}_{-} - \partial_{-} A_{+} +
\partial_j A_j \right) & = & - \Delta^d_\perp A_{-} + j^{+} +
\Lambda\ ,\\ 
\left( 2\partial_{+} \partial_{-} - \Delta^d_\perp
\right){A}_{i} & = & \partial_{i}\left( \partial_{+}{A}_{-} +
\partial_{-}A_{+} - \partial_j A_j \right) + {j}^{i}\ ,
\label{eqAi} \\
A_{+} & = & \alpha \Lambda\ ,
\end{eqnarray}
where $\Delta^d_{\perp} = (\partial_j)^2 $ denotes the Laplace
operator in $d>2$ dimensions. As the independent modes we take
the modified canonical momentum 
\begin{eqnarray}
\Pi = \partial_{+}A_{-} - \partial_i A_i -\partial_{-}A_{+}
\end{eqnarray}
and the gauge fields
\begin{eqnarray}
C_{i} & = & A_i - \partial_i \frac{1}{\Delta^d_\perp} * (\Pi +
2 \partial_j A_j) \ , \label{alpWeyfromCitoAi}\\
C_{-} & = & A_{-} - \partial_{-}\frac{1}{\Delta^d_\perp} * (\Pi
+ 2 \partial_j A_j) \, \label{alpWeyfromCmtoAm}
\end{eqnarray} 
thus the dynamical equations of motion are
\begin{eqnarray}
\left( 2\partial_{+} \partial_{-} - \Delta_\perp \right){C}_i
& = & j^i - 2 \partial_i \frac{1}{\Delta_\perp^d}*
(\partial_{-} j^{-} + \partial_j j^j) \ , \label{alpWeyHameqCi}\\ 
\partial_{+} C_{-} & = &- \frac{1}{\Delta_\perp^d} * \left(
\partial_i j^i + \partial_{-} j^{-} \right)\ , 
\label{alpWeyHameqCm}\\ 
\partial_{+} \Pi & = & - \alpha (\Delta_\perp^d)^2 C_{-}+   j^{-}
+ \alpha \Delta_\perp^d j^{+}.\label{alpWeyHameqPi}
\end{eqnarray}
The canonical Hamiltonian density can be expressed in terms of
the independent modes
\begin{eqnarray}
{\cal H}^{eff}_{\alpha Weyl} & = & \frac 1 2 (\partial_i C_j)^2 +
\frac{\alpha}{2} \left[ \Delta_\perp^d C_{-} - j^{+}\right]^2 -
  \Pi \frac{1}{\Delta_\perp^d} *  \left[\partial_i j^{i} + \partial_{-}
j^{-} \right] \nonumber\\
& - & C_i \left[j^i - 2\partial_i \frac{1}{\Delta_\perp^d} *
(\partial_{-}j^{-} + \partial_j j^j) \right] - C_{-} j^{-}\ ,
\label{alpWeyeffham} 
\end{eqnarray}
and the non-vanishing commutators at LF are 
\begin{eqnarray}
\left [\Pi (x^{+},\vec{x}), C_{-}(x^{+},\vec{y}) \right] & = &-
i \delta^{d+1}(\vec{x} - \vec{y}) \ , \label{alpWeycomPiCm}\\ 
\left [2 \partial_{-} C_i (x^{+},\vec{x}), C_j(x^{+},\vec{y})
\right] & = & - i \delta_{ij} \delta^{d+1}(\vec{x} -\vec{y})
\label{alpWeycomCiCj}. 
\end{eqnarray}
Just like in the low-dimensional case, we have the
instantaneous current interaction which, during the Wick
contractions, modifies the perturbative gauge field propagator
from the chronological product form
\begin{eqnarray}
D_{\mu \nu}(x-y) = i \langle 0 | T^{+} \ \bar{A}_{\mu}(x)
\bar{A}_{\nu}(y) |0\rangle - \alpha g_{-\mu} g_{-\nu}
\delta^{d+2}(x-y)\ , \label{alpWeyperprop} 
\end{eqnarray}
where we have introduced the notation
\begin{eqnarray}
\bar{A}_{i}& = & C_i - \partial_i \frac{1}{\Delta_\perp^d} *
(\Pi + 2 \partial_j C_j)\ ,\\ 
\bar{A}_{-}& = & C_{-} - \partial_{-} \frac{1}{\Delta_\perp^d}
* (\Pi + 2 \partial_j C_j)\ ,\\ 
\bar{A}_{+}& = & - {\alpha}  \Delta_\perp^d C_{-}.
\end{eqnarray}
The free propagators for independent modes are calculated
directly from the Fourier representations
\begin{eqnarray}
C_{-} (x) & = & c_{-}(\vec{x}) = \int_{-\infty}^{\infty}
\frac{d^dk_\perp}{(2 \pi)^d} \int_{0}^{\infty}
\frac{dk_{-}}{2\pi}  \ \left[e^{-i \vec{k}\cdot
\vec{x}} a(\vec{k}) + e^{+i \vec{k}\cdot \vec{x}}
a^{\dagger}(\vec{k})\right]\ , \\
C_{i}(x) & = & \int_{-\infty}^{\infty} \frac{d^dk_\perp}{(2
\pi)^d} \int_{0}^{\infty} \frac{dk_{-}}{2\pi\ 2k_{-}}
 \ \left[e^{-i {k}\cdot{x}} c_i(\vec{k}) + e^{+i
{k}\cdot {x}} c^{\dagger}_i(\vec{k})\right]_{k_{+} =
\frac{k_\perp^2}{2k_{-}}}\ , \\
\Pi(x) & = & \pi(\vec{x}) - \alpha x^{+}(\Delta_\perp^d)^2
c_{-}(\vec{x})\ , \\ 
\pi(\vec{x}) & = & \int_{-\infty}^{\infty} \frac{d^dk_\perp}{(2
\pi)^d} \int_{0}^{\infty} \frac{dk_{-}}{2\pi}
 \ \left[e^{-i \vec{k}\cdot
\vec{x}} p(\vec{k}) + e^{+i \vec{k}\cdot \vec{x}}
p^{\dagger}(\vec{k})\right],
\end{eqnarray}
where the non-vanishing commutators for the creation and annihilation
operators are\footnote{We recognize that the present operators $a^{\dag}$
and ${p}^{\dag}$ are trivial generalizations of the respective
operators in 1+1 dimensions. Free field physical states,
selected by the weak Gauss law, can be created by
$c_i^{\dag}$ and $p^{\dag}$ excitations, where the former would
be the positive norm photon states while the latter would be
the accompanying  zero norm states.}
\begin{eqnarray}
\left[ a(\vec{k}), p^{\dagger}(\vec{k'})\right] & = & 
\left[ a^{\dagger}(\vec{k}), p(\vec{k'})\right] = i (2\pi)^{d+1}
\delta^{d+1}(\vec{k} - \vec{k'})\ , \label{alpWeycomrelcp}\\ 
\left[ c_{i}(\vec{k}), c^{\dagger}_j(\vec{k'})\right] & =
&(2\pi)^{d+1} \ 2 k_{-} \ \delta_{ij}\delta^{d+1}(\vec{k} -
\vec{k'}),
\end{eqnarray}
where $\delta^{d+1}(\vec{k}) = \delta^d(k_\perp)\delta(k_{-})$.
Determination of free propagators for independent modes is
rather simple 
\begin{eqnarray}
\langle 0 | T\ C_{-}({x}) \Pi({y})|0\rangle & = &
E_F^{1}(x_L -y_L)\delta^d(x_\perp - y_\perp) \ ,\\
\langle 0 | T\ \Pi({x}) \Pi({y})|0\rangle & = & \alpha
(\Delta_\perp^d)^2 E_F^{2}(x_L -y_L)
\delta^d(x_\perp - y_\perp) \ , \\ 
\langle 0 | T\ C_{i}({x}) C_j({y})|0\rangle & = &
\delta_{ij} D_F^{d+2}(x-y)\ ,
\end{eqnarray}
where the covariant Feynman propagator function $D^{d+2}_F(x)$
is defined in Appendix \ref{dodfenGrefun}, while the
calculation of the perturbative gauge field propagator is quite
tedious; here is the final result presented in concise Fourier
representation 
\begin{equation}
D_{\mu \nu}({x}-{y}) = \int \frac{d^{d+2} k}{(2 \pi)^{d+2}}
{e^{-i {k}\cdot 
({x}-{y})}} \left\{\frac{1}{k^2 + i \epsilon} \left[ g_{\mu \nu}
- \frac{(k_\mu N_\nu + k_\nu N_\mu)}{k_{+}
+ i\ \epsilon' \ {\rm sgn}(k_{-})}\right] - \alpha \frac{k_\mu k_\nu}{[k_{+} 
+ i\ \epsilon' \ {\rm sgn}(k_{-})]^2}\right\}, \label{31clasweylprop}
\end{equation}
which evidently has a regular limit $d \rightarrow 2$, where it
gives the expected  result (\ref{naiveAAprop}).

\section{General axial gauge }\label{sectiongenaxigau}

\setcounter{equation}{0}

In this section we would like to discuss the general axial
gauge condition imposed on the gauge field potential 
$n^\mu A_\mu = 0 $, where the axial gauge vector has the form
$n^{+} = 1, n^{-} = - \alpha, n^\perp = 0$. This general choice
will allow us to analyse and compare within the LF canonical
formalism different gauge conditions: 
\begin{itemize}
\item the LF-Weyl  - for  $\alpha = 0$ ;
\item the temporal Minkowski - for $\alpha = - 1$ ;
\item the spatial Minkowski  - for $\alpha = 1$.
\end{itemize}
Also the verification whether the limit $\alpha \rightarrow \pm
\infty$, can be considered as a possible limiting procedure
leading to the LC gauge $A_{-}= 0$. The present form of the
gauge condition $n^\mu A_\mu = A_{+} - \alpha A_{-} = 0 $ can
be implemented either explicitly or via the Lagrange multiplier
field; below we take the first possibility. Also here we
discuss two cases in 1+1 dimensions and in 3+1 dimensions
because they have quite different physical interpretations.
Again we deal explicitly only with the gauge field sector
because the discussion of the fermions and interactions with
them would follow along the lines described in Section
\ref{rozdzialcechLor}.\\ 

\subsection{Gauge fields in 1+1 dimensions}

Now the Lagrangian density is greatly simplified
\begin{eqnarray}
{\cal L}_{gen axi}^{1+1} & = &  \frac 1 2
\left(\partial_{+} {A}_{-} - \alpha \partial_{-}{A}_{-}\right)^2 -
+ {A}_{-} ({j}^{-} + \alpha {j}^{+}), 
\label{genaxiLagralpha} 
\end{eqnarray}
and generates only one Euler-Lagrange equation
\begin{eqnarray}
(\partial_{+} - \alpha \partial_{-})^2   A_{-} & = & 
j^{-} + \alpha j^{+}\ ,
\end{eqnarray}
which is equivalent to the system of the first order equations
\begin{eqnarray}
\partial_{+} \Pi & = & \alpha \partial_{-} \Pi + j^{-} + \alpha
j^{+} \ , \label{genaxi11alphaELeqPi}\\ 
\partial_{+} A_{-} & = & \Pi + \alpha
\partial_{-} A_{-}. \label{genaxi11alphaELeqAm}
\end{eqnarray}
When these equations are assumed as the Hamilton equations
with the Hamiltonian density
\begin{eqnarray}
{\cal H}_{can}^{1+1} & = & \Pi^{-}\partial_{+}A_{-}  - {\cal L}_{temp}
 =  \frac 1 2 (\Pi^{-})^2  - \alpha  A_{-}\partial_{-} \Pi^{-} +  
- \alpha (\partial_i A_{-})^2 - A_{-} (j^{-} + \alpha j^{+}) \
,   \label{genaxi11alphacanHam} 
\end{eqnarray}
then the non-vanishing Poisson bracket is canonical and the
quantum commutator is 
\begin{equation}
\left[ \Pi (x^{-}), A_{-}(y^{-}) \right] =  - i
\delta({x}^{-} - {y}^{-})\label{genaxi11alphaPiAmDB}.
\end{equation}
Inspecting the Hamiltonian we find no instantaneous
interactions of currents, thus the perturbative and free
propagators will coincide here. The quantum free fields can be
given as 
\begin{eqnarray}
\Pi(x) & = & \pi(\alpha x^{+} + x^{-}) \ , \\
A_{-}(x) & = & a_{-}(\alpha x^{+} + x^{-}) + x^{+} \pi(\alpha
x^{+} + x^{-})\ ,
\end{eqnarray}
where the fields $\pi(x)$ and $a_{-}(x)$ were defined by
(\ref{Fouram11alpWey}) and (\ref{Fourpi11alpWey}) in Subsection
\ref{11classWeylsect}. Therefore the whole discussion given
there applies also here and in order to avoid unnecessary
repetitions, we just take the results
\begin{eqnarray}
\langle 0 | T\ A_{-}(x) \Pi(y)|0\rangle & = & E^1_F
[x^{+}-y^{-}, \alpha (x^{+} - y^{+}) + x^{-} - y^{-}] \ ,\\ 
\langle 0 | T\ A_{-}({x}) A_{-}({y})|0\rangle & = &
  E^2_F[x^{+}-y^{-}, \alpha (x^{+} - y^{+}) + x^{-} - y^{-}].
\end{eqnarray}
Next we easily find the form of the perturbative propagators
\begin{eqnarray}
D_{--}(x) & = & \langle 0 | T\ A_{-}({x}) A_{-}(0)|0\rangle = i
\int_{-\infty}^{\infty} \frac{d^2{k}}{(2 \pi)^2} e^{-i 
{k}\cdot x} \frac{1}{[k_{+} - \alpha k_{-} + i \epsilon \ {\rm
sgn}\ (k_{-})]^2} \ ,\nonumber\\
&&\\
D_{+-}(x) & = & \alpha \langle 0 | T\ A_{-}({x}) A_{-}(0)|0\rangle = i
\int_{-\infty}^{\infty} \frac{d^2{k}}{(2 \pi)^2} e^{-i 
{k}\cdot x} \frac{\alpha}{[k_{+} - \alpha k_{-} + i \epsilon \ {\rm
sgn}\ (k_{-})]^2}\ , \nonumber\\
&&\\
D_{++}(x) & = & \alpha^2 \langle 0 | T\ A_{-}({x}) A_{-}(0)|0\rangle = i
\int_{-\infty}^{\infty} \frac{d^2{k}}{(2 \pi)^2} e^{-i 
{k}\cdot x} \frac{\alpha^2}{[k_{+} - \alpha k_{-} + i \epsilon \ {\rm
sgn}\ (k_{-})]^2}\ ,\nonumber\\
&&
\end{eqnarray}
which  can be written concisely as the Fourier integral
\begin{eqnarray}
D_{\mu \nu}(x) & = & i 
\int_{-\infty}^{\infty} \frac{d^2{k}}{(2 \pi)^2} \frac{e^{-i
{k}\cdot x}}{ 2 k_{+} k_{-} + i\epsilon}\left[
- g_{\mu \nu} + \frac{n_\mu k_\nu  + n_\nu k_\mu}{k_{+} -
\alpha k_{-} + i \epsilon \ {\rm 
sgn}\ (k_{-})} \nonumber\right.\\
&&\left.- n^2 \frac{k_\mu k_\nu }{[k_{+} -
\alpha k_{-} + i \epsilon \ {\rm 
sgn}\ (k_{-})]^2}\right] \ .
\end{eqnarray}
We see that this propagator exists for all finite values of
$\alpha$ with the ML-prescription for spurious poles.  Thus
even for the spatial axial gauge we can produce a gauge field
propagator with the causal spurious poles when quantizing
canonically at LF, contrary to the ET formalism where such case
is not possible \cite{Landshoffaxial}. When one takes the limit
$\alpha \rightarrow \pm \infty$, the causal nature of spurious
poles in the above propagator is lost and one again has the
LC-gauge case with the CPV poles.\\

\subsection{Gauge fields in 3+1 dimensions}

Now we would like to analyse the general axial gauge condition
in the more physically relevant case of 3+1 dimensions. Here
again the gauge condition is implemented  explicitly in the
Lagrangian density 
\begin{eqnarray}
{\cal L}_{alpha} & = & (\partial_{+} A_i - \alpha \partial_{i} {A}_{-})
( \partial_{-} {A}_i - \partial_i {A}_{-}) + \frac 1 2
\left(\partial_{+} {A}_{-} - \alpha \partial_{-}{A}_{-}\right)^2 -
\frac 1 4 \left( \partial_i {A}_j- \partial_j {A}_i\right)^2 
\nonumber\\
&& + {A}_{-} ({j}^{-} + \alpha {j}^{+}) + {A}_i j^i\ ,
\label{genaxiLagralpha'} 
\end{eqnarray}
where the external currents $j^\mu$ describe interactions with
the charged matter. The canonical analysis starts with the
Euler-Lagrange equations 
\begin{eqnarray}
\left[ (\partial_{+} - \alpha \partial_{-})^2 + 2 \alpha
\Delta_\perp \right] A_{-} & = & \left( \partial_{+} + \alpha
\partial_{-} \right)  \partial_i A_i  + j^{-} + \alpha j^{+}\ ,\\
\left( 2\partial_{+} \partial_{-} - \Delta_\perp \right){A}_{i}
& = & \partial_{i}\left[ (\partial_{+} + \alpha \partial_{-}
){A}_{-} - \partial_{j}{A}_{j}\right] + {j}^{i}\ ,
\end{eqnarray}
which are equivalent to the Hamilton equations
\begin{eqnarray}
\partial_{+} \Pi & = & \alpha \partial_{-} \left( \Pi + 2
\partial_i A_i \right) - 2 \alpha \Delta_\perp A_{-} + j^{-} +
\alpha j^{+} \ , \label{genaxialphaELeqPi}\\ 
\left( 2\partial_{+} \partial_{-} - \Delta_\perp \right){A}_{i}
& = & \partial_{i}\left( \Pi + 2 \alpha \partial_{-} A_{-}
\right) + j^{i} \ , \label{genaxialphaELeqAi}\\ 
\partial_{+} A_{-} & = & \Pi + \partial_i A_i + \alpha
\partial_{-} A_{-} \label{genaxialphaELeqAm}\ .
\end{eqnarray}
The canonical Hamiltonian density is
\begin{equation}
{\cal H}_{can}  =   \frac 1 2 (\Pi)^2 + \frac 1 2 (\partial_i A_j)^2  
+ \Pi \partial_i A_i - \alpha A_{-}\partial_{-} (\Pi + 2 \partial_i
A_i) - \alpha (\partial_i A_{-})^2 - A_{-} (j^{-} + \alpha
j^{+}) - A_i j^i\label{genaxigenaxihamcan}
\end{equation}
with the Dirac brackets
\begin{eqnarray}
\left\{ \Pi (\vec{x}), A_{-}(\vec{y}) \right\}_{DB} & = & -
\delta^3(\vec{x} - \vec{y})\ , \label{genaxialphaPiAmDB}\\
\left\{ 2 \partial_{-} A_i (\vec{x}), A_j(\vec{y}) \right\}_{DB}
& = & - \delta_{ij} \delta^3(\vec{x} - \vec{y})\ ,
\label{genaxialphaAiAjDB} 
\end{eqnarray}
while other brackets vanish. When trying to separate these independent
variables into the independent modes we encounter the problem
of inverting the differential operator at LF
$\Delta_{\alpha}^{-} = 2 \alpha \partial^2_{-} - \Delta_\perp$
which, only for $\alpha < 0$, is an elliptic operator with the
regular Green function (defined in 
Appendix \ref{InvLaploperdod}). Thus we 
have to choose only negative values of $\alpha$ which is
equivalent to the selection of temporal axial
gauges.\footnote{The cases of spatial gauges ($\alpha > 0$)
will not be discussed here and are left for future
investigations, while the null gauge ($\alpha = 0$) is singular
at 3+1 dimensions and needs some infrared regularizations.}
Now we define independent modes 
\begin{eqnarray}
\Lambda & = &  \Pi - 2 \alpha \partial_{-}^2
\frac{1}{\Delta_\alpha^{-}}* \left[ \Pi - 2 \alpha \partial_{-} 
A_{-} +  2 \partial_j A_j \right]\ ,\\
C_{-} & = & A_{-} + \partial_{-} \frac{1}{\Delta_\alpha^{-}}*
\left[ \Pi - 2 \alpha \partial_{-} A_{-} + 2 \partial_j A_j
\right] \ ,\\ 
C_i & = & A_i + \partial_i \frac{1}{\Delta_\alpha^{-}}* \left[
\Pi - 2 \alpha \partial_{-} A_{-} + 2 \partial_j A_j \right],
\end{eqnarray}
which satisfy dynamical equations of motion
\begin{eqnarray}
(\partial_{+} - \alpha \partial_{-} )C_{-} & = &
\frac{1}{\Delta_\alpha^{-}} * \left[ \partial_k j^k +
\partial_{-} (j^{-} + \alpha j^{+})\right]\ , \\
(\partial_{+} - \alpha \partial_{-} ) \Lambda & = & 2 \alpha
\Delta_{\alpha}^{-} C_{-} + j^{-} + \alpha j^{+}
- 2 \alpha \partial_{-} \frac{1}{\Delta_\alpha^{-}} *
\left[   \partial_k j^k  + \partial_{-} (j^{-} + \alpha j^{+})
\right]\ , \\
(2 \partial_{+} \partial_{-} - \Delta_\perp ) C_i & = & j^i
+ 2 \partial_i  \frac{1}{\Delta_{\alpha}^{-}}* \left[
\partial_k j^k + \partial_{-} (j^{-} + \alpha j^{+}) \right]
\ , 
\end{eqnarray} 
and have non-vanishing commutators at LF
\begin{eqnarray}
2 \partial_{-}^x \left[ C_i(\vec{x}), C_j(\vec{y})\right] & = &
-i \delta_{ij} \delta^3(\vec{x} - \vec{y})\ ,\\
\left[ \Lambda(\vec{x}), C_{-}(\vec{y})\right] & = &
- i \delta^3(\vec{x} - \vec{y}).
\end{eqnarray}
The Hamiltonian density (\ref{genaxigenaxihamcan}), when
expressed in term of these modes
\begin{eqnarray}
{\cal H}_{can} & = & \frac 1 2 (\partial_i C_j)^2 + \alpha \Lambda
\partial_{-} C_{-} - \alpha C_{-} \Delta^{-}_\alpha C_{-}
 - j^i\left[ C_i +  \partial_i  
\frac{1}{\Delta_\alpha^{-}}* \left[ \Lambda - 2 \alpha \partial_{-}
C_{-} + 2 \partial_j C_j \right] \right] \nonumber\\
& - & (j^{-} + \alpha j^{+}) 
\left[ C_{-} +  \partial_{-} 
\frac{1}{\Delta_\alpha^{-}}* \left[ \Lambda - 2 \alpha \partial_{-}
C_{-}  + 2 \partial_j C_j\ , 
\right]\right]
\end{eqnarray}
contains no direct interactions of currents, therefore the
perturbative propagators are the chronological product of
free fields. For free fields we write  
\begin{eqnarray}
C_{-}(x) & = & c_{-}(\alpha x^{+} + x^{-}, x_\perp)\ ,\\
\Lambda(x) & = & 2 \alpha x^{+} \Delta_{\alpha}^{-} c_{-}(\alpha
x^{+} + x^{-}, x_\perp) + \lambda(\alpha x^{+} + x^{-},
x_\perp)\ ,\\ 
C_i(x)& = & c_i(x),
\end{eqnarray}
where 
\begin{eqnarray}
c_{-}(\vec{x}) & = & \int_{-\infty}^{\infty} \frac{d^2k_\perp}{(2
\pi)^2} \int_{0}^{\infty} \frac{dk_{-}}{2\pi} \ \left[e^{-i
\vec{k}\cdot \vec{x}} a(\vec{k}) + e^{+i \vec{k}\cdot \vec{x}}
a^{\dagger}(\vec{k})\right]\ ,\\
\lambda(\vec{x}) & = &\int_{-\infty}^{\infty} \frac{d^2k_\perp}{(2
\pi)^2} \int_{0}^{\infty} \frac{dk_{-}}{2\pi} \ \left[e^{-i
\vec{k}\cdot \vec{x}} p(\vec{k}) + e^{+i \vec{k}\cdot \vec{x}}
p^{\dagger}(\vec{k})\right]\ ,\\
c_{i}(x) & = & \int_{-\infty}^{\infty} \frac{d^2k_\perp}{(2
\pi)^2} \int_{0}^{\infty} \frac{dk_{-}}{2\pi\ 2k_{-}}
 \ \left[e^{-i {k}\cdot{x}} c_i(\vec{k}) + e^{+i
{k}\cdot {x}} c^{\dagger}_i(\vec{k})\right]_{k_{+} =
\frac{k_\perp^2}{2k_{-}}}\ ,
\end{eqnarray}
with the commutators for creation and annihilation operators
\begin{eqnarray}
\left[ a(\vec{k}), p^{\dagger}(\vec{k'})\right] & = & 
\left[ a^{\dagger}(\vec{k}), p(\vec{k'})\right] = i (2\pi)^{3}
\delta^{3}(\vec{k} - \vec{k'})\ , \label{genaxicomrelcp}\\ 
\left[ c_{i}(\vec{k}), c^{\dagger}_j(\vec{k'})\right] & =
&(2\pi)^{3} \ 2 k_{-} \ \delta_{ij}\delta^{3}(\vec{k} -
\vec{k'})\ ,
\end{eqnarray}
while other commutators vanish. Now we easily find the relevant
chronological products
\begin{eqnarray}
\langle 0 | T\ C_{-}({x}) \Lambda({y})|0\rangle & = &
E_{\alpha F}^{1}(x_L - y_L)\delta^2(x_\perp - y_\perp) \ ,\\ 
\langle 0 | T\ \Lambda({x}) \Lambda({y})|0\rangle & = &
- 2 \alpha \Delta_{\alpha}^{-}E_{\alpha F}^{2}(x_L - y_L)
\delta^2(x_\perp - y_\perp) \ ,\\
\langle 0 | T\ C_{i}({x}) C_j({y})|0\rangle & = &
\delta_{ij} D_F^{4}(x-y),
\end{eqnarray}
where $E^{1,2}_{\alpha F} (x_L) = E^{1,2}_F(x^{+}, x^{-} +
\alpha x^{+})$.
The perturbative propagators contain the 
following linear combinations of independent modes:
\begin{eqnarray}
\Pi & = & \Lambda - 2 \alpha \partial_{-}^2
\frac{1}{\Delta_\alpha^{-}}* \left[ \Lambda - 2 \alpha
\partial_{-} C_{-} + 2 \partial_j C_j \right] \ ,
\label{genaxigenaxirevPi} \\ 
A_{-} & = & C_{-} + \partial_{-} \frac{1}{\Delta_\alpha^{-}}*
\left[ \Lambda - 2 \alpha \partial_{-} C_{-} + 2 \partial_j C_j
\right]\ ,  \label{genaxigenaxirevAm}\\ 
A_i & = & C_i + \partial_i \frac{1}{\Delta_\alpha^{-}}* \left[
\Lambda - 2 \alpha \partial_{-} C_{-} + 2 \partial_j C_j
\right] \ , \label{genaxigenaxirevAi}
\end{eqnarray}
and we find after some algebra
\begin{eqnarray}
\langle 0 | T\ A_{-}({x}) A_{-}({y})|0\rangle & = &
i 2 \partial_{-}^x \partial_{-}^x \left( D_F * E_{\alpha
F}^2\right) (x-y)\ ,\\
\langle 0 | T\ A_{i}({x}) A_{-}({y})|0\rangle & = &
i 2 \partial_{i}^x (\alpha \partial_{-}^x + \partial_{+})
 \left( D_F * E_{\alpha F}^2\right) (x-y)\ , \\
\langle 0 | T\ A_{i}({x}) A_{j}({y})|0\rangle & = & \delta_{ij}
D_F(x-y) + i 2 \alpha \partial_{i}^x \partial_{i}^x \left( D_F
* E_{\alpha F}^2\right) (x-y).
\end{eqnarray}
Next, using the relation $A_{+} = \alpha A_{-}$, we find the
Fourier representation for all components of the gauge field
propagator 
\begin{eqnarray}
\langle 0 | T\ A_{\mu}({x}) A_{\nu}({y})|0\rangle & = & 
i \int \frac{d^{4} k}{(2 \pi)^{4}} {e^{-i {k}\cdot({x} -
{y})}}\left\{ \frac{1}{2 k_{+} k_{-} - k_\perp^2 + i \epsilon}
\left[ - g_{\mu \nu} + \frac{(k_\mu n_\nu + k_\nu n_\mu)}{k_{+}
- \alpha k_{-} + i\ \epsilon' \ {\rm sgn}(k_{-})}\right]
\right. \nonumber\\ 
&&\left. - n^2 \frac{k_\mu k_\nu}{[k_{+} - \alpha k_{-}+ i\
\epsilon' \ {\rm sgn}(k_{-})]^2} \right\}. \label{31genaxiprop}
\end{eqnarray}
Thus we have found the perturbative gauge field propagator with
the ML-prescription for spurious poles for all temporal gauges.
Now the LF Weyl gauge can be understood as the limit $\alpha
\rightarrow 0$ taken in (\ref{31genaxiprop}).
 One can compare all LF Weyl propagators 
obtained via different routines: the massive electrodynamics
(\ref{31massWeylpropag}) when $m^2 \rightarrow 0$, the class of
the LF Weyl gauges (\ref{31clasweylprop}) when $\alpha
\rightarrow 0$, and the above general axial gauge
(\ref{31genaxiprop}) when $\alpha \rightarrow 0$ leading always
to the same result 
\begin{equation}
\langle 0 | T\ A_{\mu}({x}) A_{\nu}({y})|0\rangle_{LF Weyl}  =  
i \int \frac{d^{4} k}{(2 \pi)^{4}} \frac{e^{-i
{k}\cdot({x}-{y})}}{2 k_{+} k_{-} - k_\perp^2 +
i \epsilon} \left[ - g_{\mu \nu} 
+ \frac{(k_\mu N_\nu + k_\nu N_\mu)}{k_{+} + i\ \epsilon' \
{\rm sgn}(k_{-})}\right] \label{genLFWeylpropag}.
\end{equation}
The other passage limit to null gauge, when $\alpha \rightarrow \pm
\infty $  changes again the ML-prescription into the CPV for the
LC-gauge propagator. \\

\section{Flow covariant gauge}\label{modLorchap}
 
\setcounter{equation}{0}

The analysis presented in Section \ref{rozdzialcechLor} can be
considered as an infrared regularized model which in the limit
$m^2 \rightarrow 0$ leads to the QED for the class of covariant
Lorentz gauges $\partial_\mu A^\mu = \alpha \Lambda $.
Therefore here, instead of repeating the previous analysis
for explicitly massless vector gauge fields, we decided to study
another class of gauge conditions: the flow covariant gauge
$\partial_{+} A_{-} + \alpha \partial_{-}A_{+} - \alpha
\partial_\perp A_\perp = 0$ which for $\alpha = 1$ produces the
results for the Lorentz gauge while for $\alpha = 0$ it describes
the LC-gauge $A_{-} = 0$.\footnote{Strictly speaking it leads
to the modified LC-gauge $\partial_{+}A_{-} = 0$ which should
have the same perturbative Feynman rules as the LC-gauge.}

\subsection{Model in 1+1 dimensions} 

First we would like to discuss the electrodynamics with charged
fermion fields in 1+1 dimensions which is classically 
described by the Lagrangian density 
\begin{equation}
{\cal L}^{1+1 QED }_{flow}  = 
\frac 1 2 \left(\partial_{+} {A}_{-} -
\partial_{-}{A}_{+}\right)^2 
+  \bar{\psi} \left( i \gamma^\mu \partial_\mu - e \gamma^\mu
{A}_{\mu} - M\right) \psi + 
{\Lambda}\left( \partial_{+} A_{-} + \alpha \partial_{-} A_{+} 
\right) \ ,
\end{equation}
where the Lagrange multiplier field $\Lambda$ implements the
flow covariant gauge. Here we explicitly see that the flow
gauge is attainable, by some suitable gauge transformation, for
all values of the gauge parameter $\alpha \neq 1$. Thus we expect
that we may encounter expressions  singular at $\alpha =
-1$ during the canonical procedure and in final results. \\

\subsubsection{Gauge field sector}

According to our previous routine, we start with the
sector of gauge field coupled with the arbitrary external sources
$j^\mu$
\begin{eqnarray}
{\cal L}^{1+1}_{mLor} & = & \frac 1 2 \left(\partial_{+}
{A}_{-} - \partial_{-}{A}_{+}\right)^2 + {A}_{-} {j}^{-} +
{A}_{+} {j}^{+} + \Lambda \left( \partial_{+} A_{-} + \alpha
\partial_{-} A_{+} \right)\ ,
\end{eqnarray}
and here we have the Euler-Lagrange equations for the gauge potential
and the Lagrange multiplier fields
\begin{eqnarray}
\partial_{+}\left( \partial_{+}A_{-} - \partial_{-} A_{+} +
\Lambda \right)  & = &  j^{-}\ , \label{11mlorELeqAm}\\ 
\partial_{-}\left( \partial_{-}A_{+} - \partial_{+} A_{-} +
\alpha \Lambda \right)  & = &  j^{+} \ ,\label{11mlorELeqAp}\\ 
\partial_{+} {A}_{-} & = & - \alpha \partial_{-} A_{+} 
 \label{11mlorcovgauge}.
\end{eqnarray}
However two fields can be parameterized
\begin{eqnarray}
\Lambda & = & \frac{1}{1 + \alpha} \left( \Pi^{-} +
\frac{1}{\partial_{-}} * j^{+} \right)\ , \\
A_{+} & = & \frac{\alpha}{(1 + \alpha)^2}
\frac{1}{\partial_{-}} \left( \alpha \Pi^{-} -
\frac{1}{\partial_{-} }* j^{+}\right)\ ,
\end{eqnarray}
by means of the canonical conjugate momentum field $\Pi^{-}$
and effectively there are only two dynamical equations of
motion: 
\begin{eqnarray}
\partial_{+} \Pi^{-} & = & j^{-} \ ,\label{11mLorhameqPim}\\
\partial_{+} A_{-} & = & \frac{\alpha}{(1 + \alpha)^2} \left(
\alpha \Pi^{-} - \frac{1}{\partial_{-} }* j^{+}\right)\ ,
\label{11mLorhameqAm}
\end{eqnarray}
with the canonical Poisson-Dirac bracket at LF
\begin{equation}
\left \{{A}_{-}(x^{+}, x^{-}), {\Pi}^{-}(x^{+}, y^{-})
\right \}_{DB} =  \delta({x}^{-} - {y}^{-}) \ ,
\label{11mLorbracket}
\end{equation}
while the canonical Hamiltonian density is
\begin{eqnarray}
{\cal H}^{1+1 can}_{flow} & = & \Pi^{-} \partial_{-}A_{+}
- {\cal L}^{1+1}_{flow} = \frac{1}{2(1 + \alpha)^2} \left(
\alpha \Pi^{-} - \frac{1}{\partial_{-}} * j^{+} \right)^2 -
A_{-} j^{-}. 
\end{eqnarray}
Now we can give the effective Lagrangian density for the gauge
field sector 
\begin{eqnarray}
{\cal L}^{1+1 eff}_{flow} & = & \Pi^{-} \partial_{+} A_{-}
- \frac{1}{2(1 + \alpha)^2} \left( \alpha \Pi^{-} -
\frac{1}{\partial_{-}} * j^{+} \right)^2 + A_{-} j^{-}\ ,
\end{eqnarray}
and then easily incorporate fermions back into the dynamical
system.  

\subsubsection{Interaction with fermion fields}

When the gauge fields are described by effective independent
variables, the system containing complete dynamics of fermions and their
interaction with gauge fields is given by the Lagrangian
density 
\begin{eqnarray}
\widetilde{\cal L}^{1+1 QED}_{flow} & = & \Pi^{-} \partial_{+} A_{-} +
\sqrt{2} i \psi^{\dag}_{+} \partial_{+} \psi_{+} 
- \frac{1}{2(1 + \alpha)^2} \left( \alpha \Pi^{-} + e
\sqrt{2}\frac{1}{\partial_{-}} * (\psi^{\dag}_{+} \psi_{+})
 \right)^2 \nonumber\\
& + & 
\sqrt{2} i \psi^{\dag}_{-} \partial_{-} \psi_{-} - e \sqrt{2} 
\psi^{\dag}_{-} \psi_{-} A_{-} - M \left(\psi^{\dag}_{-}
\gamma^0 \psi_{+} + \psi^{\dag}_{+}\gamma^0 \psi_{-}\right). 
\end{eqnarray}
As it usually happens in the LF formalism, the fermion fields 
$\psi_{-}$ i $\psi^{\dag}_{-}$ are non-dynamical and as dependent
variables can be expressed by other fields
\begin{eqnarray}
\psi_{-} & = & \frac{1}{\sqrt{2}}\ \frac{1}{i\partial_{-} - e
A_{-}}* M \gamma^0 \psi_{+}\ ,\\ 
\psi_{-}^{\dag} & = & M \frac{1}{\sqrt{2}} \ \xi^{\dag} *
\frac{1}{i\partial_{-} - e A_{-}} \gamma^0.
\end{eqnarray}
In this way, we have arrived at the Hamiltonian density which
depends solely on the independent dynamical fields (which are
already independent modes) 
\begin{eqnarray}
{\cal H}^{1+1  QED}_{flow} & = & 
\frac{1}{2(1 + \alpha)^2} \left( \alpha \Pi^{-} + e
\sqrt{2}\frac{1}{\partial_{-}} * (\psi^{\dag}_{+} \psi_{+})
 \right)^2 + \frac{M^2}{\sqrt{2}} \psi^{\dag}_{+} \frac{1}{i
\partial_{-} - e A_{-}}* \psi_{+}. \label{11flowtotalHamilt}
\end{eqnarray}
Now the canonical quantization is immediate, one takes the
properly ordered expression (\ref{11flowtotalHamilt}) as the quantum
Hamiltonian density and non-vanishing (anti)commutation relations
\begin{eqnarray}
\left [ \Pi^{-} (x^{+}, {x}^{-}), A_{-}(x^{+}, {y}^{-}) \right
] & = & i \delta({x}^{-} - {y}^{-})\ , \label{11PiAmcommut}\\
\left \{ \psi^{\dag}_{+}(x^{+}, {x}^{-}), \psi_{+}(x^{+},
{y}^{-}) \right \} & = & \frac{\Lambda_{+}}{\sqrt{2}} \delta({x}^{-} -
y^{-})\label{11psipsidagantcom}. 
\end{eqnarray}
Here the equations for quantum operators will have the same
functional form as their classical \linebreak
counterparts\footnote{The
expression $\psi^{\dag}_{+}(x) \psi_{+}(x)$ is singular for
quantum fields operators and needs some regularization. If one
takes the splitting-point method which is compatible with the
local gauge symmetry $\lim_{\eta \rightarrow
0}\psi^{\dag}_{+}(x^{+}, x^{-} - \eta) e^{- i e \int_{x^{-} -
\eta}^{x^{-} + \eta} d\xi A_{-}(x^{+}, \xi)} \psi_{+}({x^{-} +
\eta})$, then the equations for $\Pi^{-}$, $\psi_{+}$ i
$\psi^{\dag}_{+}$ are modified. However, this would lead us
beyond the scope of this paper and no such regularization 
will be discussed hereafter.} 
\begin{eqnarray}
\partial_{+} \Pi^{-} & = &  e \frac{M^2}{\sqrt{2}} \psi^{\dag}_{+}
* \frac{1}{i\partial_{-} - e A_{-}} \ \ \
\frac{1}{i\partial_{-} - e A_{-}} * \psi_{+}\ ,\\
\partial_{+} A_{-} & = & \frac{\alpha}{(1 + \alpha)^2} \left[
\alpha \Pi^{-} + e \sqrt{2} \frac{1}{\partial_{-}} *
(\psi^{\dag}_{+} \psi_{+}) \right]\ ,\\
i \sqrt{2}\partial_{+} \psi_{+} & = &   \frac{M^2}{\sqrt{2}} 
\frac{1}{i\partial_{-} - e A_{-}} * \psi^{\dag}_{+} - \frac{e
\sqrt{2}}{(1 + \alpha)^2} \psi^{\dag}_{+}
\frac{1}{\partial_{-}} * \left[ \alpha \Pi^{-} + e \sqrt{2}
\frac{1}{\partial_{-}} * (\psi^{\dag}_{+} \psi_{+}) \right],
\\
\hspace{-20pt} - i \sqrt{2}\partial_{+} \psi_{+}^{\dag} & = & 
\frac{M^2}{\sqrt{2}} \psi^{\dag}_{+} * \frac{1}{i\partial_{-} -
e A_{-}} - \frac{e \sqrt{2}}{(1 + \alpha)^2} \psi^{\dag}_{+}
\frac{1}{\partial_{-}} * \left[ \alpha \Pi^{-} + e \sqrt{2}
\frac{1}{\partial_{-}} * (\psi^{\dag}_{+} \psi_{+})
\right].
\end{eqnarray} 
 
\subsubsection{Perturbation theory}

In the interaction representation, the field operators have 
free dynamics generated by the free Hamiltonian which in
the present model is
\begin{equation}
{\cal H}^{1+1 flow}_{0} =  
\frac{\alpha^2}{2(1 + \alpha)^2} \left( \Pi^{-} \right)^2
+ \frac{M^2}{\sqrt{2}} \psi^{\dag}_{+} \frac{1}{i
\partial_{-} }* \psi_{+} \label{11freeHamilt}.
\end{equation}
The remaining part of the Hamiltonian is taken as the
interaction Hamiltonian
\begin{eqnarray}
{\cal H}^{1+1  QED}_{int} & = & {\cal H}^{1+1  QED}_{flow}
- {\cal H}^{1+1 flow}_0 =  e \frac{\sqrt{2} \alpha^2}{(1 +
\alpha)^2} \Pi^{-}\frac{1}{\partial_{-}} *
(\psi^{\dag}_{+}\psi_{+})  \nonumber \\ 
&+ & e^2 \frac{1}{(1 + \alpha)^2} \left(\frac{1}{\partial_{-}}
* (\psi^{\dag}_{+}\psi_{+})\right)^2 - \frac{M^2}{\sqrt{2}}
\psi^{\dag}_{+} \left[\frac{1}{i \partial_{-} } - \frac{1}{i
\partial_{-} - e A_{-}}\right]* \psi_{+} \ ,\label{11intHamilt}
\end{eqnarray}
and we see that there is a direct instantaneous interaction of
currents which will modify the perturbative propagators from
the primary form of Wick's contractions of effective
gauge field potentials
\begin{equation}
{\cal D}_{\mu \nu}(x-y) = i \left< 0 \left| T \bar{A}_{\mu}(x) 
\bar{A}_{\nu}(y)\right| 0 \right> + \frac{g_{\mu -}g_{\nu -}}{
(1 + \alpha)^2}  \frac{1}{\partial_{-}^2}*\delta(x-y)
\end{equation}
where
\begin{eqnarray}
\bar{A}_{-} & = & A_{-}\ ,\\
\bar{A}_{+} & = & - \frac{\alpha}{(1+ \alpha)^2}
\frac{1}{\partial_{-}} \ast \Pi^{-}.
\end{eqnarray}
We need the free propagator for independent modes and they can
be obtained quite easily
\begin{eqnarray}
\langle 0 | T^{+} \ \Pi^{-}(x) A_{-}(y) | 0 \rangle & = & 
E_F^1(x-y)\ , \\
\langle 0 | T^{+} \ A_{-}(x) A_{-}(y) | 0 \rangle & = & 
\frac{\alpha^2}{(1 + \alpha)^2} E_F^2(x-y)\ , \\
\left\langle 0 \left| T^{+} \ \psi_{+}(x) \psi_{+}^{\dag}(y)
\right| 0 \right\rangle & = & 
i \sqrt{2} \Lambda_{+} \partial_{-}^x
\Delta_F^{2}(x-y, M^2)\ ,
\end{eqnarray}
where $\Delta_F^{2}(x, M^2)$ is given in Appendix
\ref{dodfenGrefun}. Then one finds the perturbative propagator
for the gauge fields
\begin{eqnarray}
D_{--}(x) & = & \frac{\alpha^2}{(1 + \alpha)^2} E_F^2(x)\ , \\
D_{+-}(x)  & = & -\frac{\alpha}{(1 + \alpha)^2}
\frac{1}{\partial_{-}} \ast E_F^1(x)\ , \\
D_{++}(x) & = & \frac{1}{(1 + \alpha)^2}
\frac{1}{\partial_{-}^2}*\delta(x-y) \ ,
\end{eqnarray}
which has non-causal structure for $(+)$ components. This is
connected with the infrared singularities of covariant
massless fields in the 1+1 dimensional LF formulation
\cite{HagenYee76}. Specially 
the LC-gauge limit $\alpha \rightarrow 0$ produces the
gauge field propagator with the CPV prescription for spurious
pole and no causal pole at all.\\

\subsection{Higher-dimensional model}

The failure of the 1+1 dimensional approach towards
perturbative propagators with causal poles indicates the danger
of the infrared singularities also in the 3+1 dimensions.
Therefore here we introduce the dimensional regularization by
taking $d = D-1 > 2$ transverse coordinates 
$x_\perp = \{ x_2, \ldots, x_{d+1} \}$. The gauge field sector
described by the Lagrangian density
\begin{eqnarray}
{\cal L}^{D+1\ gauge}_{flow} & = &
\left(\partial_{+} {A}_i - \partial_i
{A}_{+}\right)\left(\partial_{-} {A}_i - \partial_i
{A}_{-}\right)  +  \frac 1 2 \left(\partial_{+} {A}_{-} -
\partial_{-}{A}_{+}\right)^2 - \frac 1 4 \left( \partial_i {A}_j
- \partial_j {A}_i\right)^2 \nonumber\\
& + & {A}_{-} {j}^{-} + {A}_{+} {j}^{+} + {A}_i {j}^i +
\Lambda \left( \partial_{+} A_{-} + \alpha \partial_{-} A_{+} -
\alpha \partial_i A_i\right), \label{mLor31Lagrden1}
\end{eqnarray}
still contains constraints; therefore, omitting all details which
will be presented elsewhere, we can use the effective description 
\begin{eqnarray}
\widetilde{\cal L}^{D+1 \ gauge}_{flow} & = & \partial_{-} C_i
\partial_{+} C_i - \frac 1 2 \left( \partial_i{C}_j \right)^2
- (1 + \alpha)\partial_{+} \lambda \partial_{-} \phi 
 + \alpha \partial_i \lambda \partial_i \phi \nonumber\\
& - & \frac{1}{2} \left(\alpha \lambda - \frac{1}{1 + \alpha}
\frac{1}{\partial_{-}}* j^{+}\right) ^2 
- \left( \partial_i C_i + \Delta_\perp \phi \frac{\alpha }{1 +
\alpha} \right) \frac{1}{\partial_{-}}* j^{+}\\
&- &\phi \left( \partial_{-} j^{-} + \partial_{i} j^{i} \right)
+ j^i C_i \nonumber
\end{eqnarray}
where the independent modes $(C_i,\phi, \lambda)$ are defined
by primary fields 
\begin{eqnarray}
\phi & = & \frac{1}{\Delta^d_\perp} \ast \left[(1 + \alpha)
(\partial_i A_i - \partial_{-}A_{+}) - \alpha \Lambda -
\frac{1}{\partial_{-}} \ast j^{+}\right]\ ,\\
\lambda & = & \Lambda - \frac{1}{1 + \alpha}
\frac{1}{\partial_{-}}* j^{+}\ , \\
C_i & = & A_i - \partial_i \phi. 
\end{eqnarray}
Next we add directly the fermion contributions 
and as the starting point for the quantization of the complete
model, we take the effective Lagrangian density
\begin{eqnarray}
{\cal L}^{D+1}_{flow} & = & \partial_{-} C_i \partial_{+} C_i -
(1 + \alpha) \partial_{+} \lambda \partial_{-} \phi 
 + i \sqrt{2} \psi^{\dag}_{+} \partial_{+} 
\psi_{+} + \sqrt{2} \psi^{\dag}_{-} \left( i \partial_{-} - e
\partial_{-} \phi\right)\psi_{-}  \nonumber\\
& - & \frac 1 2 \left( \partial_i{C}_j \right)^2
+ \alpha \partial_i \lambda \partial_i \phi + \frac{1}{2}
\left(\alpha \lambda + \frac{1}{1 + \alpha}
\frac{1}{\partial_{-}}* J^{+}\right) ^2 \nonumber\\ 
& + &\left( \partial_i C_i + \Delta_\perp \phi \frac{\alpha }{1
+ \alpha} \right) \frac{1}{\partial_{-}}* J^{+} - \xi^{\dag}
\psi_{-} - \psi_{-}^{\dag}\xi \ , \label{Lagr31mLorQED} 
\end{eqnarray}
where
\begin{eqnarray}
\xi & = & \left(- i \partial_i \alpha^i + M \beta \right)\psi_{+}
+ e \left (C_i + \partial_i \phi\right ) \alpha^i\psi_{+}\ ,\\ 
\xi^{\dag} & = & \left ( i \partial_i\psi^{\dag}_{+} \alpha^i +
M \psi_{+}^{\dag}\beta \right) + e \psi_{+}^{\dag} \alpha^i
\left(C_i + \partial_i \phi\right) \ ,\\
J^{+} & = & e\sqrt{2} \psi_{+}^{\dag} \psi_{+}.
\end{eqnarray}
After removing the dependent fermion fields
$\psi_{-}$ and  $\psi^{\dag}_{-}$
\begin{eqnarray}
\psi_{-} & = & \frac{1}{\sqrt{2}}\
\frac{1}{i\partial_{-} - e \partial_{-} \phi}*\xi\ ,\\
\psi_{-}^{\dag} & = & \frac{1}{\sqrt{2}} \ \xi^{\dag} *
\frac{1}{i\partial_{-} - e \partial_{-} \phi},
\end{eqnarray}
we have the Hamiltonian density 
\begin{eqnarray}
{\cal H}_{flow}^{D+1} & = & \frac 1 2 \left( \partial_i{C}_j
\right)^2 - \alpha \partial_i \lambda \partial_i \phi +
\frac{1}{2} \left(\alpha \lambda + \frac{1}{1 + \alpha}
\frac{1}{\partial_{-}}* J^{+}\right) ^2 \nonumber\\
&- & \left( \partial_i C_i + \Delta_\perp \phi \frac{\alpha }{1
+ \alpha} \right) \frac{1}{\partial_{-}}* J^{+} +
\frac{1}{\sqrt{2}} \xi^{\dag} \frac{1}{i \partial_{-} - e
\partial_{-} \phi}* \xi \ ,\label{31mLortotalHamilt} 
\end{eqnarray}
which depends only on the independent modes, and the
non-vanishing (anti)commutators at LF are 
\begin{eqnarray}
2\partial^x_{-}\left [ C_i (\vec{x}), C_j(\vec{y})\right ] & =
& - i \delta_{ij}\delta^3(\vec{x} - \vec{y})\ ,
\label{31mLorCiCjcommut}\\ 
(1 + \alpha)\partial^x_{-}\left [ \phi (\vec{x}), \lambda(\vec{y})\right ]
& = & i \delta^3(\vec{x} - \vec{y}) \ ,
\label{31mLorphilambdacommut}\\ 
\left \{ \psi^{\dag}_{+}(\vec{x}), \psi_{+}(\vec{y}) \right \}
& = & \frac{1}{\sqrt{2}} \Lambda_{+} \delta^3(\vec{x} - \vec{y})\
\label{31mLorpsipsidagantcom}. 
\end{eqnarray}
Therefore we have the field equations
\begin{eqnarray}
(2 \partial_{+} \partial_{-} - \Delta_\perp) {C}_{i} & = & -e
\frac{1}{\sqrt{2}} \xi^{\dag} * \frac{1}{i\partial_{-} - e
\partial_{-} \phi}\alpha^i \psi_{+} -   e
\frac{1}{\sqrt{2}} \psi^{\dag}_{+} \alpha^i
 \frac{1}{i\partial_{-} - e \partial_{-} \phi} * \xi\nonumber\\
&& - \frac{1}{\partial_{-}}* \partial_iJ^{+}\ , \label{31mLoreffeqCi} \\ 
\left[(1 + \alpha) \partial_{+} \partial_{-} -
\alpha\Delta_\perp \right]{\phi} & = & \alpha^2 \lambda +
\frac{\alpha}{1 + \alpha} \frac{1}{\partial_{-}}*J^{+}\ ,
\label{31mLoreffeqphi}\\ 
\left[(1 + \alpha) \partial_{+} \partial_{-} -\alpha
\Delta_\perp \right]{\lambda} & = & - \frac{e}{\sqrt{2}}
\partial_i \left[ \xi^{\dag} * \frac{1}{i\partial_{-} - e 
\partial_{-} \phi}\alpha^i \psi_{+}+  \psi^{\dag}_{+} \alpha^i
 \frac{1}{i\partial_{-} - e \partial_{-} \phi} * \xi
\right]\nonumber\\
& - & \frac{e}{\sqrt{2}} \partial_{-} \left[ \xi^{\dag} *
\frac{1}{i\partial_{-} - e \partial_{-} \phi} \
\frac{1}{i\partial_{-} - e \partial_{-} \phi} * \xi 
\right]\nonumber\\
& - &  \Delta_\perp  \frac{\alpha}{1 + \alpha} 
\frac{1}{\partial_{-}}* J^{+} \ ,\label{31mLoreffeqlambda2}\\ 
i \sqrt{2} \partial_{+} \psi & = & \frac{1}{\sqrt{2}}
\left[ - i \partial_i \alpha^i + M \beta + e \left (C_i +
\partial_i \phi\right ) \alpha^i\right] 
 \frac{1}{i\partial_{-} - e \partial_{-} \phi} * \xi\nonumber\\
& - & e \sqrt{2} \frac{1}{1 + \alpha} \psi_{+}
\frac{1}{\partial_{-}} * \left(\alpha \lambda + \frac{1}{1 + \alpha} 
\frac{1}{\partial_{-}}* J^{+}\right)\nonumber\\
& + & e \sqrt{2} \psi_{+} \frac{1}{\partial_{-}} *
\left( \partial_i C_i + \Delta_\perp \phi \frac{\alpha }{1 +
\alpha} \right)\ , 
\label{31mLoreffeqpsi}\\ 
- i \sqrt{2} \partial_{+} \psi^{\dag} & = &
\frac{1}{\sqrt{2}}\xi^{\dag} * \left[ i
\stackrel{\leftarrow}{\partial_i} \alpha^i + M \beta 
+ e \left (C_i + \partial_i \phi\right ) \alpha^i\right]
 \frac{1}{i\partial_{-} - e \partial_{-} \phi} \nonumber\\
& - & e \sqrt{2} \frac{1}{1 + \alpha} \psi_{+}^{\dag}
\frac{1}{\partial_{-}} * \left(\alpha \lambda + \frac{1}{1 +
\alpha} \frac{1}{\partial_{-}}* J^{+}\right)\nonumber\\
& + & e \sqrt{2} \psi_{+}^{\dag} \frac{1}{\partial_{-}} *
\left( \partial_i C_i + \Delta_\perp \phi \frac{\alpha }{1 +
\alpha} \right) \ ,
\label{31mLoreffeqpsidag}
\end{eqnarray} 
which describe the quantum theory in the Heisenberg picture. 
We also notice that for the complete interacting theory one
can define a free field
\begin{equation}
\Lambda = \lambda - \frac{1}{1 + \alpha}\frac{1}{\partial_{-}}*
J^{+}\ , 
\end{equation}
which satisfies noncovariant dynamical equation
\begin{equation}
\left[(1 + \alpha) \partial_{+} \partial_{-} -
\alpha\Delta_\perp \right]{\Lambda}  =  0\ ,
\end{equation}
but has two extra non-vanishing q-number commutators 
\begin{eqnarray}
(1+ \alpha)\left [\partial_{-} \Lambda (\vec{x}),
\psi_{+}(\vec{y})\right ] & = & e \delta^3(\vec{x} - \vec{y})
\psi_{+}(\vec{x}) 
\ , \label{31mLorpsiLambdacommut}\\ 
(1+ \alpha)\left [\partial_{-} \Lambda (\vec{x}),
\psi_{+}^{\dag}(\vec{y})\right ]
& = & - e \delta^3(\vec{x} - \vec{y}) \psi_{+}^{\dag} (\vec{x})
\label{31mLordagpsiLambdacommut}. 
\end{eqnarray}
These properties  show that the $\Lambda$ field can be
taken for the specification of physical states via the
condition 
\begin{equation}
\langle phys' | \Lambda(x) |phys \rangle = 0 \label{D1flowphyscond}
\end{equation}
just like in the ET formalism \cite{Lautrup66},
\cite{Nakanishi1972} . However in the
perturbation calculations, the property of free propagation
is far less important than the presence of q-number commutators
which can be nontrivial obstacles for Wick's contractions.
Therefore when one works with $\Lambda$, then one 
should take other fermion fields
\begin{eqnarray}
\chi(x) & = & e^{ie \phi(x)} \psi_{+}(x)\\
\chi^{\dag}(x) & = & \psi^{\dag}_{+}(x) e^{-ie \phi(x)} 
\end{eqnarray}
which already commute with $\Lambda$. However, for these new
field operators the Hamiltonian density operator changes
drastically (the field $\phi$ decouples from the fermion currents)
\begin{equation}
{\cal H}^{LC}_{total}  =  
  \frac 1 2 \left( \partial_i{C}_j \right)^2
- \alpha \partial_i \Lambda \partial_i \phi + \frac{1}{2}
\left(\alpha \Lambda + \frac{1}{\partial_{-}}* J^{+}_{\chi}\right) ^2 
-  \partial_i C_i  \frac{1}{\partial_{-}}* J^{+}_{\chi}
 + \frac{1}{\sqrt{2}} \xi_{LC}^{\dag} \frac{1}{i
\partial_{-} }* \xi_{LC} \label{31mLortotalLCHamilt},
\end{equation}
where now we have
\begin{eqnarray}
J^{+}_{\chi} & = & \sqrt{2} \chi^{\dag} \chi\ , \\
\xi_{LC} & = & \left(-i \partial_i \alpha^i + M \beta + e C_i
\alpha^i \right) \chi\ ,\\ 
\xi_{LC}^{\dag} & = & \left (i \partial_i\chi^{\dag} \alpha^i +
M \chi^{\dag}\beta \right) + e \chi^{\dag} \alpha^i C_i .
\end{eqnarray}
One can verify that this new system describes the QED for the 
LC-gauge condition and the perturbative gauge field propagator would have
the CPV-prescription for spurious poles. On the one hand this shows
how different gauges can be linked together at the level of 
quantum field operators, while on the other hand this explains
how easily the causal poles may be replaced by the CPV ones.\\ 

\subsubsection{Perturbation theory}

The perturbation theory is formulated in the interaction
representation with the free Hamiltonian density
\begin{eqnarray}
{\cal H}_{0}^{D+1} = \lim_{e\rightarrow 0} {\cal
H}_{flow}^{D+1} & = & \frac{\alpha}{2} \lambda^2 + \frac 1 2 
\left(\partial_j C_i\right) ^2 
- \alpha \partial_i \phi \partial \lambda
 + \frac{1}{\sqrt{2}} \xi^{\dag}_0
\frac{1}{i \partial_{-} }* \xi_0 \label{31mLorfreeHamilt}\ ,
\end{eqnarray}
where
\begin{eqnarray}
\xi_0 & = & \left(- i \partial_i \alpha^i + M \beta
\right)\psi_{+} \, \\
\xi^{\dag}_0 & = & \left ( i \partial_i\psi^{\dag}_{+} \alpha^i
+ M \psi_{+}^{\dag}\beta \right)\ ,
\end{eqnarray}
and the interaction Hamiltonian density can be divided into
two parts
\begin{equation}
{\cal H}^{D+1}_{int} = {\cal H}_{flow}^{D+1} - {\cal
H}_{0}^{D+1} = {\cal 
H}^1_{int} + {\cal H}^2_{int}\ ,
\end{equation}
where 
\begin{eqnarray}
{\cal H}^1_{int} & = &    \frac{1}{\sqrt{2}} \xi^{\dag}_0 \left( \frac{1}{ i
\partial_{-} - e \partial_{-} \phi} - \frac{1}{i \partial_{-}}
\right)\ast \xi_0 \nonumber\\
& + & \frac{e}{\sqrt{2}} \psi^{\dag}_{+} \alpha^i (C_i +
\partial_{i} \phi)\frac{1}{ i \partial_{-} - e \partial_{-}
\phi} \ast \xi_0 + \frac{e}{\sqrt{2}} \xi^{\dag}_{0} \frac{1}{ i
\partial_{-} - e \partial_{-} \phi} \alpha^i\ast (C_i + \partial_{i}
\phi) \psi_{+}\nonumber\\
& + & \frac{e^2}{\sqrt{2}} \psi^{\dag}_{+} \alpha^j (C_j +
\partial_{j} \phi)\frac{1}{ i \partial_{-} - e \partial_{-}
\phi} \alpha^i \ast (C_i + \partial_{i} \phi) \psi_{+} 
\ , \label{31mlorintHam1}\\
{\cal H}^2_{int} & = & - \left( \partial_i C_i + \frac{\alpha}{1
+ \alpha} \Delta_\perp \phi - \frac{\alpha}{1
+ \alpha}\lambda \right )\frac{1}{\partial_{-}}*
J^{+} + \frac 1 2 \left(\frac{2}{1 + \alpha}\right)^2 \left(
\frac{1}{2\partial_{-}}* J^{+}\right)^2.\label{31mlorintHam2}
\end{eqnarray}
We notice a close analogy between the present case and the
model from Section \ref{rozdzialcechLor}, therefore here we will
point out only the differences between these two cases. The
first one comes from the free equations of motion 
\begin{eqnarray}
\left[(1 + \alpha) \partial_{+} \partial_{-} - \alpha
\Delta_\perp \right]{\lambda} & = & 0\ ,\\ 
\left[(1 + \alpha) \partial_{+} \partial_{-} - \alpha
\Delta_\perp \right] {\phi} & = &  \alpha^2 \lambda \ ,
\end{eqnarray}
where here we have the multipole non-covariant dynamical
field $\phi$ which has no true Fourier representation. However,
from the commutation relations (\ref{31mLorCiCjcommut}),
(\ref{31mLorphilambdacommut}) and
(\ref{31mLorpsipsidagantcom}) one can derive the form of
chronological products 
\begin{eqnarray}
\left< 0 \left| T C_i(x) C_j(y)\right| 0 \right> & = &  
\delta_{ij} D_F(x-y)\ ,\\
\left< 0 \left| T \phi(x) \lambda(y)\right| 0 \right> & = & 
- G^1_{\alpha F}(x-y)\ ,\\
\left< 0 \left| T \phi(x) \phi(y)\right| 0 \right> & = &
\alpha^2  G^2_{\alpha F}(x-y)\ ,\\
\left< 0 \left| T \psi_{+}(x) \psi_{+}^{\dag}(y)\right| 0
\right> & = & i \sqrt{2} \Lambda_{+} \partial_{-}^x
\Delta_F(x-y, M^2)\ ,
\end{eqnarray}
where the covariant $D_F(x)$ and the non-covariant
$G^{1,2}_{\alpha F}(x)$ massless Feynman propagator functions
are defined in Appendix \ref{dodfenGrefun}. The combinations of
independent modes which are coupled with  fermion currents are
\begin{eqnarray}
\bar{A}_{+}& = & \frac{1}{\partial_{-}}* \left[  \partial_i
C_i - \frac{\alpha}{1 +\alpha} \lambda + \frac{\alpha}{1
+\alpha}  \Delta_\perp \phi \right ]\ ,
\label{31mLordefbarAp}\\
\bar{A}_{-} & = & \partial_{-} \phi  \ ,\label{31mLordefbarAm}\\
\bar{A}_i & = & C_i + \partial_{i} \phi \ ,\label{31mLordefbarAi}
\end{eqnarray}
with the chronological products given by
\begin{eqnarray}
\left< 0 \left| T \bar{A}_i(x) \bar{A}_j(y)\right| 0 \right> & = &  
\delta_{ij} D_F(x-y) + \alpha^2 \partial_i^x \partial_j^y
G^2_{\alpha F}(x-y)\ ,\\
\left< 0 \left| T \bar{A}_{-}(x) \bar{A}_j(y)\right| 0 \right> & = &  
\alpha^2 \partial_{-}^x \partial_j^y G^2_{\alpha F}(x-y)\ ,\\
\left< 0 \left| T \bar{A}_{-}(x) \bar{A}_{-}(y)\right| 0 \right> & = &  
\alpha^2 \partial_{-}^x \partial_{-}^y G^2_{\alpha F}(x-y)\ ,\\
\left< 0 \left| T \bar{A}_{+}(x) \bar{A}_{-}(y)\right| 0 \right> & = &  
- \frac{\alpha}{1 + \alpha}\left[ G^1_{\alpha F}(x-y)
+ \alpha^2 \Delta_\perp G^2_{\alpha F}(x-y)\right] \ ,\\
\left< 0 \left| T \bar{A}_{+}(x) \bar{A}_{i}(y)\right| 0 \right> & = &  
i (1 -\alpha ) \partial_{+}^x \partial_i^{y} (G^1_{\alpha
F}*D_F) (x-y) + \alpha^2 \partial_{+}^x \partial_i^{y}
G^2_{\alpha F}(x-y)\ , \\
\left< 0 \left| T \bar{A}_{+}(x) \bar{A}_{+}(y)\right| 0
\right> & = &  i 2 (1- \alpha) \partial_{+}^x \partial_{+}^{y} (G^1_{\alpha
F}*D_F) (x-y)
+ \alpha^2 \partial_{+}^x \partial_{+}^{y} G^2_{\alpha F}(x-y)
\nonumber\\ 
&&+ \frac{i}{(1 + \alpha)^2} \frac{1}{\partial_{-}^2} * 
\delta(x-y)\ , 
\end{eqnarray}
where no explicit inverse Laplace operator appear; thus the
limit $d \rightarrow 2$ can be taken already at the level of
independent modes and the free propagators in 3+1 dimensions
have the Fourier representation
\begin{eqnarray}
&&\hspace{-20pt}\left< 0 \left| T \bar{A}_{\mu}(x) \bar{A}_{\nu}(y)\right| 0
\right>  = i \int \frac{d^4k}{(2\pi)^4}
{e^{-i k\cdot (x-y)}}\left\{\frac{1}{2 k_{+} k_{-} - k_\perp +
i\epsilon}
\left[ - g_{\mu \nu} + \frac{k_{+}(1-\alpha)
[n_\nu^{LC} k_\mu + n_\mu^{LC} k_{\nu}]
 }{(1 + \alpha)k_{+} k_{-} - \alpha k_\perp^2 + 
i\epsilon '} \right] \right.\nonumber\\
&&\hspace{165pt} \left.+ \  \frac{\alpha^2 k_\nu k_\mu}{[(1 + \alpha)k_{+} 
k_{-} - \alpha k_\perp^2 + i\epsilon ']^2} 
- \frac{n_\nu^{LC} n_\mu^{LC}}{(1 + \alpha)^2} 
{\rm CPV}\frac{1}{k_{-}^2}\right\}\ ,
\end{eqnarray}
where $n_{+}^{LC} = 1, n_{-}^{LC} = n_{i}^{LC} = 0$. However,
the Wick contractions of gauge fields and direct interactions
of currents will give rise to the modification of perturbative
propagators
\begin{equation}
D_{\mu \nu}^{flow}(x-y) = i \left< 0 \left| T \bar{A}_\mu(x)
\bar{A}_\nu(y)\right| 0 \right> + \frac{g_{\mu -} g_{\nu
-}}{(1 + \alpha)^2} \frac{1}{\partial_{-}^2} * 
\delta(x-y)\ ,
\end{equation} 
which no longer have any spurious pole with CPV prescription
\begin{eqnarray}
D_{\mu \nu}^{flow}(x) & = & -\int \frac{d^4k}{(2\pi)^4} e^{-i k\cdot
x} \left\{\frac{1}{2 k_{+} k_{-} - k_\perp^2 + i\epsilon}
\left[ - g_{\mu \nu} + \frac{k_{+}(1-\alpha) [n_\nu^{LC} k_\mu +
n_\mu^{LC} k_{\nu}] }{(1 + \alpha)k_{+} k_{-} - \alpha
k_\perp^2 + i\epsilon '} \right] \right . \nonumber\\
&& \left.+  \frac{\alpha^2 k_\nu k_\mu}{[(1 + \alpha)k_{+} 
k_{-} - \alpha k_\perp^2 + i\epsilon ]^2} \right\}.
\label{31pertgaugeprop} 
\end{eqnarray}
The cancellation of non-covariant terms during the Wick
contractions of fermion fields is identical with that in
Section \ref{rozdzialcechLor}, therefore, quoting those
previous results, we say that the perturbative calculations
based on the interaction Hamiltonians (\ref{31mlorintHam1}),
(\ref{31mlorintHam2}) and canonical free propagators are
equivalent to those with covariant interaction vertices and
covariant perturbative propagators. In the perturbative
gauge field propagator (\ref{31pertgaugeprop}) with all causal
poles we can take the appropriate limits to the Lorentz gauge
($\alpha \rightarrow 1$) 
\begin{eqnarray}
D_{\mu \nu}^{Lor}(x) & = & - \int \frac{d^4k}{(2\pi)^4} e^{-i k\cdot
x} \frac{1}{2 k_{+} k_{-} - k_\perp^2 + i\epsilon}
\left[ - g_{\mu \nu} +
\frac{k_\nu k_\mu}{[2 k_{+} k_{-} - \alpha k_\perp^2 + i\epsilon
]} \right] \label{31pertLorprop} 
\end{eqnarray}
and to the LC-gauge ($\alpha \rightarrow 0$)\footnote{The
ML-prescription is written 
here in the equivalent (in the sense of distributions) form
$\frac{1}{[k_{-}]_{ML}} = \frac{k_{+}}{k_{+} k_{-} + i \epsilon'}$.}
\begin{eqnarray}
D_{\mu \nu}^{flow}(x) & = & - \int \frac{d^4k}{(2\pi)^4} e^{-i k\cdot
x} \frac{1}{2 k_{+} k_{-} - k_\perp^2 + i\epsilon}
\left[ - g_{\mu \nu} + \frac{k_{+} [n_\nu^{LC} k_\mu +
n_\mu^{LC} k_{\nu}] }{k_{+} k_{-} + i\epsilon '} \right].
\label{31pertLCprop} 
\end{eqnarray}
When we compare (\ref{31pertLorprop}) with the double limit
($\alpha \rightarrow 0$ and $m^2 \rightarrow 0$) of
(\ref{procavectpropagator}) we find that they are the same and 
also coincide with the ET propagator. The expression
(\ref{31pertLCprop}) represents the ML-prescription for the
LC-gauge within the LF procedure. Contrary to other attempts
\cite{McCartorRob}, \cite{Soldati} which used two front
surfaces for free gauge field sector, the present results are
valid for the perturbative QED with fermions. \\

\section{Electrodynamics of charged scalar fields}

\setcounter{equation}{0}

In previous sections we have discussed electrodynamics of
charged fermions which describes the phenomena where electrons
interact with photons. However it is also interesting to
consider charged matter consisting of other fields. In the
usual ET approach, scalar fields are treated as the simplest
possible choice for matter fields. In the LF formalism,
electrodynamics of scalar fields is far from being trivial.
Below we will discuss only one choice of gauge fixing
condition, the LF Weyl gauge in 1+1 dimensions, which has particularly
strange properties. Electromagnetic currents 
built from scalar fields contain derivatives of matter fields
$j^{scalar}_{\mu} = e \phi^{\dag}
\stackrel{\leftrightarrow}{\partial}_\mu \phi$, which are
dramatically  different from the fermion currents. At LF this means
that the complete dynamics of scalar fields directly manifests
itself in coupling with the gauge fields. While such phenomenon
has been recognized a long ago \cite{LCkomutatory2}, it is only
here that   a solution to this problem has been found. \\ 

\subsection{LF Weyl gauge in 1+1 dimensions}

When the LF Weyl gauge $A_{+} = 0$ is strongly imposed on gauge
fields, the electrodynamics of scalar fields is described by the
Lagrangian density
\begin{eqnarray}
{\cal L}^{1+1\ scalar}_{Weyl} & = & \frac 1 2 \left(\partial_{+} {A}_{-}
\right)^2 + \left(\partial_{-} - i e A_{-}\right) \phi^{\dag}
\partial_{+} \phi + \partial_{+}\phi^{\dag} \left(\partial_{-} + i e
A_{-}\right) \phi  - m^2 \phi^{\dag} \phi
\end{eqnarray} 
and its equations of motion (Euler-Lagrange equations) are
\begin{eqnarray}
\partial_{+}^2{A}_{-} & = & - i e \left( \phi^{\dag} \partial_{+} \phi 
-  \partial_{+} \phi^{\dag} \phi \right)\ ,  \label{11scalarweylELeqAm}\\
2 \left(\partial_{-} + i e A_{-}\right) \partial_{+}\phi & = &
- ie \phi \partial_{+} A_{-} - m^2 \phi\ , \label{11scalarweylELeqphi}\\
2 \left(\partial_{-} - i e A_{-}\right) \partial_{+}\phi^{\dag} & = &
ie \phi^{\dag} \partial_{+} A_{-} - m^2
\phi^{\dag}.\label{11scalarweylELeqphid} 
\end{eqnarray}
This means that all fields are dynamical variables, but as
usually, the Gauss law 
\begin{equation}
G = \partial_{-} \partial_{+} A_{-} - i e \left( \phi^{\dag}
\partial_{-} \phi - \phi \partial_{-} \phi^{\dag}\right) + 2
e^2 \phi^{\dag} \phi A_{-} 
\end{equation}
is missing and has to be added as an extra postulate of the
physical quantum theory. The canonical momenta conjugated to
all dynamical fields
\begin{eqnarray}
\Pi^{-} & = &\partial_{+} A_{-} \ ,\label{11scalarweyldefPim}\\
\Pi_{\phi} & = & \left(\partial_{-} - i e A_{-}\right)
\phi^{\dag} \ , \label{11scalarweyldefPiphi}\\
\Pi_{\phi^{\dag}} & = & \left(\partial_{-} + i e A_{-}\right)
\phi \ , \label{11scalarweyldefPiphid}
\end{eqnarray}
show the presence of the second class primary constraints
(according to Dirac's nomenclature) which are characteristic
for LF dynamics of relativistic particles. However, contrary to our
previous treatment of similar constraints, due to the presence
of other fields we cannot discard (\ref{11scalarweyldefPiphi}
and \ref{11scalarweyldefPiphid}) as superficial constraints. 
This makes a sharp distinction between these expressions and 
canonical momenta for $A_i$ in QED or even for free scalar
fields. In our case, Dirac's method of quantization for
constrained systems would give nonzero brackets (commutators)
for scalar fields and $\Pi^{-}$ or between 
two momenta $\Pi^{-}$. All these brackets and also the one for
scalar fields would depend functionally on $A_{-}$. Thus
everyone should understand those who decided to choose the
LC-gauge $A_{-} =0$ where all above problems totally
disappear.\\ 
Here we follow our method of dealing with constrained system
where we pay special attention to equations of motion. 
Eqs. (\ref{11scalarweylELeqAm}-\ref{11scalarweylELeqphid})
can be rewritten as the first order equations in $\partial_{+}$
\begin{eqnarray}
\partial_{+} \Pi^{-} & = & - i e \left( \phi^{\dag} \partial_{+} \phi 
-  \partial_{+} \phi^{\dag} \phi \right) \ ,
\label{11scalarweylELeqPim}\\
2 \left(i\partial_{-} - e A_{-}\right) \partial_{+}\phi & = & e
\phi \Pi^{-} - i m^2 \phi
\ , \label{11scalarweylELeqphi'}\\
2 \left(i\partial_{-} + e A_{-}\right) \partial_{+}\phi^{\dag} & = &
- e \phi^{\dag} \Pi^{-} - i m^2 \phi^{\dag}\ ,
\label{11scalarweylELeqphid'}\\
\partial_{+} A_{-}& = & \Pi^{-}.
\end{eqnarray}
Then using the Green function  $(i\partial_{-} + e
A_{-})^{-1}[x^{-}, y^{-}]$\footnote{The perturbative
definition of this Green function and the notation for
convolutions of integral operators, which is specially suitable
for the further analysis, are given in Appendix 
\ref{dodscalfun}.}   for the covariant partial derivative
\begin{equation}
  \left(i\partial_{-} -  e A_{-}\right)^x
(i\partial_{-} - e A_{-})^{-1}[x^{-}, y^{-}]  = 
-  \left(i\partial_{-} + e A_{-}\right)^y
(i\partial_{-} - e A_{-})^{-1}[x^{-}, y^{-}] 
  =  \delta(x^{-}-y^{-})\ ,
\end{equation}
we can transform these equations into the form of
Hamilton equations of motion
\begin{eqnarray}
\partial_{+}\phi(x) & = & \frac 1 2 \int dy^{-} (i\partial_{-} - e
A_{-})^{-1}[x^{-},y^{-}]\left(e \phi \Pi^{-} -i
m^2\phi\right)(x^{+}, y^{-})\ ,\label{11scalarweylHeqphi}\\ 
\partial_{+}\phi^{\dag}(x) & = & \frac 1 2 \int dy^{-} \left( e
\phi^{\dag} \Pi^{-} + i m^2 \phi^{\dag}\right) (x^{+}, y^{-})
(i\partial_{-} - e A_{-})^{-1}[y^{-}, x^{-}] \ , 
\label{11scalarweylHeqphid} \\
\partial_{+} \Pi^{-}(x) & = & \frac{e}{2} \phi^{\dag}(x) \int
dy^{-} (i\partial_{-} - e A_{-})^{-1}[x^{-}, y^{-}]\left( -i e
\phi \Pi^{-} - m^2 \phi\right)(x^{+}, 
y^{-})\nonumber\\ 
&+& \frac{e}{2} \phi(x) \int dy^{-} \left(i e \phi^{\dag} \Pi^{-} -
m^2 \phi^{\dag}\right) (x^{+}, y^{-})  (i\partial_{-} - e
A_{-})^{-1}[y^{-}, x^{-}]\ ,
 \label{11scalarweylHeqPim}\\
\partial_{+} A_{-}(x) & = & \Pi^{-}(x) \ ,
\label{11scalarweylHeqAm}
\end{eqnarray}
or in the self-explanatory matrix notation (which will be used hereafter)
\begin{eqnarray}
\partial_{+}\phi & = & \frac 1 2  (i\partial_{-} - e
A_{-})^{-1} \ast \left(e \phi \Pi^{-} - i
m^2\phi\right)\label{11scalarweylHeqphi'} \ , \\ 
\partial_{+}\phi^{\dag} & = & \frac 1 2  \left( e \phi^{\dag}
\Pi^{-} + i m^2 \phi^{\dag}\right) \ast (i\partial_{-} - e
A_{-})^{-1}\label{11scalarweylHeqphid'} \ , \\
\partial_{+} \Pi^{-} & = & \frac{e}{2} \phi^{\dag}
(i\partial_{-} - e A_{-})^{-1} \ast \left(-i e \phi \Pi^{-} -
m^2 \phi\right) + \frac{e}{2} \left(i e \phi^{\dag} \Pi^{-} - m^2
\phi^{\dag} \right) \ast (i\partial_{-} - e A_{-})^{-1}
\phi\nonumber \ , \\ 
&&\label{11scalarweylHeqPim'}\\ 
\partial_{+} A_{-} & = & \Pi^{-}. \label{11scalarweylHeqAm'}
\end{eqnarray}
Now we encounter a real problem because, on the one hand, these 
equations evidently describe  interacting fields, while on the
other hand,  the canonical Hamiltonian density has a free field form
\begin{eqnarray}
{\cal H}_{can}^{QED} & = & \Pi^{-} \partial_{+} A_{-} +
\Pi_{\phi^{\dag}} \partial_{+}\phi^{\dag} + \Pi_{\phi}
\partial_{+}\phi - {\cal L}_{Weyl} = \frac 1 2 (\Pi^{-})^2 +
m^2 \phi^{\dag}\phi \label{11scalarweylcanHam} 
\end{eqnarray}
and does not depend on the coupling constant $e$. Even more, the
generator of translations in the direction $x^{-}$ is not a
kinematical operator but depends  on interactions
\begin{eqnarray}
P_{-} & = & \int dy^{-} \left[ \Pi^{-}\partial_{-} A_{-} + \Pi_{\phi^{\dag}}
\partial_{-}\phi^{\dag} + \Pi_{\phi} \partial_{-}\phi \right]
\nonumber\\ 
& = & \int dy^{-} \left[ \Pi^{-}\partial_{-} A_{-} +
(\partial_{-} - i e A_{-})\phi \partial_{-}\phi^{\dag} + 
(\partial_{-} - i e A_{-})\phi^{\dag} \partial_{-}\phi
\right].
\end{eqnarray}
These observations are apparently in a conflict with the
original Dirac analysis of the front form of dynamics
\cite{Dirac1949}. The only way out of these problems is to
allow the Dirac brackets to contain interactions and it really
happens here because  we have non-vanishing expressions
\begin{eqnarray}
\left\{ \Pi^{-} (x^{+}, x^{-}), A_{-}(x^{+},y^{-}) \right\}_{DB}
& = & - \  \delta(x^{-} - y^{-})\label{11scalarweylDBPiAm} \ , \\
\left\{ \Pi^{-} (x^{+}, x^{-}), \phi(x^{+},y^{-}) \right\}_{DB}
& = & \ - \frac{e}{2} \phi(x^{+}, x^{-})(i\partial_{-} - e
A_{-})^{-1}[x^{-}, y^{-}]\ , \\ 
\left\{ \Pi^{-} (x^{+}, x^{-}), \phi^{\dag}(x^{+},y^{-}) \right\}_{DB}
& = &  - \frac{e}{2} \phi^{\dag}(x^{+}, x^{-})(i\partial_{-} - e
A_{-})^{-1}[x^{-}, y^{-}]\ , \\
\left\{ \phi (x^{+}, x^{-}), \phi^{\dag}(x^{+},y^{-}) \right\}_{DB}
& = & - \frac{i}{2} (i\partial_{-} - e A_{-})^{-1}[x^{-},
y^{-}] \label{11scalarweylDBphiphid}\ , \\ 
\left\{ \Pi^{-} (x^{+}, x^{-}), \Pi^{-}(x^{+},y^{-}) \right\}_{DB}
& = & - i \frac{e^2}{2} \phi^{\dag}(x^{+}, x^{-})(i\partial_{-} - e
A_{-})^{-1}[x^{-}, y^{-}]\phi(x^{+}, y^{-}) \nonumber\\ 
& +& i  \frac{e^2}{2} \phi^{\dag}(x^{+}, y^{-})(i\partial_{-} - e
A_{-})^{-1}[y^{-}, x^{-}]\phi(x^{+}, x^{-}). 
\end{eqnarray}
Now it is not difficult to show that these brackets lead to the
correct equations of motion for all field variables
(\ref{11scalarweylHeqPim}-\ref{11scalarweylHeqphid}) and also
give the expected translations in the direction $x^{-}$ 
\begin{eqnarray}
\left\{ \Pi^{-} (x), P_{-}(x^{+}) \right\}_{DB} & = &
\partial_{-} \Pi^{-}(x)\ , \\
\left\{ A_{-} (x), P_{-}(x^{+}) \right\}_{DB} & = &
\partial_{-} A_{-}(x)\ ,\\
\left\{ \phi (x), P_{-}(x^{+}) \right\}_{DB} & = &
\partial_{-} \phi(x)\ , \\
\left\{ \phi^{\dag} (x), P_{-}(x^{+}) \right\}_{DB} & = &
\partial_{-} \phi^{\dag}(x).
\end{eqnarray}
If one would like to take these brackets as a basis for
respective quantum commutators, while defining the canonical
quantum theory, then one would end up with a hopeless problem
of the perturbative calculations. When the interaction is not
located in the Hamiltonian but in commutators, then the
definition of the interaction representation is not unique, if
at all possible.  

\subsection{Dressed scalar fields}

In our previous analysis we have found that LF dynamics allows
for using various forms of fields which differ in both the 
evolution equations and the commutator relations (e.g. the Lagrange
multiplier fields in Section \ref{modLorchap}). Thus we expect
that there are also various forms of scalar fields, one of them
may have free commutators. Usually at LF, the dressed scalar 
fields have the field-dependent phase factor like 
\begin{eqnarray}
\widetilde{\phi} = \exp\left\{ - i\ e (\partial_{-})^{-1} \ast A_{-}
\right\}{\phi} \label{11scalarLCdefphi}\ , \\ 
\widetilde{\phi}^{\dag} = {\phi}^{\dag} \exp\left\{i \ e
 A_{-} \ast (\partial_{-})^{-1}  \right\} \label{11scalarLCdefphid}.
\end{eqnarray}
However this really means a gauge transformation to the
LC-gauge and evidently is not a true solution for 
the LF Weyl gauge problem. Therefore we take another
possibility and define a new scalar field $\varphi$ 
\begin{equation}
\phi  =  {\cal W}_{-1}[\widehat{a}] \ast \varphi \ ,
\label{11scalarWeylphialpha}
\end{equation}
where the integral operator ${\cal W}_{- 1}[\widehat{a}]$ 
is defined in Appendix \ref{dodscalfun}. For simplicity,
 the other scalar field $\phi^{\dag}$ is not
changed.\footnote{Also another choices of dressed scalar fields
are possible, specially the symmetrical choice for both fields.
These issues will be discussed elsewhere.} 
Now one can easily check that for the pair of scalar fields
$\varphi$ and $\phi^{\dag}$, their Dirac bracket has already
a free form
\begin{equation}
\left\{ \varphi (x^{+}, x^{-}), \phi^{\dag}(x^{+},y^{-})
\right\}_{DB} = - \frac{i}{2} (i\partial_{-})^{-1}(x^{-} -
y^{-}). 
\end{equation}
Encouraged by this result we find the scalar field contribution
to the Lagrangian 
\begin{eqnarray}
L^{scalar} & = & \partial_{+} \phi^{\dag} \ast \left ( \partial_{-}
+ i e A_{-} \right ) \phi + \left ( \partial_{-} - i e A_{-}
\right )\phi^{\dag} \ast \partial_{+} \phi - m^2 \phi^{\dag}
\ast \phi = \nonumber\\ 
& = & \partial_{-} \phi^{\dag} \ast \partial_{+} \varphi +
\partial_{+}\phi^{\dag} \ast\partial_{-}\varphi - m^2
\phi^{\dag} \ast {\cal W}_{-1}[\widehat{a}] \ast \varphi
\nonumber\\ 
&& + i e \phi^{\dag} \ast \partial_{+} A_{-} {\cal W}_{-1}
[\widehat{a}] \ast \varphi
\end{eqnarray}
and then calculate the canonical momenta conjugated to scalar fields
\begin{eqnarray}
\Pi_{\varphi} & = & \partial_{-} \phi^{\dag}\ ,\\
\Pi_{\phi^{\dag}} & = & \partial_{-} \varphi\ , 
\end{eqnarray}
which are already free. This means that we have again trivial
primary constraints for these momenta which can be discarded
in the Hamiltonian approach. Also the third momentum (for
$A_{-}$ gauge field) has been changed 
\begin{equation}
\Pi  =  \partial_{+} A_{-} - \partial_{-} A_{+} + i e
\phi^{\dag}  {\cal W}_{-1}[\widehat{a}] \ast \varphi \ ,
\end{equation}
but there are no constraints connected with this mode, so this
change introduces no problems into further analysis. Now the
complete Hamiltonian for the LF Weyl gauge is 
\begin{eqnarray}
\widetilde{H}_{can}^{QED} & = & \frac 1 2 \int dx^{-} \ \left( \Pi - i e
\phi^{\dag} {\cal W}_{-1}[\widehat{a}] \ast \varphi \right)^2 +
m^2 \phi^{\dag} \ast {\cal W}_{ - 1}[\widehat{a}] \ast \varphi
\end{eqnarray}
and evidently it contains interactions, while the translation
generator 
\begin{equation}
P_{-} = \frac 1 2 \int dx^{-} \Pi \ast \partial_{-} A_{-} + 2
\partial_{-} \phi^{\dag} \ast \partial_{-}\varphi
\end{equation}
is kinematical. Finally, we give nonzero commutators at LF
\begin{eqnarray}
2 \partial_{-}^x \left[ \phi^{\dag}(x^{+}, x^{-}),
\varphi(x^{+}, y^{-}) \right]& = & i \delta(x^{-} - y^{-})\ , \\
\left[ \Pi(x^{+}, x^{-}),
A_{-}(x^{+}, y^{-}) \right] & = & - i \delta(x^{-} - y^{-})\ , 
\end{eqnarray}
which also have canonical free forms. Therefore we see that the
redefinition (\ref{11scalarWeylphialpha}) of the scalar field
leads to the canonical description of scalar QED at the LF.

\subsection{Perturbation theory}

Now we can define the interaction representation with the free
evolution of field operators which is given by the free
Hamiltonian $H_0$ 
\begin{equation}
H_0 = \frac 1 2 \int dx^{-} \ \Pi^2 + m^2 \phi^{\dag} \ast
\varphi.
\end{equation}
The evolution of states will be given by the interaction 
Hamiltonian $H_I$
\begin{eqnarray}
H_I  =  \widetilde{H}_{can}^{QED} - H_0 
& = & \frac 1 2 \left( i e \phi^{\dag}_{0} 
{\cal W}_{-1}[\widehat{a}]\ast \phi_{-1}\right)^2 - i e
\phi^{\dag}_{0}\Pi \ast {\cal W}_{-1}[\widehat{a}]\ast
\phi_{-1}\nonumber\\ 
& + & m^2 \phi^{\dag}_{0} \ast \left({\cal W}_{-1}[\widehat{a}]
-1\right) \ast \phi_{-1}.
\end{eqnarray}
At low orders (in $e$) of the LF perturbation,  different
contributions generated  by the above Hamiltonian formally sum up
to the covariant results, which can be obtained within the ET
approach.\footnote{The ET interaction Hamiltonian also contains
{\it one} noncovariant term which cancels the
noncovariant part of the second time derivative of gauge field
propagator \cite{ItzykZub85}.} However, if we want to prove the
equivalence of both perturbations for all orders then we need to
follow the methods used earlier in Section \ref{rozdzialcechLor}.
In order to separate the effects of scalar and gauge field
contractions, we divide the interaction Hamiltonian into two parts
\begin{eqnarray}
H_I & = & H_{I}^1 + H_{I}^2\ , \\
H_I^1 & = & \phi^{\dag}_{0} \ast \left\{- i e \Pi + m^2
\widehat{a}\right\} \ast {\cal
W}_{-1}[\widehat{a}]\ast \phi_{-1}\ , \\
H_I^2 & = & \frac 1 2 \int dx^{-} \left( i e \phi^{\dag}_{0} 
{\cal W}_{-1}[\widehat{a}] \ast \phi_{-1}\right)^2 = \frac 1 2
\int dx^{-} J_\Pi^2. 
\end{eqnarray}
The functional form of the Wick theorem \cite{SchwingerFunct}
\begin{equation}
T \exp -i \int \sigma J_\Pi = \  :  \exp - i \int J_\Pi \left(
\sigma + i \frac{\delta}{\delta \sigma}\right) :\ = \exp - \frac i 2 \int
J_\Pi^2 \ :\exp -i \int \sigma J_\Pi:\ ,
\end{equation}
indicates that the Hamiltonian $H_I^2$ can be equivalently
substituted by the linear term $\int \sigma J_\Pi$, where the new
field $\sigma$ has non-zero Wick's contractions
\begin{equation}
\left \langle 0 \left | T^{+} \sigma(x) \sigma(y)\right | 0
\right \rangle  = i \delta^2(x-y).
\end{equation}
Then from the canonical propagators
\begin{eqnarray}
\left \langle 0 \left | T^{+} \Pi(x) \Pi(y)\right | 0
\right \rangle & = & 0\ , \\ 
\left \langle 0 \left | T^{+} A_{-} (x) \Pi(y)\right | 0
\right \rangle & = & E_F^1(x-y)\ ,
\end{eqnarray}
we find the Wick contractions for the modified field
$\widetilde{\Pi}^{-} = \Pi^{-} + \sigma$ 
\begin{eqnarray}
\left \langle 0 \left | T^{+} \widetilde{\Pi}^{-}(x)
\widetilde{\Pi}^{-}(y) \right | 0 \right \rangle & = & i
\delta^2(x-y)\ ,\\ 
\left \langle 0 \left | T^{+} A_{-} (x) \widetilde{\Pi}^{-}
(y)\right | 0 \right \rangle & = & E_F^1(x-y). 
\end{eqnarray} 
Thus effectively we can take the equivalent interaction
Hamiltonian which is bilinear in scalar fields
\begin{equation}
H_I^{eff}  =  \phi^{\dag}_{0} \ast \left\{- i e \widetilde{\Pi} + m^2
\widehat{a}\right\} \ast {\cal W}_{-1}[\widehat{a}]\ast
\varphi = \phi^{\dag}_{0} \ast \widetilde{H}_{cov}\ast {\cal 
W}_{-1}[\widehat{a}]\ast \varphi.
\end{equation}
Having the Wick contraction for the pair of scalar
fields 
\begin{equation}
\left \langle 0 \left | T^{+} \phi^{\dag}(x) \varphi(y) \right
| 0 \right \rangle  =  \Delta_F(x-y,m^2)\label{scalarfieldprop}
\end{equation}
we can write the  functional form of the Wick
contractions for scalar fields
\begin{eqnarray}
&&\hspace{-35pt} T^{+} \exp -i \left\{ \phi^{\dag} \ast
\widetilde{H}_{cov}\ast {\cal W}_{-1}[\widehat{a}] \ast
\varphi \right\} = \nonumber\\
&&\hspace{75pt}  =  \exp -i \left\{\left(\phi^{\dag} 
+ \frac{\delta}{\delta \varphi} \ast \Delta_F  \right) \ast
\widetilde{H}_{cov}\ast {\cal W}_{-1}[\widehat{a}] \ast
\left(\varphi + \Delta_F\ast \frac{\delta}{\delta
\phi^{\dag}}\right)\right\} = \nonumber\\ 
& & \hspace{75pt}= \ : \exp -i \left\{\phi^{\dag} \ast \widetilde{H}_{cov}\ast
{\cal W}_{-1}[\widehat{a}] \ast \left(1 + i \Delta_F\ast
\widetilde{H}_{cov}\ast {\cal W}_{-1}[{a}]\right)^{-1} \ast
\varphi\right\}\ :  \times  \nonumber\\ 
&&\hspace{95pt} \times \ \  \exp - {\rm Tr} \ln \left(1 + i \Delta_F \ast
\widetilde{H}_{cov} \ast {\cal W}_{-1}[\widehat{a}] \right)\ .
\end{eqnarray}
In  $\widetilde{H}_{cov}$ there is the integral operator $m^2
(i\partial_{-})^{-1}$ which has a convolution either with the field
$\phi^{\dag}$ 
\begin{equation}
\phi^{\dag}_{0} \ast m^2 (i\partial_{-})^{-1} = - 2i \partial_{+}
\phi^{\dag}_{0}
\end{equation}
or with the propagator function $\Delta_F$
\begin{equation}
\Delta_F \ast m^2 (i\partial_{-})^{-1} = - 2i \partial_{+}
\Delta_F \stackrel{\leftarrow}{\partial_{+}} + i (i\partial_{-})^{-1} .
\end{equation}
Now it is quite easy to check the following factorization:
\begin{equation}
1 - i \Delta_F\ast \widetilde{H}_{cov}\ast {\cal
W}_{-1}[\widehat{a}] = \left(1 - i \Delta_F\ast 
{H}_{cov}\right) \ast {\cal W}_{-1}[\widehat{a}]\ , 
\end{equation}
where  
\begin{equation}
H_{cov} = - i e \widetilde{\Pi} - 2 i e
\stackrel{\leftarrow}{\partial_{+}} A_{-}\ , 
\end{equation}
and next we can write the expression for the effective scalar field
contractions
\begin{eqnarray}
T^{+} \exp -i \left\{\phi^{\dag} \ast \widetilde{H}_{cov}\ast {\cal
W}_{-1}[\widehat{a}] \ast \varphi \right\} = \ : \exp -i \left
\{\phi^{\dag} \ast {H}_{cov}
\ast \left(1 + \Delta_F\ast {H}_{cov}\ast \right)^{-1} \ast
\varphi\right\}\: \times\nonumber\\ 
 \hspace{140pt}\times  \exp - {\rm Tr} \ln \left(1 + i \Delta_F \ast
{H}_{cov} \right) \times \exp - {\rm Tr} \ln{\cal
W}_{-1}[\widehat{a}] .
\end{eqnarray}
In this way, we have found the effective Feynman rules for
scalar and gauge fields in the usual form of perturbative
propagators (\ref{scalarfieldprop}) and
(\ref{genLFWeylpropag}), respectively. The vertices are given by 
\begin{equation}
H_{cov} = - i e \partial_{+}A_{-} - 2 i e
\stackrel{\leftarrow}{\partial_{+}} A_{-}
\end{equation}  
so at any vertex we have the factor $(2 p_{+} - k_{+})$, 
where $k_{+}$ is the momentum  of gauge field line and $p_{+}$
is the momentum  of scalar field line at its $\phi^{\dag}$
end.\footnote{Due to the momentum conservation at each vertex, 
this is equivalent to the symmetrical rules where momenta of
both scalar lines are taken, for example see
\cite{BjorkenDrell1964}.} It is very interesting that here
for the scalar field QED all noncovariant contributions boil
down to the closed loop factor 
\begin{equation}
\exp - {\rm Tr} \ln{\cal W}_{-1}[\widehat{a}] \approx
\exp  {\rm Tr} \ln\left[i \partial_{-} - e A_{-}\right]
\end{equation}
and according to the same arguments that we have used for the
fermion field case, we can omit these contributions completely.

\newpage

\part{Finite volume QED}

 	\section{LF Weyl gauge}\label{chapfvWeyl}

\setcounter{equation}{0}

\subsection{The DLCQ method}

Notation for models in a finite volume of LF are generally
given in \cite{PrzNausKall}. Below, for convenience, we present
some of the basic steps of the DLCQ method underling these points
which are different from \cite{PrzNausKall}.\\
We choose the space of LF as a "hypertorus": $-L < x^- < L$ i
$-L_{\perp} < x_{\perp} < L_{\perp}$ and impose periodic
boundary conditions for boson fields. For fermions we choose 
antiperiodic boundary conditions and this choice means that,
contrary to bosons, fermions will solely have modes with all
nonzero components of momentum.\\
Zero modes of boson fields can be discussed using the
classification introduced in \cite{KaP94}. We denote the
full gauge field as $V_\mu({\vec x})$ and this can be expanded
in Fourier modes. We distinguish, respectively, the {\it
simple} zero mode 
\begin{equation}
\int_{-L}^{+L} \frac{dx^-}{2L} V_\mu(x^-,x_\perp)
\end{equation}
and the {\it normal mode} gauge field:
\begin{equation}
A_\mu ({\vec x}) \equiv V_\mu ({\vec x}) -
\int_{-L}^{+L} \frac{dx^-}{2L} V_\mu(x^-,x_\perp).
\end{equation}
The latter degrees of freedom are known to represent the
usual propagating photons in the light-cone representation,
and as such we reserve the symbol $A_\mu$ to denote them. From
the simple zero mode one can build the totally space-independent
{\it global} zero modes
\begin{equation}
q_\mu = \int_{-L}^{L} \int_{-L_\perp}^{L_\perp}
\frac{dx^- d^2x_\perp}{8 L L_\perp^2} V_\mu ({\vec x}) \; ,
\end {equation}
where in future we shall write $d^3x$ for $dx^- d^2x_{\perp}$
and suppress the limits of integration. Evidently, the $q_\mu$
are $0+1$ dimensional fields, namely quantum mechanical
variables - thus the notation $q$. 
Finally, the simple and
global zero modes can be used to build modes 
with {\it no} $x^-$-dependence but {\it no constant} part in $x_\perp$,
i.e. the {\it proper} zero modes:
\begin{equation}
a_\mu (x_\perp) = \int_{-L}^{+L} \frac{dx^-}{2L} V_\mu(x^-,x_\perp)
 - q_\mu
\;.
\end{equation}
When the decomposition of modes is complete we shall refer
to one of the normal $A_\mu$, proper zero mode $a_\mu$ and
global zero mode $q_\mu$ sectors. \\
For all the above sectors one needs the corresponding delta functions.
We adopt the notation that the {\it periodic}  three-dimensional
delta function is represented as $\delta^{(3)}({\vec x} - {\vec y})$,
which includes the zero modes.
For the {\it antiperiodic} delta function, a subscript `a' is
appended: $\delta_a^{(3)}({\vec x} - {\vec y})$. The explicit
difference between these two objects can be easily seen by
expanding in discrete Fourier modes.
Next we distinguish the delta functions appropriate
for each mode sector for periodic functions. Thus in the
normal mode sector we must subtract the $x^-$ independent part
of $\delta^{(3)}$, and so define
\begin{equation}
\delta^{(3)}_n ({\vec x} - {\vec y}) \equiv
\delta^{(3)}({\vec x} - {\vec y}) -
{1 \over {2L}} \delta^{(2)} (x_\perp - y_\perp)
\; .
\end{equation}
In the proper zero mode sector
we must subtract the overall two-dimensional volume factor
in defining the relevant delta distribution:
\begin{equation}
\delta_p^{(2)}(x_\perp - y_\perp) \equiv \delta^{(2)}(x_\perp -
y_\perp) - {1\over {4L_\perp^2}}\; .
\end{equation}
\noindent
When reaching a canonical formulation, the QED Lagrangian,
expressed in terms of the complete fields $V_\mu$, $\psi$ and
$\psi^{\dagger}$, takes the standard form 
\begin{equation}
L = \int d^3{x} \left[ - \frac 1 4 (\partial_\mu V_\nu
- \partial_\nu V_\mu)(\partial^\mu V^\nu
- \partial^\nu V^\mu) + \bar{\psi}(i \gamma^{\mu} \partial_\mu -
e \gamma^\mu V_\mu - M)\psi\right] \;.\label{QEDLagr}
\end{equation}
Once the boson field is decomposed into its different sectors,
$
V_\mu(\vec{x}) = A_\mu (\vec{x}) + a_\mu(x_\perp) + q_\mu \;,
$ the Lagrangian Eq.(\ref{QEDLagr}) breaks into three parts
\begin{equation}
L = L_{nm}  + L_{pzm} + L_{gzm} \;,\label{defLagrQED}
\end{equation}
where
\begin{eqnarray}
L_{nm} & = & \int d^3{x} \left[ - \frac 1 4 (\partial_\mu A_\nu
- \partial_\nu A_\mu)(\partial^\mu A^\nu
- \partial^\nu A^\mu) + \bar{\psi}(i \gamma^\mu \partial_\mu -
e \gamma^\mu A_\mu - M)\psi\right] \ , \label{nmLagr}\\
L_{pzm} & = & \int d^3{x} \left[ - \frac 1 4 (\partial_\mu a_\nu
- \partial_\nu a_\mu)(\partial^\mu a^\nu
- \partial^\nu a^\mu) - e \bar{\psi} \gamma^\mu \psi a_\mu
\right]\label{pzmLagr}\ , \\
L_{gzm} & = & (8LL_\perp^2) \frac 1 2 (\partial_{+}
q_{-})^2 - e q_\mu \int d^3{x} \bar{\psi} \gamma^\mu
\psi \;.
\label{gzmLagr}
\end{eqnarray}
These formulas in turn can be simplified by decomposing
the total electromagnetic current,
${\cal J}^\mu = - e {\bar \psi} \gamma^\mu \psi$, into its
normal mode and proper and global zero mode parts:
${\cal J}^\mu = J_{nm}^\mu + J_{pzm}^\mu + Q^\mu
$.\footnote{This notation is different from \cite{PrzNausKall}
in order to avoid collision with external 
currents $j^{\mu}$ and fermion currents $J^\mu$.} Fermion
current, while being bilinear in fermion fields, satisfies
periodic boundary conditions and its global zero modes are
given by 
\begin{eqnarray}
Q^{\pm} & = & -  e\sqrt{2} \int \frac{d^3x}{8L\ L^2_\perp}
\psi^{\dagger}_{\pm}(\vec{x})\psi_{\pm}(\vec{x})\ , \\
Q^i & = & - e \int \frac{d^3x}{8L\ L^2_\perp} \left[
\psi^{\dagger}_{+}(\vec{x})\alpha^{i}\psi_{-}(\vec{x}) +
\psi^{\dagger}_{-}(\vec{x})\alpha^{i}\psi_{+}(\vec{x}) \right].
\label{fvWeyldefqi} 
\end{eqnarray}
The proper zero modes are
\begin{eqnarray}
J_{pzm}^{\pm}(x_\perp) & = & - e\sqrt{2} \int_{-L}^{L}
\frac{dx^{-}}{2L}\psi^{\dagger}_{\pm}(\vec{x}) \psi_{\pm}
(\vec{x})- Q^{\pm}\ , \\ 
J_{pzm}^{i}(x_\perp) & = & -  e \int_{-L}^{L} \frac{dx^{-}}{2L}
\left[ \psi^{\dagger}_{+}(\vec{x})\alpha^{i}\psi_{-}(\vec{x}) + 
\psi^{\dagger}_{-}(\vec{x})\alpha^{i}\psi_{+}(\vec{x})\right] -
Q^i. \label{fvWeyldefji}
\end{eqnarray}
Finally we can give the normal modes for fermion currents as
follows: 
\begin{eqnarray}
J_{nm}^{\pm}(\vec{x}) & = & -  
e\sqrt{2} \psi^{\dagger}_{\pm}(\vec{x})\psi_{\pm}(\vec{x})
- J^{\pm}_{pzm}(x_\perp) - Q^{\pm} \ , \\
J_{nm}^{i}(\vec{x}) & = & - e \left[
\psi^{\dagger}_{+}(\vec{x})\alpha^{i}\psi_{-}(\vec{x}) + 
\psi^{\dagger}_{-}(\vec{x})\alpha^{i}\psi_{+}(\vec{x})\right] -
J_{pzm}^{i}(x_\perp) - Q^{i}\;.
\end{eqnarray}

\subsection{Canonical formalism for LF Weyl gauge}

The Weyl gauge condition can be imposed strongly, namely $A_+ =
a_+ = q_+ = 0$ at the classical level. As usual, this means
that Gauss' law (given explicitly later) appears as a constraint
to be imposed on the physical states. 
Our procedure for carrying the canonical procedure is as
follows: we will analyse different subsystems, where only one
field sector is treated in terms of independent degrees of
freedom while the remaining fields are regarded as non-dynamical
external fields and/or currents. Then we will exchange
non-dynamical modes for effective interactions of dynamical ones
and will give a simpler (though nonlocal) Lagrangian where
only dynamical fields are present. 
This will result in a sequence of equivalent effective
Lagrangians which contain fewer modes but have the same
Euler-Lagrange equations  as those which are generated
by the primary Lagrangian, provided the constraint equations are
implemented for non-dynamical fields.
This procedure is based on the observation that different
Lagrangians can lead to the same system of Euler-Lagrange
equations though they may have very different constraint
structure. One can feel free to choose the most
suitable one for carrying out the canonical quantization
procedure. 

\subsubsection{ Proper zero mode sector}

The Lagrangian Eq.(\ref{pzmLagr}) can be written explicitly in
LF coordinates
\begin{equation}
L^{pzm}_{Weyl} = \int d^3{x} \left[
- \partial_i a_{-} \partial_{+}a_i + \frac 1 2 (
\partial_{+}a_{-})^2 - \frac
1 4 (\partial_{i}a_j - \partial_{j}a_i)^2
+ J^\perp_{pzm} a_\perp + J^{-}_{pzm} a_{-}
\right]\;,\label{fvWeyldefLagrpzm} 
\end{equation}
where the Weyl gauge condition $a_+ = 0$ has been explicitly
imposed. Because we are interested here in boson fields
we will treat fermion currents as arbitrary external
currents, however without introducing any distinction in
notation we hope that this will cause no misunderstandings.
Thus the Lagrangian (\ref{fvWeyldefLagrpzm}) leads to the
classical equations of motion 
\begin{eqnarray}
\partial^2_{+} a_{-} & = & \partial_i \partial_{+} a_{i} +
J_{pzm}^{-} \ , \label{fvWeylELeq1}\\ 
- \partial_i \partial_{+} a_{-} & = &\Delta_\perp a_i -
\partial_i \partial_{k} a_{k}  + J_{pzm}^i \ , \label{fvWeylELeq2}
\end{eqnarray}
which impose the parameterization of the field $a_i$ 
\begin{equation}
a_i = - \frac{1}{\Delta_\perp} \ast
\left (J_{pzm}^i + \frac{1}{2L}\partial_i \pi\right).
\label{fvWeyltransvproj} 
\end{equation}
This new field $\pi$ has the Dirac bracket 
\begin{equation}
\left \{\pi(x_\perp), a_{-}(y_\perp) \right \}_{DB}
=  - \delta_p^{(2)} (x_\perp - y_\perp)\ , 
\end{equation}
which is the only nonzero bracket in this sector. The 
canonical Hamiltonian contains the effective nonlocal terms
\begin{eqnarray}
H^{pzm}_{Weyl} & = & \frac{2L}{2} \int d^2x_\perp 
J_{pzm}^i \frac{1}{\Delta_{\perp}} \ast J_{pzm}^i - \ \int
d^2x_\perp \pi \frac{1}{\Delta_{\perp}} \ast 
\partial_i J_{pzm}^i - 2L \ \int d^2x_\perp a_{-}
J_{pzm}^{-} . 
\end{eqnarray}
One can check that the effective equations of motion which
follow from the above Hamiltonian and bracket agree with the
Euler-Lagrange Eqs.(\ref{fvWeylELeq1},
\ref{fvWeylELeq2}). Therefore our 
Hamiltonian and brackets describe the same classical system as
the primary Lagrangian Eq.(\ref{fvWeyldefLagrpzm}) and one can
give an equivalent Lagrangian density 
\begin{equation}
{\cal L}^{pzm}_{eff} = \frac{1}{2L} \pi \partial_{+} a_{-} -
{\cal H}_D^{pzm} = \frac{1}{2L} \pi \partial_{+} a_{-} +
\frac{1}{2L} \pi \frac{1}{\Delta_{\perp}} \ast \partial_i J_{pzm}^i +
a_{-} J_{pzm}^{-} - \frac 1 2 J_{pzm}^i
\frac{1}{\Delta_{\perp}} \ast J_{pzm}^i \;,
\label{fvWeylpzmefflagr}
\end{equation}
which directly leads to the correct bracket  and equations of motion.

\subsubsection{Normal mode sector}

In the second step, we analyse the sector of normal modes
of gauge field potentials ${A}_\mu$ treating normal modes
of electromagnetic currents $J_{nm}^\mu$ as arbitrary external
sources. From the Lagrangian (\ref{nmLagr}) we take these terms
which contain ${A}_\mu$
\begin{equation}
{\cal L}_{Weyl}^{nm} = \partial_{+} {A}_i \left(\partial_{-}
{A}_i - \partial_i A_{-}\right)
+ \frac 1 2 \left(\partial_{+} {A}_{-} \right)^2 - \frac 1 4
\left( \partial_i {A}_j -
\partial_j {A}_i\right)^2 + {A}_{-} {J}^{-} + {A}_i {J}^i \;,
\end{equation}
and find the Euler-Lagrange equations of motion
\begin{eqnarray}
\partial_{+} \left( \partial_{+} {A}_{-}-
\partial_i {A}_{i} \right)  & = & J_{nm}^{-} \ ,\\ 
(2 \partial_{+} \partial_{-} - \Delta_\perp) {A}_{i}
& = & \partial_{i} \left( \partial_{+} {A}_{-}
- \partial_j {A}_{j} \right) + J_{nm}^{i}.
\end{eqnarray}
We see that these equations have the same structure as
the respective equations in Section \ref{chapivWeyl} in the limit
$\alpha \rightarrow 0$, therefore we can adopt our previous
results here. Thus we take a new field $\Pi = \partial_{+}A_{-}
- \partial_i {A}_i$ and get the diagonal structure of Dirac
brackets 
\begin{eqnarray}
\left \{{A}_{-}(\vec{x}), \Pi(\vec{y}) \right \}_{DB} & =&
\delta_n^{(3)} (\vec{x} - \vec{y})\ , \\
2\partial_{-}^x \left \{{A}_{i}(\vec{x}), {A}_j (\vec{y})
\right \}_{DB} & = & - \delta_{ij}
\delta_n^{(3)}(\vec{x}-\vec{y}) \ , 
\end{eqnarray}
while all other brackets vanish. The canonical Hamiltonian
\begin{equation}
{\cal H}_{Weyl}^{nm} = \frac 1 2 (\Pi)^2 + \Pi \partial_i {A}_i + \frac 1
2 \left( \partial_i {A}_j \right)^2 - {A}_{-} J_{nm}^{-} - {A}_i
J_{nm}^i
\end{equation}
generates equations of motion which are equivalent to the
previous ones and
can be used for defining an effective Lagrangian
\begin{eqnarray}
{\cal L}^{nm}_{Weyl} & = & \partial_{+} {A}_{i}\partial_{-}
{A}_{i} + \Pi \partial_{+} {A}_{-} - {\cal H}_{Weyl}^{nm}\nonumber \\
& = & \partial_{+} {A}_{i}\partial_{-} {A}_{i} - \frac 1 2
\left( \partial_i {A}_j\right)^2 - \frac 1 2 (\Pi)^2 + \Pi
\left( \partial_{+} {A}_{-} - \partial_i {A}_i \right) + {A}_{-}
J_{nm}^{-} + {A}_i J_{nm}^i . \label{fvWeylnmefflagr}
\end{eqnarray}
Having analysed the canonical structure of the gauge field
sector, one can substitute the Lagrangian Eq.(\ref{pzmLagr}) by
Eq.(\ref{fvWeylpzmefflagr}) and the boson part of Eq.(\ref{nmLagr}) by
Eq.(\ref{fvWeylnmefflagr}) and instead of the total Lagrangian
Eq.(\ref{defLagrQED}), one can work with the effective Lagrangian
\begin{eqnarray}
\widetilde{\cal L}^{eff}_{Weyl} & = &
\partial_{+} {A}_{i}\partial_{-} {A}_{i} -
\frac 1 2 \left( \partial_i {A}_j\right)^2 - \frac 1 2 (\Pi)^2 +
\Pi \left( \partial_{+} {A}_{-} - \partial_{i} {A}_{i} \right)
+ \frac{1}{2L}  \pi \partial_{+} {a}_{-}
\nonumber \\
& + & \frac 1 2 \left( \partial_{+}
q_{-}\right)^2
 +  i \sqrt{2} {\psi_{+}}^{\dagger} \partial_{+} \psi_{+} + i
\sqrt{2} {\psi_{-}}^{\dagger} \partial_{-} \psi_{-} + i
{\psi_{-}}^{\dagger} \alpha^i \partial_{i} \psi_{+} + i
{\psi_{+}}^{\dagger} \alpha^i \partial_{i} \psi_{-}\nonumber\\
&- & M {\psi_{+}}^{\dagger} \gamma^0 \psi_{-} - M
{\psi_{-}}^{\dagger} \gamma^0 \psi_{+}-  e \sqrt{2}
{\psi_{-}}^{\dagger} \psi_{-}{V}_{-} \label{efflagr1}
\label{prelimeffLag} \\
&-& eV^{\prime}_i \left({\psi_{+}}^{\dagger} \alpha^i \psi_{-}+
{\psi_{-}}^{\dagger} \alpha^i \psi_{+}\right)- \frac 1 2 J_{pzm}^i
\left( \frac{1}{\Delta_\perp} \ast J_{pzm}^i\right) \; . \nonumber
\end{eqnarray}
In this expression,
\begin{equation}
V^\prime_i = {A}_i - \frac{1}{2L} \partial_i
\frac{1}{\Delta_{\perp}} \ast \pi + q_{i},
\end{equation}
namely it is the original $V_i$ but with its proper zero mode
$a_i$ expressed as in Eq.(\ref{fvWeyltransvproj}) and the new
"current-current" interaction subtracted.\footnote{This term
has been explicitly reintroduced as the last
term in Eq.(\ref{prelimeffLag}).}  The decomposition of $V_-$
remains unchanged.

\subsubsection{Fermion sector}

In the next step, we take the fermion part of the effective
Lagrangian Eq.(\ref{efflagr1})
\begin{eqnarray}
{\cal L}^{fer}_{Weyl} & = & i \sqrt{2} {\psi_{+}}^{\dagger}
\partial_{+} \psi_{+} + i \sqrt{2} {\psi_{-}}^{\dagger}
\partial_{-} \psi_{-} + i {\psi_{-}}^{\dagger} \alpha^i
\partial_{i} \psi_{+} + i {\psi_{+}}^{\dagger} \alpha^i
\partial_{i} \psi_{-}\nonumber\\ 
&- & M {\psi_{+}}^{\dagger} \gamma^0 \psi_{-} - M
{\psi_{-}}^{\dagger} \gamma^0 \psi_{+}-  e \sqrt{2}
{\psi_{-}}^{\dagger} \psi_{-}{V}_{-} \label{fermionlagr}\\
& - &e{V}_i^{\prime}
\left({\psi_{+}}^{\dagger} \alpha^i \psi_{-}+
{\psi_{-}}^{\dagger} \alpha^i \psi_{+}\right)- \frac 1 2 j^i
\left( {\cal G}_{(\perp)}[0] \ast j^i\right) \;.\nonumber
\end{eqnarray}
The non-dynamical modes are the fermion field $\psi_{-}$
and the global zero mode $q_{i}$.
These are determined by the following differential equation and global
integral condition:
\begin{equation}
\left( i \sqrt{2} \partial_{-} - e \sqrt{2} V_{-} \right)
\psi_{-} = - i \alpha^i \partial_i \psi_{+} + M \gamma^0
\psi_{+} \ \ + \ e \alpha^i \left( V_i - \frac{1}{\Delta_\perp}
\ast J_{pzm}^{i} \right) \psi_{+} \ , \label{eqpsim}
\end{equation}
\begin{equation}
 0 = Q^i = \frac{1}{8LL^2_\perp} \int \ d^3 {x}\ (
\psi^{\dagger}_{+} \alpha^i \psi_{-} + \psi^{\dagger}_{-} \alpha^i
\psi_{+})(\vec{x}).
\end{equation}
The first equation leads to 
\begin{equation}
\psi_{-}(\vec{x}) = \frac{1}{\sqrt{2}} \frac{1}{i\partial_{-} -
e V_{-}} \ast \xi (\vec{x}) - \frac{e}{\sqrt{2}}
\frac{1}{\Delta_{\perp}} \ast (J_{pzm}^i (x_\perp) - q_{i} )
\alpha^i \frac{1}{i\partial_{-} - e V_{-}} \ast
\psi_{+} (\vec{x}) \;, \label{prsolpsim} 
\end{equation}
where
\begin{eqnarray}
\xi(\vec{x}) & = & \left[ M \gamma^0 - i\alpha^i \partial_i +
e \alpha^i{A}_i(\vec{x}) - e \alpha^i
\frac{1}{2L} \partial_i
\left( \frac{1}{\Delta_\perp} \ast \pi\right)(\vec{x})
\right] \psi_{+}(\vec{x})\label{defxi} \; .
\end{eqnarray}
Note that these are not yet the solutions for the dependent fermion
field $\psi_{-}$ because these fields appear
also on the right-hand side in the zero mode currents $J^i_{pzm}$.
However, one can introduce them into the definition of $J^i_{pzm} +
Q^i$ (see Eqs.(\ref{fvWeyldefqi},\ref{fvWeyldefji}))
\begin{equation}
J^i_{pzm}(x_\perp) + Q^i
= - \frac{e}{2 L} \Gamma^i(x_\perp) + \frac{e^2}{2L}
{\cal M}^{ik}(x_\perp) \left[ \left ( \frac{1}{\Delta_\perp}\ast
J^k_{pzm}\right)(x_\perp) - q_{k}\right] \;,
\end{equation}
where
\begin{eqnarray}
\Gamma^i(x_\perp) & = & \frac{1}{\sqrt{2}} \int dx^{-}
\psi_{+}^{\dagger} (\vec{x}) \alpha^i \left(\frac{1}{i
\partial_{-} - e V_{-}} \ast \xi \right) (\vec{x}) \nonumber\\ 
& + &   \frac{1}{\sqrt{2}} \int dx^{-}
 \xi^{\dagger}(\vec{x}) \alpha^i \left(\frac{1}{i
\partial_{-} - e V_{-}} \ast
 \psi_{+} \right) (\vec{x})\ ,\\
{\cal M}^{ij}(x_\perp)
&= & \delta^{ij} \sqrt{2} \int dx^{-}
 \psi^{\dagger}_{+} (\vec{x}) \left(\frac{1}{i
\partial_{-} - e V_{-}} \ast
 \psi_{+} \right) (\vec{x}) = \delta^{ij} {\cal M}^2(x_\perp)
\; .
\label{gamM}
\end{eqnarray}
Now, from the constraint $Q^i = 0$ one gets the differential equation
\begin{equation}
\left[ \Delta_{\perp} - \frac{e^2}{2L} {\cal M}^2(x_\perp) \right]
\left[\left(\frac{1}{\Delta_{\perp}} \ast J_{pzm}^i\right)
(x_\perp) - q_{i}\right] = - \frac{e}{2L} \Gamma^i (x_\perp),
\label{formalQisol} 
\end{equation}
which has a formal solution
\begin{equation}
\left( \frac{1}{\Delta_{\perp}} \ast J_{pzm}^i \right)(x_\perp)
- q_{i} = - \frac{e}{2L} \int d^2y_\perp {\cal
G}_{(\perp)}[x_\perp, y_\perp; {\cal M}^2] \Gamma^i(y_\perp)
\end{equation}
in terms of the functional ${\cal G}_{(\perp)}[x_\perp,
y_\perp; {\cal M}^2] $ introduced in Eq.(\ref{eq:Gperp}).
Finally, we may express the non-dynamical fermion field as
\begin{equation}
\psi_{-}(\vec{x})  = \frac{1}{\sqrt{2}} \left(\frac{1}{i
\partial_{-} - e V_{-}} \ast \xi\right)(\vec{x}) + \frac{e^2}{2
\sqrt{2} L} \left( {\cal G}_{(\perp)}[{\cal M}^2] \ast
\Gamma^i\right) (x_\perp) \left( \frac{1}{i \partial_{-} - e
V_{-}} \ast \psi_{+}\right)(\vec{x}) \;, \label{prsolpsi}
\end{equation}
whereby we obtain
\begin{equation}
H^{fer}_D = \frac{1}{\sqrt{2}} \int d^3x \xi^{\dagger}(\vec{x})
\left(\frac{1}{i \partial_{-} - e V_{-}} \ast \xi\right)
(\vec{x}) + \frac 1 2 \frac{e^2}{2L} \int d^2x_\perp 
\Gamma^i (x_\perp) \left( {\cal G}_{(\perp)} [{\cal M}^2] \ast
\Gamma^i \right) (x_\perp)\label{Dhamglob}
\end{equation}
as the Dirac Hamiltonian for unconstrained fields.
Just as before we can give the effective
classical
Lagrangian for the
fermions
\begin{eqnarray}
{\cal L}_{eff}^{fer} & = & (\partial_{+} \psi_{+})
\Pi_{\psi_{+}} - {\cal H}^{fer}_D = i
\sqrt{2}\psi^{\dagger}_{+} \partial_{+} \psi_{+} \nonumber\\ 
& - & 
 \frac{1}{\sqrt{2}} \xi^{\dagger} \left( \frac{1}{i
\partial_{-} - e V_{-}}\ast \xi \right) -\frac 1 2 \frac{e^2}{4L^2} \Gamma^i
\left({\cal G}_{(\perp)} [{\cal M}^2] \ast \Gamma^i \right)
\label{efflagrfer}.
\end{eqnarray}

\subsection{Quantum theory}
Having eliminated all non-dynamical fields we may now proceed
by substituting the fermion part of Eq.(\ref{efflagr1}) by that 
given in Eq.(\ref{efflagrfer}). We thereby obtain the total
effective Lagrangian 
\begin{eqnarray}
{\cal L}_{eff} & = & \partial_{+} {A}_{i}\partial_{-} {A}_{i} -
\frac 1 2 \left( \partial_i {A}_j\right)^2 - \frac 1 2 (\Pi)^2 +
\Pi \left( \partial_{+} {A}_{-} - \partial_{i} {A}_{i} \right)
- \frac{1}{2L} \pi  \partial_{+} {a}_{-} +
\frac 1 2 \left( \partial_{+}
q_{-}\right)^2 \nonumber\\
& +& i \sqrt{2}\psi^{\dagger}_{+} \partial_i \psi_{+} -
\frac{1}{\sqrt{2}}  \xi^{\dagger} \left( \frac{1}{i
\partial_{-} - e V_{-}} \ast \xi \right)
-\frac 1 2 \frac{e^2}{4L^2} \Gamma^i \left({\cal
G}_{(\perp)} [{\cal M}^2] \ast \Gamma^i \right)
\label{effQEDLagr}
\end{eqnarray}
as the starting point for the canonical quantisation procedure.
It generates the Euler-Lagrange equations 
of motion which agree with the dynamical equations of the
primary Lagrangian Eq.(\ref{QEDLagr}) with all non-dynamical equations
(formally) implemented.
The canonical quantisation is simple here and one gets
the equal-$x^{+}$ quantum commutation
relations
\begin{eqnarray}
\left [{\Pi}(\vec{x}), {A}_{-}(\vec{y}) \right ] & = & - i
\delta_n^{(3)} (\vec{x} - \vec{y})\ ,\label{comPiA-}\\
\left \{ \psi_{+}^{\dagger}(\vec{x}),
{\psi_{+}}(\vec{y}) \right \} & = & \frac{1}{\sqrt{2}} \
\Lambda_{+} \delta^{(3)}_a(\vec{x}- \vec{y}) \ , \\ 
\left [\pi(x_\perp), {a}_{-} (y_\perp) \right ] & = & - i
\delta_p^{(2)} (x_\perp - y_\perp) \ ,\label{comaLa-}\\
2 \partial_{-}^x\left [{A}_{i}(\vec{x}), {A}_j (\vec{y}) \right ] & = &
-i \delta_{ij} \delta_n^{(3)} (\vec{x} - \vec{y})
\ ,\\
\left[ p^{-}, {q}_{-} \right ] & = & - i \ , 
\end{eqnarray}
and the quantum Hamiltonian, like ${\cal L}_{eff}$, which comes from
${\cal H}_D^{pzm}$, ${\cal H}_D^{nm}$ and $ H_D^{fer}$,
\begin{eqnarray}
H_{eff} & =& \int d^3{x}\left[ \frac 1 2 (\Pi(\vec{x}))^2 + \Pi
\partial_i {A}_i (\vec{x}) + \frac 1 2 \left( \partial_i
{A}_j(\vec{x}) \right)^2 \right] + \frac{1}{16 L L^2_\perp}
(p^{-})^2  \nonumber\\
& + & \frac{1}{\sqrt{2}} \int d^3{x} \xi^{\dagger}(\vec{x})
\left(\frac{1}{i \partial_{-} - e V_{-}} \ast \xi\right) (\vec{x})
+  \frac 1 2 \frac{e^2}{2L} \int d^2x_\perp \Gamma^i (x_\perp)
\left( {\cal G}_{(\perp)} [{\cal M}^2] \ast \Gamma^i \right)
(x_\perp),\nonumber\\
 \label{quanHam}
\end{eqnarray}
tacitly defining the ordering. One can show that, due to the
effective equations of motion, the  Gauss law operator 
\begin{equation}
G(\vec{x}) = \partial_{-} \Pi(\vec{x}) + 2 \partial_{-} \partial_i
{A}_i (\vec{x}) - \Delta_\perp {A}_{-}(\vec{x}) -
\Delta_\perp {a}_{-} (x_\perp) - e \sqrt{2}
\psi^{\dagger}_{+}(\vec{x}) \psi_{+}(\vec{x})
\label{gaussop}
\end{equation}
is $x^{+}$-independent. It leads to a classical first class
constraint, $G \simeq 0$; namely it must annihilate physical
states in the quantum theory. Furthermore, it is intimately
connected to the residual gauge symmetry 
\begin{eqnarray}
{A}_i(\vec{x})^h & = & \Omega_h {A}_i(\vec{x})
\Omega^{\dagger }_h = {A}_i(\vec{x})  - \partial_i {h}(\vec{x})\ , \\
{A}_{-}(\vec{x})^h & = & \Omega_h {A}_{-}(\vec{x})
\Omega^{\dagger }_h = {A}_i(\vec{x})  - \partial_{-} h(\vec{x})\ , \\
\Pi(\vec{x})^h & = & \Omega_h \Pi(\vec{x}) \Omega^{\dagger }_h
= \Pi(\vec{x})  + \Delta_\perp h(\vec{x})\ , \\
\psi_+(\vec{x})^h & = & \Omega_h \psi_+(\vec{x}) \Omega^{\dagger }_h
= e^{ i e h({\vec x})} \psi_+ (\vec{x})\ , \\
{a}_{-}(x_\perp)^h & = & \Omega_h {a}_{-}(x_\perp)
\Omega^{\dagger }_h = {a}_{-}(x_\perp)  \ , \\
\pi(x_\perp)^h & = & \Omega_h \pi(x_\perp)
\Omega^{\dagger }_h = \pi(x_\perp) + \Delta_\perp
\int_{-L}^{L} dy^{-}\ h(y^{-}, x_\perp)\ ,\\
{p^{-}}^h & = & \Omega_h p^{-}
\Omega^{\dagger }_h = p^{-}\ ,
\end{eqnarray}
where 
\begin{equation}
\Omega_h = e^{i \int d^3{\vec x} h({\vec x}) \ G(\vec{x})}\ .
\end{equation}

\subsubsection{Translation Generators}

With the effective Lagrangian density Eq.(\ref{effQEDLagr}) one
can calculate the canonical momentum-energy tensor
\begin{eqnarray}
T^{\nu\mu} & = & \frac{\delta {L}_{eff}}{\delta \ (\partial_{\nu} {A}_i)}
\partial^{\mu} {A}_i + \partial^{\mu} \psi_{+}
\frac{\delta {L}_{eff}}{\delta \
(\partial_{\nu} \psi_{+})} + \frac{1}{2L} \frac{\delta {L}_{eff}}{\delta \
(\partial_{\nu} \pi)} \partial^{\mu} \pi
\nonumber\\
& + & \frac{1}{2L} \frac{\delta {L}_{eff}}{\delta \
(\partial_{\nu} a_{-})} \partial^{\mu} a_{-} + \frac{1}{8 L
L_\perp^2} \frac{\delta {L}_{eff}}{\delta \ (\partial_{\nu}
q_{-})} \partial^{\mu} q_{-} - g^{\nu \mu} {\cal L}_{eff}.
\end{eqnarray}
Then from the generators of translations $P^\mu = \int d^3{x}
\ T^{+\mu}(\vec{x})$, the spatial translations are
\begin{eqnarray}
P^{i} &= &\int d^3{x} \left[ - \partial_{-}{A}_k
\partial_{i} {A}_k - {\Pi} \partial_{i}
{A}_{-} - i \sqrt{2} \psi^{\dagger}_{+} \partial_{i}
\psi_{+} - \frac{1}{2L}\pi \partial_i {a}_{-}
\right] \label{Momi} \ , \\
P^{+} & = & \int d^3{x} \left[ \partial_{-} {A}_k
\partial_{-} {A}_k + {\Pi} \partial_{-}
{A}_{-} + i \sqrt{2} \psi^{\dagger}_{+} \partial_{-}
\psi_{+}\right]. \label{Momplus}
\end{eqnarray}
Now from the (anti)commutation relations
Eqs.(\ref{comPiA-}-\ref{comaLa-}) one can recover
the correct Heisenberg relations for
all dynamical quantum fields $\varphi_J = ({A}_k,
{A}_{-}, {\Pi}, \pi, {a}_{-})$
\begin{eqnarray}
\partial_{i} \varphi_J & = & - i \left[ P^{i},\varphi_J
\right]\  ,\\
\partial_{-} \varphi_J & = & i \left[ P^{+},\varphi_J
\right]\ ,
\end{eqnarray}
and this confirms the translation invariance of QED in the Weyl
gauge. We note that the generators $P^{+}$ and $P^{i}$ are not
invariant under the residual gauge transformation with
gauge function $h({\vec x})$:
\begin{eqnarray}
P^{+}_h & = & P^{+} + \int d^3{x} \ G(\vec{x}) \partial_{-}
h(\vec{x}) \ , \\
P^{i}_h  & = &
P^{i} - \int d^3{x} \ G(\vec{x}) \partial_{i} h(\vec{x})
\ , 
\end{eqnarray}
and this is connected with the lack of gauge invariance of the
canonical energy-momentum tensor. We return to this below.

\subsubsection{Implementation of  Gauss' Law} 

As we have mentioned earlier, the physical quantum gauge system
has to satisfy Gauss' law \cite{GaussLawlit}. However this
cannot be implemented 
{\em strongly} as the condition for quantum field operators
which would be incompatible with the commutation relations
but rather as the condition for physical states     
\begin{equation}
G(\vec{x})  |{\rm{phys}}\rangle = 0 
\label{fvWeylgauss}. 
\end{equation}
In \cite{PrzNausKall} the method of 
quantum mechanical gauge fixing \cite{LNOT94} has been used and
details of this formalism are given therein. Here we will give
only some most important results. Two gauge transformations are
defined with the help of the Gauss law operator 
\begin{eqnarray}
U_1[\vartheta] &= &
\exp\left( - i \int d^3 x g({\vec x}) \vartheta[{\vec x};
A_-]\right) \ ,\\ 
U_2[\eta] & = & \exp\left (i\int d^2x_\perp \rho_2(x_\perp) \eta[x_\perp;
\pi]\right)\ ,
\end{eqnarray}
where
\begin{eqnarray} 
\vartheta [{\vec x};A_-] &= &\left( \frac{1}{\partial_{-}}
\ast  A_{-}\right) (\vec{x}) \label{fvWeylfirsttrans} \ , \\
g({\vec x})& \equiv& G({\vec x}) - \partial_- \Pi({\vec x})
= 2 \partial_- \partial_i A_i - \Delta_\perp A_- - 
 \Delta_\perp a_-  - e \sqrt{2} \psi^\dagger_+ \psi_+ \ ,  \\
\eta[x_\perp; \pi] & = & -\frac{1}{2L} \left( \frac{1}{\Delta_\perp}
\ast \pi \right)(x_\perp) \ ,\\
 \rho_2& = &\frac{e\sqrt{2}}{2L} \int dx^- \psi^\dagger_+ \psi_+ \ .
\end{eqnarray}
They allow us to express the Gauss law condition
(\ref{fvWeylgauss}) in two pairs
\begin{eqnarray}
\Pi |{\rm{phys}}^\prime\rangle & =& 0 \label{fvWeylnewgauss1} \
, \\
\int { {dx^-}\over{2L}} g({\vec x}) |{\rm{phys}}^\prime\rangle
& = & 0 \ , \label{fvWeylnewgauss2} 
\end{eqnarray}
where
\begin{equation}
 |{\rm{phys}}^\prime \rangle \equiv  U_1  |{\rm{phys}}\rangle,
\end{equation}
and 
\begin{eqnarray}
a_-(x_\perp) |{\rm{phys''}} \rangle & =& 0 \ , 
\label{fvWeylaminusconstr}\\ 
Q |{\rm{phys''}} \rangle & = &\int d^2 x \rho_2(x_\perp)  
 |{\rm{phys}''} \rangle =0 \label{fvWeylglobQ}\ ,
\end{eqnarray}
where 
\begin{equation}
|{\rm{phys''}} \rangle \equiv U_2 |{\rm{phys}}^\prime \rangle.
\end{equation}
Here we have the neutrality condition (\ref{fvWeylglobQ}) as the
only one constraint which survives from the infinite number of
primary constraints which have no physical meaning. \\
Also the {\it physical Hamiltonian} can be defined as it acts
on physical states
\begin{equation}
U_2U_1 H U_1^\dagger U_2^\dagger |{\rm{phys}}''\rangle =  
H_{fin}^{Weyl} |{\rm{phys}}''\rangle 
\end{equation}
and due to (\ref{fvWeylnewgauss1} and \ref{fvWeylaminusconstr})
the field operators $\Pi$ and $a_{-}$ can be omitted as  cyclic
variables 
\begin{eqnarray}
H_{fin}^{Weyl} & =& \int d^3x\left[ \frac 1 2
\left( \partial_j A_j - \sqrt{2} e \frac{1}{\partial_{-}} \ast (
\psi^\dagger_+\psi_+)\right)^2 +
\frac 1 4 \left( \partial_i A_j -\partial_j A_i\right)^2
 \right] \nonumber\\ 
& + &\frac{1}{16L L^2_\perp} (p^{-})^2 + \frac{1}{\sqrt{2}} \int
d^3{x} \chi^{\dagger}(\vec{x}) \left(\frac{1}{i \partial_{-}
- e \left[q^\prime_{-} - \frac{1}{\Delta_{\perp}} \ast \rho_2\right]} \ast
\chi\right) (\vec{x}) \nonumber \\ 
& + &\frac 1 2 \frac{e^2}{2L} 
\int d^2x_\perp \Gamma^{\prime \prime i} (x_\perp)
\left( {\cal G}_{(\perp)} [{\cal M}^{\prime \prime 2}] \ast
\Gamma^{\prime \prime i} \right)
(x_\perp)  \;. \label{fvWeylfinalHamil}
\end{eqnarray} 
The operators $\Gamma''$ and ${\cal M''}$ follow directly from 
the respective operators $\Gamma$ and ${\cal M}$ in which 
$\xi$ is replaced by 
\begin{equation}
\chi = [ m\gamma_0-i\alpha^i\partial_i +e \alpha^i A_i] \psi_+
\end{equation}
and $V_-$ by
\begin{equation}
q^\prime_- - \frac{1}{\Delta_{\perp}} \ast \rho_2\ ,
\end{equation}
where
\begin{equation}
q_-' = q_- - \frac{e}{2L}\frac{1}{\Delta_{\perp}} \ast \delta^2(0).
\end{equation}
This redefinition of $q_-'$ can be viewed as an infinite
renormalization of the field, which does not change the commutators.\\
Similarly for the generator of other translations we have
\begin{eqnarray}
U_2 U_1 P^i U_1^\dagger U_2^\dagger & = &
- \int d^3 x \left( \partial_- A_j \partial_i A_j
+i\sqrt{2} \psi^\dagger_+ \partial_i \psi_+ \right) 
\equiv P^i_{{fin}}\ ,\\
U_2 U_1 P^+ U_1^\dagger U_2^\dagger
& = & \int d^3 x \left( \partial_- A_j \partial_- A_j
+i\sqrt{2} \psi^\dagger_+ \partial_- \psi_+ \right)  
\equiv P^+_{{fin}}\ , 
\end{eqnarray}
which means their invariance in the subspace of physical
states.\\  
Also we stress that in the physical operators, the dependence on
the proper zero mode $a_{-}$ has totally disappeared which
contradicts the naive interpretation of this mode
as a gauge-invariant physical field.\\

\section{Covariant gauge}

\setcounter{equation}{0}

In the finite volume LF the Lorentz covariant gauge condition
$\partial_\mu V^\mu = 0$ would be equivalent to three
independent conditions for each sector
\begin{equation}
\partial_\mu V^\mu = 0 \Longrightarrow
\left\{ \begin{array}{l}
\partial_+ A_{-} + \partial_{-} A_{+} - \partial_i A_{i}= 0 \\
\partial_+ a_{-} - \partial_i a_i = 0\\
\partial_+ q_{-} = 0\ .
\end{array}\right.
\end{equation}
However the last condition for global zero modes is not 
{\it attainable} by means of any gauge transformation because 
mode $q_{-}$ is {\it gauge-invariant}.\footnote{This resembles
the situation in the LC-gauge in the sector of zero modes,
where conditions $a_{-}= q_{-} = 0$ are not attainable as gauge
conditions \cite{KaP94}, \cite{LCgaugeQED}.} Therefore here as
the Lorentz covariant gauge we 
take the following conditions:
\begin{eqnarray}
\partial_+ A_{-} + \partial_{-} A_{+} - \partial_i A_{i}& = & 0
\ , \\
\partial_+ a_{-} - \partial_i a_i & = & 0\ , \\
q_{+} &= &0\ ,
\end{eqnarray}
with the LF Weyl gauge for global zero modes. In Lagrangians
for  different gauge field modes the first two
conditions will be implemented by means of the Lagrange
multipliers $\Lambda_{nm}$ and $\Lambda_{pzm}$, respectively,
while the third one will be imposed explicitly.\\

\subsection{Normal mode sector}

In the sector of normal modes we start with the 
Lagrangian density 
\begin{eqnarray}
{\cal L}_{cov}^{nm} & = &
\left(\partial_{+} {A}_i - \partial_i
{A}_{+}\right)\left(\partial_{-} {A}_i - \partial_i
{A}_{-}\right)  +  \frac 1 2 \left(\partial_{+} {A}_{-} -
\partial_{-}{A}_{+}\right)^2 - \frac 1 4 \left( \partial_i
{A}_j -  \partial_j {A}_i\right)^2\nonumber\\
&  + &{A}_{-} {J}^{-}_{nm} + 
{A}_{+} {J}^{+}_{nm} + {A}_i {J}^i_{nm} +  {\Lambda}_{nm}\left( \partial_{+} 
A_{-} + \partial_{-} A_{+} - \partial_\perp A_\perp\right). 
\end{eqnarray}
Next we introduce the canonical momentum conjugated with $A_{-}$
\begin{equation}
{\Pi}^{-}= \partial_{+} {A}_{-} - \partial_{-} {A}_{+} +
\Lambda_{nm}\ ,
\end{equation}
and then modify the Lagrange multiplier field
\begin{equation}
\lambda_{nm} = \Lambda_{nm} - \frac{1}{2\partial_{-}}*
J^{+}_{nm} \ ,
\end{equation}
in order to separate dynamical equations 
\begin{eqnarray}
\left[2 \partial_{+} \partial_{-} - \Delta_\perp\right]
{\lambda}_{nm} & = & \Delta_\perp \frac{1}{2\partial_{-}}* 
J_{nm}^{+} + \partial_{-} J_{nm}^{-} + \partial_i J_{nm}^{i} \
, \label{fvLorHameqlambda} \\ 
(2 \partial_{+} \partial_{-} - \Delta_\perp){\Pi}^{-} & = & 
 2 \partial_{-} J_{nm}^{-} + \partial_i J_{nm}^{i}\ , 
\label{fvLorHameqmodPim}\\ 
(2 \partial_{+} \partial_{-} - \Delta_\perp) {A}_{i} & = &
\partial_i \lambda_{nm} + J_{nm}^i + \frac{1}{2\partial_{-}}*
J_{nm}^{+} \ , \label{fvlorHameqmodAi}
\end{eqnarray}
from the constraints
\begin{eqnarray}
A_{-} & = & \frac{1}{\Delta_\perp}*\partial_{-} \left[\Pi^{-} +
\partial_i A_i - 2\lambda_{nm} \right]\ , \label{fvLorconstrmodAm}\\ 
A_{+} & = & \frac{1}{2\partial_{-}}*\left[ \partial_i A_i +
\lambda_{nm} - \Pi^{-} + \frac{1}{2\partial_{-}}* J_{nm}^{+}
\right] \  \label{fvLorconstrmodAp}. 
\end{eqnarray}
From the canonical Hamiltonian density 
\begin{eqnarray}
{\cal H}^{nm}_{cov} & = & (\partial_{+} {A}_{-}) {\Pi}^{-} +
(\partial_{+} {A}_{i}) {\Pi}^i - {\cal L} = \frac 1 2
\left({\Pi}^{-} - \lambda_{nm} - 
\frac{1}{2\partial_{-}}* J_{nm}^{+}\right )^2 + \frac 1 4 \left(
\partial_i {A}_j - \partial_j {A}_i\right)^2 \nonumber \\
& + &  \left(\lambda_{nm} + 
\frac{1}{2\partial_{-}}* J_{nm}^{+} \right) \partial_i A_i - 
\partial_{-}\left[ \Pi^{-} + \partial_i A_i - 2
\lambda_{nm} \right] \frac{1}{\Delta_\perp} * J_{nm}^{-} - {A}_i J_{nm}^i  
\end{eqnarray}
and the dynamical equations of motion we find the Dirac brackets
\begin{eqnarray}
2\partial^x_{-}\left \{{A}_{i}(\vec{x}), {\Pi}^{-}(\vec{y})
\right \}_{DB} & = & \partial_i^x \delta^3(\vec{x} - \vec{y})
\ , \\
2\partial_{-}^x\left \{{\Pi}^{-}(\vec{x}), {\Pi}^{-}(\vec{y})
\right \}_{DB} & = &  \Delta_\perp \delta^3(\vec{x} - \vec{y})
\ , \\
2 \partial_{-}^x \left \{{A}_{i}(\vec{x}), {A}_j (\vec{y})
\right \}_{DB} & = & - \delta_{ij} \delta^3(\vec{x} - \vec{y})
\ , \\
2\partial_{-}^x \left \{ \lambda_{nm}(\vec{x}), {A}_i (\vec{y})
\right \}_{DB} & = & - \partial_{i}^x \delta^3(\vec{x} -
\vec{y}) \ ,
\end{eqnarray}
while all other brackets vanish. However, another choice of field
variables 
\begin{eqnarray}
\phi & = & \frac{1}{\Delta_\perp} \ast \left( \Pi^{-} - 2
\lambda_{nm} + \partial_i A_i\right)\ , \\
C_i & = & A_i - \partial_i \frac{1}{\Delta_\perp} \ast
\left(\Pi + \partial_j A_j\right) = A_i - \partial_i \phi \ ,
\end{eqnarray}
leads to  a simpler structure of Dirac brackets
\begin{eqnarray}
2\partial^x_{-}\left \{ C_i (\vec{x}), C_j(\vec{y})\right
\}_{DB} & = & - \delta_{ij}\delta^3(\vec{x} - \vec{y}) \ ,
\label{fvcovDBCiCj}\\
2 \partial^x_{-}\left \{ \phi (\vec{x}), \lambda_{nm} (\vec{y})
\right \}_{DB} & = & \delta^3(\vec{x} - \vec{y}) \ ,
 \label{fvcovDBlambdaphi}
\end{eqnarray}
and dynamical equations are separated
\begin{eqnarray}
\left[2 \partial_{+} \partial_{-} - \Delta_\perp\right] {\phi}
& = & \lambda_{nm} - \frac{1}{2\partial_{-}}*J_{nm}^{+} \ ,
\label{fvLorHameqphi}\\ 
\left[2 \partial_{+} \partial_{-} - \Delta_\perp\right]
{\lambda}_{nm} & = & \Delta_\perp \frac{1}{2\partial_{-}}*
J_{nm}^{+} + \partial_{-} J_{nm}^{-} + \partial_i J_{nm}^{i}
\ , \label{fvLorHameqlambda2}\\ 
(2 \partial_{+} \partial_{-} - \Delta_\perp) {C}_{i} & = &
J_{nm}^i + \partial_i \frac{1}{\partial_{-}}* J_{nm}^{+}.
\label{fvLorHameqCi} 
\end{eqnarray} 
Next from the Hamiltonian density
\begin{eqnarray}
{\cal H}^{nm}_{cov} & = & \frac{1}{2} \left( \lambda_{nm} -
\partial_i C_i - \frac{1}{2\partial_{-}}* J_{nm}^{+}\right) ^2 
+ \frac 1 4 \left( \partial_i{C}_j - \partial_j {C}_i\right)^2 +
\nonumber \\ 
&+ & \left(\Delta_\perp \phi + \partial_i C_i\right)
\left(\lambda_{nm} + \frac{1}{2\partial_{-}}* J_{nm}^{+}\right) + \phi
\left( \partial_{-} J_{nm}^{-} + \partial_{i} J_{nm}^{i}
\right) - J_{nm}^i C_i
\end{eqnarray}
and the diagonal Dirac brackets (\ref{fvcovDBCiCj},
\ref{fvcovDBlambdaphi}) we can construct the effective 
Lagrangian density for the sector of normal modes
\begin{eqnarray}
{\cal L}^{nm, \ eff}_{cov} & = & \partial_{-} C_i \partial_{+} C_i
- \frac 1 2 \left( \partial_i{C}_j \right)^2 - 2\partial_{+}
\lambda_{nm} \partial_{-} \phi + \partial_i \lambda_{nm} \partial_i \phi
-\frac{1}{2} \left(\lambda_{nm} - \frac{1}{2\partial_{-}}*
J_{nm}^{+}\right)^2 \nonumber\\
& - & \left(2 \partial_i C_i + \Delta_\perp \phi
 \right) \frac{1}{2\partial_{-}}*
J_{nm}^{+}
- \phi \left( \partial_{-} J_{nm}^{-} + \partial_{i} J_{nm}^{i} \right)
+ J_{nm}^i C_i .\label{fvLoreffLagrnm}
\end{eqnarray}
	
\subsection{Proper zero mode sector}\label{fvcorpropzersec}

This sector is specially interesting because both its 
Lagrangian density
\begin{equation}
{\cal L}_{cov}^{pzm}  =  \partial_i a_{-} \left(\partial_i a_{+} - 
\partial_{+} a_i \right) + \frac 1 2 \left( \partial_{+} a_{-}
\right)^2 - \frac 1 4 \left( \partial_i a_j - \partial_j a_i
\right)^2  +  a_\mu J^\mu_{pzm} 
+\Lambda_{pzm}( \partial_{+} a_{-} -
\partial_i a_i )  
\end{equation}
and the Euler-Lagrange equations  
\begin{eqnarray}
\partial_{+}^2 a_{-} & = & \partial_i ( \partial_{+} a_i -
\partial_i a_{+} ) + J_{pzm}^{-} - \partial_{+} \Lambda_{pzm}
\ ,  \\
- \partial_{+} \partial_i a_{-} & = & \Delta_\perp a_i -
\partial_ i \partial_j a^j + J_{pzm}^{i} + \partial_{i}
\Lambda_{pzm} \ , \\
0 & = & - \Delta_\perp a_{-} + J_{pzm}^{+}\ , \\
\partial_{+} a_{-} & = & \partial_i a_i\ ,
\end{eqnarray}
apparently indicate dynamical modes. However, a
closer inspection shows that all fields are non-dynamical and
explicitly depend on arbitrary external currents $J^\mu_{pzm}$
\begin{eqnarray}
a_\mu & = & \frac{1}{\Delta_\perp}* \left(J^{pzm}_\mu +
\partial_\mu \Lambda_{pzm}\right) \ , \label{fvLorpzmamu}\\
\Lambda_{pzm} & = & - \frac{1}{\Delta_\perp}*
\left(\partial_{+} J_{pzm}^{+} + \partial_{i} J_{pzm}^{i}
\right). \label{fvLorpzmLambda}
\end{eqnarray}
Therefore there is no canonical structure here and this
resembles the massive QED in Section \ref{massbezgaugefix}. All
we can do now is to find the effective Lagrangian which properly
couples the dependent gauge potentials (\ref{fvLorpzmamu}) with
external currents. The best choice is 
\begin{equation}
{\cal L}^{pzm, eff}_{cov} = J_{pzm}^{-} \frac{1}{\Delta_\perp}
J_{pzm}^{+} + \frac 1 2 \Lambda^2_{pzm} - \partial_\mu \Lambda_{pzm}
\frac{1}{\Delta_\perp} J_{pzm}^\mu - \frac 1 2 J_{pzm}^i
\frac{1}{\Delta_\perp} J_{pzm}^{i}\label{fvLoreqefflagr2}, 
\end{equation}
where the non-dynamical field $\Lambda_{pzm}$ is left in order
to avoid the term $\partial_{+} J^{+}_{pzm}$.

\subsection{Fermion field sector}\label{fvcovfersect}

Gathering our effective Lagrangians 
(\ref{fvLoreffLagrnm}) and (\ref{fvLoreqefflagr2}) we have
the total effective Lagrangian density for QED in the Lorentz covariant
gauge
\begin{eqnarray}
{\cal L}^{eff}_{cov} & = & 
\partial_{-} C_i \partial_{+} C_i
- \frac 1 2 \left( \partial_i{C}_j \right)^2 - 2\partial_{+}
\lambda_{nm} \partial_{-} \phi + \partial_i \lambda_{nm} \partial_i \phi
-\frac{1}{2} \lambda_{nm}^2 - \frac 1 2 \left(\frac{1}{2\partial_{-}}*
J_{nm}^{+}\right)^2\nonumber\\
& + & \frac 1 2 \Lambda_{pzm}^2 + i \sqrt{2} {\psi_{+}}^{\dag}
\partial_{+} \psi_{+} + i 
\sqrt{2} {\psi_{-}}^{\dag} \partial_{-} \psi_{-} + i
{\psi_{-}}^{\dag} \alpha^i \partial_{i} \psi_{+} + i
{\psi_{+}}^{\dag} \alpha^i \partial_{i} \psi_{-}\nonumber\\
&- & M {\psi_{+}}^{\dag} \gamma^0 \psi_{-} - M
{\psi_{-}}^{\dag} \gamma^0 \psi_{+} -  e \sqrt{2}
{\psi_{-}}^{\dag} \psi_{-}{V}_{-} -  e \sqrt{2}
{\psi_{+}}^{\dag} \psi_{+}{V}_{+} \nonumber\\
&-& e{V}_i \left({\psi_{+}}^{\dag} \alpha^i \psi_{-}+
{\psi_{-}}^{\dag} \alpha^i \psi_{+}\right)- \frac 1 2 J_{pzm}^i
\frac{1}{\Delta_\perp} \ast J_{pzm}^i + J_{pzm}^{-}
\frac{1}{\Delta_\perp}  \ast J_{pzm}^{+} \;,\label{fvLorefflagrtot1}
\end{eqnarray}
where the gauge potential $V_\mu$ is given by
\begin{eqnarray}
V_i & = & {C}_i + \partial_i \phi + \partial_i 
\frac{1}{\Delta_\perp} \ast \Lambda_{pzm} +  q_{i} \ , \\
V_{-} & = & \partial_{-} \phi +  q_{-}\ , \\
V_{+} & = & \frac{1}{2\partial_{-}} \ast \left( 2\partial_i C_i
+ \Delta_\perp \phi - \lambda_{nm} \right) + 
\partial_{+} \frac{1}{\Delta_\perp} \ast \Lambda_{pzm}\ .
\end{eqnarray}
Though the equations of motion for fermion fields lead
effectively to the condition $\Lambda_{pzm} = 0$ we would like
to remove this field before analysing fermion
sector.\footnote{Its presence would generate the extra 
primary condition which contains fermion fields
$$
\Pi^{\Lambda_{pzm}} = \frac{\delta L^{tot}_{eff}}{\delta \partial_{+}
\Lambda_{pzm}} = - e \sqrt{2} \frac{1}{\Delta_{\perp}} \ast 
\int dx^{-}(\psi^{\dag}_{+} \psi_{+}).
$$}
This can be achieved by the redefinition of fermion fields
\begin{eqnarray}
\psi_{\pm}^\prime & = & \exp \left( - i e 
\frac{1}{\Delta_\perp} \ast \Lambda_{pzm}\right)
{\psi_{\pm}}  \ , \\
{\psi'}^{\dag}_{\pm} & = & {\psi}^{\dag}_{\pm} \exp \left(i e
\frac{1}{\Delta_\perp} \ast \Lambda_{pzm}\right) \ ,
\end{eqnarray}
which in (\ref{fvLorefflagrtot1}) changes gauge potentials 
into $V_\mu^{cov}$\footnote{Strictly speaking these gauge
potentials {\it do not} satisfy the Lorentz covariant gauge
condition but rather 
$$
\partial^\mu V^{cov}_\mu = \frac{1}{4 \partial_{-}} * J^{+}_{nm}.
$$
However the direct interactions of currents compensates the above
non-vanishing current term.}
\begin{eqnarray}
V^{cov}_i & = & {C}_i + \partial_i \phi +  q_{i} \ , \\
V^{cov}_{-} & = & \partial_{-} \phi +  q_{-}\ , \\
V^{cov}_{+} & = & \frac{1}{2\partial_{-}} \ast \left(
2\partial_i C_i + \Delta_\perp \phi - \lambda_{nm} \right)\ .
\end{eqnarray}
Thus the non-dynamical field $\Lambda_{pzm}$ appears only in the
term $\frac 1 2 \Lambda_{pzm}^2$ and its equation of motion is
$\Lambda_{pzm} = 0$. Therefore we can completely disregard this
field in the further analysis . \\
The fermion Lagrangian density 
\begin{eqnarray}
{\cal L}^{fer}_{cov} & = & 
i \sqrt{2} {\psi'_{+}}^{\dag} \partial_{+} \psi'_{+} + i
\sqrt{2} {\psi'_{-}}^{\dag} \partial_{-} \psi'_{-} + i
{\psi'_{-}}^{\dag} \alpha^i \partial_{i} \psi'_{+} + i
{\psi'_{+}}^{\dag} \alpha^i \partial_{i} \psi'_{-}
- \frac 1 2 \left(\frac{1}{2\partial_{-}}*
J_{nm}^{+}\right)^2\nonumber\\
&- & M {\psi'_{+}}^{\dag} \gamma^0 \psi'_{-} - M
{\psi'_{-}}^{\dag} \gamma^0 \psi'_{+} -  e \sqrt{2}
{\psi'_{-}}^{\dag} \psi'_{-}{V^{cov}}_{-} -  e \sqrt{2}
{\psi'_{+}}^{\dag} \psi'_{+}{V^{cov}}_{+} \nonumber\\
&-& e{V^{cov}}_i \left({\psi'_{+}}^{\dag} \alpha^i \psi'_{-}+
{\psi'_{-}}^{\dag} \alpha^i \psi'_{+}\right)- \frac 1 2 {J'}_{pzm}^i
\frac{1}{\Delta_\perp} \ast {J'}_{pzm}^i + {J'}_{pzm}^{-}
\frac{1}{\Delta_\perp}  \ast {J'}_{pzm}^{+} \;,\label{fvLorfermlagr1}
\end{eqnarray}
where all currents are built from the new primed fermion fields,
generates constraint equations for dependent fermion fields
$\psi_{-}'$ and global zero modes  $q_i$
\begin{eqnarray}
&&\left( i \sqrt{2} \partial_{-} - e \sqrt{2} V^{cov}_{-} -
\frac{e\sqrt{2}}{2L} \frac{1}{\Delta_\perp} * {J'}^{+}_{pzm} 
\right) \psi^\prime_{-}  =  - i \alpha^i \partial_i
\psi^\prime_{+} + m \gamma^0 \psi^\prime_{+} \nonumber \\
&&\hspace{100pt} +  \ e \alpha^i \left( V^{cov}_i - \frac{1}{\Delta}_{\perp}
\ast {J'}_{pzm}^{i} \right) \psi^\prime_{+}\ ,  \label{fvLoreqpsim}\\ 
&&0  =  \int \ d^3 {x}\ ({\psi^\prime}^{\dagger}_{+} \alpha^i
\psi^\prime_{-} + {\psi^\prime}^{\dagger}_{-} \alpha^i
\psi^\prime_{+})(x)\ , 
\end{eqnarray}
with a similar structure as their counterparts for the LF Weyl
gauge. Therefore we can incorporate those respective
results from Section  \ref{chapfvWeyl} and write
\begin{eqnarray}
&&\psi^\prime_{-}(\vec{x}) = \frac{1}{\sqrt{2}} \left(
\frac{1}{i \partial_{-} - e V_{-}'} \ast \xi^\prime \right)
(\vec{x}) + \frac{e^2}{2 \sqrt{2} L} \left( {\cal
G}_{(\perp)}[{\cal M^\prime}^2] \ast {\Gamma^\prime}^i\right)
(x_\perp) \alpha^i \left(\frac{1}{i \partial_{-} - e V_{-}'} \ast
\psi^\prime_{+}\right)(\vec{x})\ , \nonumber\\ 
\label{fvLorprsolpsi} \\
&&\frac{1}{\Delta_{\perp}} \ast {J^\prime}^i_{pzm} - q_{i} = 
- \frac{e}{2L} {\cal G}_{(\perp)}[{\cal M^\prime}^2] \ast
{\Gamma^\prime}^i\label{fvLorsolji}. 
\end{eqnarray}
Then we take the expression for currents
\begin{equation}
{J^\prime}^i_{pzm} 
= - \frac{e}{2 L} {\Gamma^\prime}^i - \frac{e^3}{(2L)^2}
{\cal M^\prime}^{2} {\cal G}_{(\perp)}[{\cal M^\prime}^2] \ast
{\Gamma^\prime}^i\;,
\end{equation}
where 
\begin{eqnarray}
{\Gamma^\prime}^i(x_\perp) & = & \frac{1}{\sqrt{2}} \int dx^{-}
{\psi^\prime}_{+}^{\dagger}(\vec{x}) \alpha^i
\frac{1}{i\partial_{-}- eV_{-}'} \ast \xi^\prime  (\vec{x})  \nonumber\\
& + &   \frac{1}{\sqrt{2}} \int dx^{-}
{\xi^\prime}^{\dagger}(\vec{x}) \ast \frac{1}{i\partial_{-}-
eV_{-}'} \alpha^i \psi_{+}^\prime (\vec{x}) \ , \\
{\cal M^\prime}^2(x_\perp) &= &  \sqrt{2} \int dx^{-}
{\psi^\prime}^{\dagger}_{+} (\vec{x}) \frac{1}{i\partial_{-}-
eV_{-}''} \ast  \psi^\prime_{+}  (\vec{x}) \ , \label{fvLorgamM}
\end{eqnarray}
and
\begin{eqnarray}
V'_{-} & = & \partial_{-} \phi + q_{-} - {e\sqrt{2}}
\frac{1}{\Delta_\perp} * {J'}^{+}_{pzm} \ , \\
\xi^\prime & = & \left[ M \gamma^0 - i\alpha^i \partial_i +
e \alpha^i \left(C_i + \partial_i \phi
\right)\right]\psi^\prime_{+} \label{fvLordefxi} \ . 
\end{eqnarray}
Next we find the effective Hamiltonian for fermion sector
\begin{eqnarray}
H^{fer}_{cov} & =&  \frac{1}{\sqrt{2}} \int d^3x
{\xi^\prime}^{\dagger}(\vec{x}) \frac{1}{i \partial_{-}-
eV_{-}'} \ast \xi^\prime (\vec{x}) + \frac 1 2 \int d^3x
\left(\frac{1}{\partial_{-}} \ast
{J'}_{nm}^{+} \right)^2\nonumber\\
& + & \frac 1 2 \frac{e^2}{2L} \int d^2x_\perp
{\Gamma^\prime}^i (x_\perp)  {\cal G}_{(\perp)} [{\cal M^\prime}^2] \ast
{\Gamma^\prime}^i  (x_\perp) + 
e \sqrt{2}
{\psi'_{+}}^{\dag} \psi'_{+}V^{cov}_{+} \ , 
\label{fvLorDhamglob}
\end{eqnarray}
and finally we write the expression for effective total Lagrangian density 
\begin{eqnarray}
{\cal L}^{eff, tot}_{cov} & = & 
\partial_{-} C_i \partial_{+} C_i
- \frac 1 2 \left( \partial_i{C}_j \right)^2 - 2\partial_{+}
\lambda_{nm} \partial_{-} \phi + \partial_i \lambda_{nm}
\partial_i \phi - \frac 1 2 \lambda_{nm}^2
-\frac{1}{2} \left( \frac{1}{2\partial_{-}}*
{J'}_{nm}^{+}\right)^2\nonumber\\
& + & \frac 1 2 (\partial_{+} q_{-})^2 + i \sqrt{2}
{\psi'_{+}}^{\dag} \partial_{+} \psi'_{+}  - e \sqrt{2}
{\psi'_{+}}^{\dag} \psi'_{+}V^{cov}_{+}
- \frac{1}{\sqrt{2}} {\xi^\prime}^{\dagger} \frac{1}{ i
\partial_{-} - eV_{-}'} \ast \xi^\prime \nonumber\\
& - & \frac 1 2 \frac{e^2}{4L^2} {\Gamma^\prime}^i (x_\perp)
{\cal G}_{(\perp)} [{\cal M^\prime}^2] \ast 
{\Gamma^\prime}^i  (x_\perp)\ .\label{fvLorlagrtot}
\end{eqnarray}

\subsection{Quantum theory and physical states}

When all constraints are solved at the classical level, then the
canonical quantization is straightforward -  one writes the
(anti)commutation relations
\begin{eqnarray}
2\partial_{-}^x\left [C_i(\vec{x}),
C_{j}(\vec{y}) \right ] & = & - i \delta_{ij} \delta_n^{(3)}
(\vec{x} - \vec{y})\ , \label{fvLorcomCiCj}\\ 
2\partial^x_{-} \left [{\phi}(\vec{x}), \lambda_{nm} (\vec{y})
\right ] & = & i \delta_n^{(3)} (\vec{x} -\vec{y}) \ ,
\label{fvLorcomphilambda}\\ 
\left \{ {\psi'}_{+}^{\dagger}(\vec{x}), {{\psi'}_{+}}(\vec{y})
\right \} & = & \frac{1}{\sqrt{2}} \ \Lambda_{+}
\delta^{(3)}_a(\vec{x}- \vec{y}) \ , \\ 
\left[ p^{-}, {q}_{-} \right ] & = & - i\ ,\label{fvLorcompmqm}
\end{eqnarray}
while other relations vanish,  and takes the quantum Hamiltonian in
its canonical form
\begin{eqnarray} 
H^{Lor}_{quan} & =& \int d^3{x}\left[ \frac 1 2 \left(
\partial_i{C}_j \right)^2 - \partial_i \lambda_{nm} \partial_i
\phi + + \frac 1 2 \lambda_{nm}^2 + \frac{1}{2} \left(
\frac{1}{2\partial_{-}}* {J'}_{nm}^{+}\right)^2 + e \sqrt{2}
{\psi'_{+}}^{\dag} \psi'_{+} V^{cov}_{+}\right] \nonumber\\ 
& + &\frac{1}{16 L L^2_\perp} (p^{-})^2 + \frac{1}{\sqrt{2}}
\int d^3{x} {\xi'}^{\dagger}(\vec{x}) \left(\frac{1}{i
\partial_{-} - eV_{-}'} \ast \xi'\right) (\vec{x})
\nonumber \\
& + &\frac 1 2 \frac{e^2}{2L} \int d^2x_\perp {\Gamma'}^i (x_\perp)
\left( {\cal G}_{(\perp)} [{\cal M'}^2] \ast {\Gamma'}^i \right)
(x_\perp), \label{fvLorquanHam}
\end{eqnarray}
The presence of extra {\it nonphysical} modes $\phi$ and 
$\lambda_{nm}$ indicates that the Hilbert space contains also
non-physical states. In order to select the physical states we
will follow the analysis presented in Section \ref{modLorchap} and
introduce first the quantum field $\Lambda_{nm}$ 
\begin{equation}
\Lambda_{nm} = \lambda_{nm} + \frac{1}{2\partial_{-}}*
{J'}^{+}_{nm}, \label{fvLordefLambdanm}
\end{equation}
which is a free field even for the fully interacting system
\begin{equation}
\left(2 \partial_{+} \partial_{-} - \Delta_\perp \right)
\Lambda_{nm} = 0
\end{equation}
and has nonzero commutators 
\begin{eqnarray}
2\partial^x_{-} \left [{\phi}(\vec{x}), \Lambda_{nm} (\vec{y})
\right ] & = & i \delta_n^{(3)} (\vec{x} -\vec{y}) \ ,
\label{fvLorcomphiLambda}\\ 
2\partial^x_{-}\left [ \Lambda_{nm} (\vec{x}),
{\psi'}_{+}(\vec{y}) \right ] & = &  e \psi'_{+}(\vec{x})
\delta_n^{(3)} (\vec{x} - \vec{y}) \ , \label{fvLorcompsiLambda}\\ 
2\partial^x_{-} \left [ \Lambda_{nm}(\vec{x}),
{\psi'}_{+}^{\dag}(\vec{y}) \right ] & = & -  e
{\psi'}_{+}^{\dag}(\vec{x}) \delta_n^{(3)} 
(\vec{x} - \vec{y}) \ \label{fvLorcompsidagLambda}.
\end{eqnarray}
The q-number commutators (\ref{fvLorcompsiLambda}) and 
(\ref{fvLorcompsidagLambda}) would be very cumbersome in
further analysis, therefore we introduce the dressed {\it
physical} fermion fields
\begin{eqnarray}
\psi_{phys}(x) & = & \exp i e  \phi(x) \psi_{+}(x)\ , \\
\psi_{phys}^{\dag}(x) & = & \psi_{+}^{\dag}(x) \exp - i e
\phi(x) \ ,
\end{eqnarray}
which already commute with $\Lambda_{nm}$ at LF. Next the quantum
Hamiltonian can be expressed in terms of these new fields
\begin{eqnarray}
H^{Lor}_{quan} & =& \int d^3{x}\left[ \frac 1 2 \left(
\partial_i{C}_j \right)^2 - \partial_i \Lambda_{nm} \partial_i
\phi + \frac{1}{2} \left(\Lambda_{nm} - \frac{1}{\partial_{-}}*
J_{phys}^{+}\right)^2 + e \sqrt{2}
\psi_{phys}^{\dag} \psi_{phys} V^{phys}_{+}\right] \nonumber\\ 
& + &\frac{1}{16 L L^2_\perp} (p^{-})^2 + \frac{1}{\sqrt{2}}
\int d^3{x} \xi_{phys}^{\dagger}(\vec{x}) \left(\frac{1}{i
\partial_{-} - eV_{-}^{phys}} \ast \xi_{phys}\right) (\vec{x})
\nonumber \\
& + &\frac 1 2 \frac{e^2}{2L} \int d^2x_\perp \Gamma_{phys}^i
(x_\perp) \left( {\cal G}_{(\perp)} [{\cal M}_{phys}^2] \ast
\Gamma_{phys}^i \right) (x_\perp)\ , \label{fvLorquanHam2}
\end{eqnarray}
where
\begin{eqnarray}
\Gamma_{phys}^i(x_\perp) & = & \frac{1}{\sqrt{2}} \int dx^{-}
\psi_{phys}^{\dagger}(\vec{x}) \alpha^i \frac{1}{i
\partial_{-}- eV_{-}^{phys}} \ast \xi_{phys} (\vec{x}) 
\nonumber\\ 
& + & \frac{1}{\sqrt{2}} \int dx^{-} \xi_{phys}^{\dagger}
(\vec{x}) \ast \frac{1}{i \partial_{-}- eV_{-}^{phys}} \alpha^i
\psi_{phys} (\vec{x})\ , \\ 
{\cal M}_{phys}^2(x_\perp) &= &  \sqrt{2} \int dx^{-}
\psi_{phys}^{\dagger} (\vec{x}) \frac{1}{\partial_{-}-
eV_{-}^{phys}} \ast  \psi_{phys}  (\vec{x}) \ , 
\label{fvLorgamMphys}
\end{eqnarray}
and
\begin{eqnarray}
J^{+}_{phys} & = & \sqrt{2} \psi_{phys}^{\dag}\psi_{phys} \ ,\\
V^{phys}_{+} & = & \frac{1}{\partial_{-}} * \partial_i C_i \ , \\
V^{phys}_{-} & = &  q_{-} - {e\sqrt{2}}
\frac{1}{\Delta_\perp} * J_{phys}^{+} \ , \\
\xi_{phys} & = & \left[ m \gamma^0 - i\alpha^i \partial_i +
e \alpha^i C_i \right]\psi_{phys} \label{fvLordefxiphys} \; . 
\end{eqnarray}
Now we are in  the position to give the condition for 
physical states 
\begin{equation}
\left \langle phys' \left| \Lambda_{nm}(x) \right| phys \right
\rangle = 0.
\end{equation}
This means that the excitations of $\phi$ cannot appear in
physical states, while an arbitrary excitations of
$\Lambda_{nm}$ (all having zero norm) can accompany physical photons. 
The Hamiltonian which effectively acts on physical states has
the form 
\begin{eqnarray}
H^{Lor}_{phys} & =& \int d^3{x}\left[ \frac 1 4 \left(
\partial_i{C}_j - \partial_j{C}_i \right)^2 + \frac{1}{2}
\left( \partial_iC_i - \frac{1}{\partial_{-}}*
J_{phys}^{+}\right)^2 \right] \nonumber\\ 
& + &\frac{1}{16 L L^2_\perp} (p^{-})^2 + \frac{1}{\sqrt{2}}
\int d^3{x} \xi_{phys}^{\dagger}(\vec{x}) \left(\frac{1}{i
\partial_{-} - eV_{-}^{phys}} \ast \xi_{phys}\right) (\vec{x})
\nonumber \\
& + &\frac 1 2 \frac{e^2}{2L} \int d^2x_\perp \Gamma_{phys}^i
(x_\perp) \left( {\cal G}_{(\perp)} [{\cal M}_{phys}^2] \ast
\Gamma_{phys}^i \right) 
(x_\perp), \label{fvLorfizHam}
\end{eqnarray}
and is identical with the Hamiltonian (\ref{fvWeylfinalHamil})
obtained for the LF Weyl gauge after the implementation of
Gauss' Law. Thus we can take the above gauge-independent
expression  as  {\it the physical
Hamiltonian} for QED at the finite volume LF. 

\section{Transverse Coulomb gauge}

\setcounter{equation}{0}

When the Coulomb condition  is imposed for
transverse components of gauge potentials $ \partial_i V_i = 0$
it effectively introduces constraints
for two sectors
\begin{equation}
\partial_i V_i(\vec{x}) = 0 \Longrightarrow
\left\{ \begin{array}{l}
\partial_i A_i (\vec{x}) = 0 \\
\partial_i a_i(x_\perp) = 0.
\end{array}\right.
\end{equation}
Therefore we have to add some gauge condition for the global
zero modes  and we choose the LF Weyl gauge $q_{+} = 0$.
Because all these conditions are non-dynamical, we will
impose them explicitly via the suitable decomposition of vector
gauge fields into transverse and longitudinal parts.

\subsection{Proper zero mode sector}

As usually, we start with the Lagrangian density
\begin{equation}
{\cal L}^{pzm}_{Coul} = \partial_ia_{-}  \partial_ia_{+}
+ \frac 1 2 \left( \partial_{+}a_{-}\right)^2 - \frac 1 2 
\left( \partial_ia^{Tr}_j\right)^2 + J_{pzm}^{+}a_{+} + J_{pzm}^i a^{Tr}_i
+ J_{pzm}^{-}a_{-}  \ , \label{fvCoulCoulpzmLagr}
\end{equation}
where the transverse components of fields are defined as
\begin{equation}
a^{Tr}_i \stackrel{df}{=} \left( \delta_{ik} - \partial_i
\partial_k \frac{1}{\Delta_\perp} *\right) a_k \ .
\end{equation}
The Euler-Lagrange equations
\begin{eqnarray}
\partial^2_{+} a_{-} & = & - \Delta  a_{+} + J_{pzm}^{-} \ ,\\
- \partial_i \partial_{+} a_{-} & = & \Delta_\perp a_i^{Tr} +
J_{pzm}^i \ , \\
0& =& - \Delta_\perp a_{-} + J_{pzm}^{+}\ ,
\end{eqnarray}
can be explicitly solved in terms of external currents
\begin{eqnarray}
a_{-} &=& \frac{1}{\Delta_\perp} * J_{pzm}^{+} \ , \\
a_{i} &=& - \frac{1}{\Delta_\perp} * J_{pzm}^{i} + \partial_i
\frac{1}{\Delta_\perp} * \left(\frac{1}{\Delta_\perp} *
\partial_j J_{pzm}^{j}\right) \ ,\\ 
a_{+} &=& \frac{1}{\Delta_\perp} * J_{pzm}^{-} - \partial_{+}
\frac{1}{\Delta_\perp} * \left(\frac{1}{\Delta_\perp} *
J_{pzm}^{+}\right)\ . 
\end{eqnarray}
So here again there are no independent proper zero modes and
there is no canonical structure in this sector. Thus we only 
write the effective Lagrangian
\begin{eqnarray}
{\cal L}^{pzm \ eff}_{Coul} & = & J_{pzm}^{+}
\frac{1}{\Delta_\perp} *J_{pzm}^{-} 
 - \frac 1 2 J_{pzm}^{i} \frac{1}{\Delta_\perp} *
J_{pzm}^{i} - \frac 1 2 \left(\frac{1}{\Delta_\perp} *
\partial_i J_{pzm}^{i}\right)^2 \nonumber\\
& - & \frac 1 2 \left( \Lambda_{pzm}\right)^2 - \partial_{+}
\Lambda_{pzm} \frac{1}{\Delta_\perp} \ast J_{pzm}^{+} \ ,
\label{fvCouleffpzmlagr2} 
\end{eqnarray}
where the subsidiary non-dynamical field $\Lambda_{pzm}$ has been
introduced to  avoid the explicit presence of $\partial_{+}
J_{pzm}^{+}$.

\subsection{Normal mode sector}

Normal modes are described by the  Lagrangian density 
\begin{eqnarray}
{\cal L}^{nm}_{Coul} & = & \partial_{i} {A}_{+} \partial_i
{A}_{-} + \partial_{-} {A}_i^{Tr} \partial_{+}{A}_{i}^{Tr} 
 +  \frac 1 2 \left(\partial_{+} {A}_{-} -
\partial_{-}{A}_{+}\right)^2 \nonumber\\
&&- \frac 1 2 \left( \partial_i {A}_j^{Tr}\right)^2 +
{A}_{-} {j}^{-} + {A}_{+} {j}^{+} + {A}_i^{Tr} {J_{nm}^{Tr}}^i
\ ,
\end{eqnarray}
which generates the Euler-Lagrange equations
\begin{eqnarray}
\left( 2\partial_{+} \partial_{-} - \Delta_\perp \right){A}_{-}
& = & \partial_{-}\left( \partial_{+}{A}_{-} + \partial_{-}
{A}_{+} \right) - J_{nm}^{+} \ , \\
\left( 2\partial_{+} \partial_{-} - \Delta_\perp \right){A}_{+}
& = & \partial_{+}\left( \partial_{+}{A}_{-} + \partial_{-}
{A}_{+} \right) - J_{nm}^{-}\ , \\
\left( 2\partial_{+} \partial_{-} - \Delta_\perp
\right){A}_{i}^{Tr} & = &  {J_{nm}^{Tr}}^{i}\ .
\end{eqnarray}
These equations describe dynamical evolution but contain also
the constraints. To see this most easily we parameterize two gauge
field potentials $A_{\pm}$ by the single field $\Phi$
\begin{eqnarray}
A_{-} & = & \partial_{-} \Phi \ , \\
A_{+} & = & - \frac{1}{\partial_{-}} * \left( \Delta_\perp \Phi
- \frac{1}{\partial_{-}} * J_{nm}^{+}\right)+ \partial_{+} \Phi
\ ,
\end{eqnarray}
which satisfies the dynamical equation of motion 
\begin{eqnarray}
\left( 2\partial_{+} \partial_{-} - \Delta_\perp
\right){\Phi} & = & - \frac{1}{\partial_{-}} * {J}^{+}_{nm}
+  \frac{1}{\Delta_\perp} *
(\partial_{-} J_{nm}^{-} + \partial_{+} J_{nm}^{+}) \ . 
\end{eqnarray}
In the canonical analysis the last equations would need a
modification for removing the term $\partial_{+} J_{nm}^{+}$.
However this would introduce such a term in the definition of
$A_{+}$, therefore, just like in the case of massive QED in Section
\ref{massbezgaugefix}, there is no unconstrained canonical
structure consistent with the primary constrained system.
Thus all  we can find is the effective 
Lagrangian density which contains dynamical fields
$\Phi$ and $A_{i}^{Tr}$
\begin{eqnarray}
{\cal L}^{nm \ eff}_{Coul} & = & \partial_{+} \partial_i{\Phi}
\partial_{-} \partial_i \Phi - \frac 1 2 \left( \Delta_{\perp} \Phi -
\frac{1}{\partial_{-}} *  J_{nm}^{+} \right)^2 +
\partial_{-} \Phi  J_{nm}^{-} + \partial_{+} \Phi J_{nm}^{+} \nonumber\\
& + & \partial_{-} {A}_i^{Tr} \partial_{+}{A}_{i}^{Tr} 
-  \frac 1 2 \left( \partial_i A_j^{Tr}\right)^2 +
{A}_i^{Tr} {J_{nm}^{Tr}}^i \label{fvCoulnmefflagr}.
\end{eqnarray}
Combining two  effective Lagrangians (\ref{fvCouleffpzmlagr2}) and
(\ref{fvCoulnmefflagr}) we obtain the total effective
Lagrangian for the complete system with fermion
fields 
\begin{eqnarray}
{\cal L}^{eff}_{Coul} & = & 
\partial_{+} {A}^{Tr}_{i}\partial_{-} {A}^{Tr}_{i} -
\frac 1 2 \left( \partial_i {A}_j^{Tr}\right)^2 + \partial_{+}
\partial_i{\Phi} 
\partial_{-} \partial_i \Phi - \frac 1 2 \left( \Delta_{\perp} \Phi -
\frac{1}{\partial_{-}} \ast J_{nm}^{+} \right)^2 \nonumber\\
& + & \frac 1 2 \left( \partial_{+} q_{-}\right)^2  
 +  i \sqrt{2} {\psi_{+}}^{\dagger} \partial_{+} \psi_{+} + i
\sqrt{2} {\psi_{-}}^{\dagger} \partial_{-} \psi_{-} + i
{\psi_{-}}^{\dagger} \alpha^i \partial_{i} \psi_{+} + i
{\psi_{+}}^{\dagger} \alpha^i \partial_{i} \psi_{-}\nonumber\\
&- & M {\psi_{+}}^{\dagger} \gamma^0 \psi_{-} - M
{\psi_{-}}^{\dagger} \gamma^0 \psi_{+} -  e \sqrt{2}
{\psi_{+}}^{\dagger} \psi_{+}{V}^{Coul}_{+}
-  e \sqrt{2}{\psi_{-}}^{\dagger} \psi_{-}{V}_{-}^{Coul}
\label{fvCoulefflagr1} \\
&- &eV^{Coul}_i\left({\psi_{+}}^{\dagger} \alpha^i \psi_{-}+
{\psi_{-}}^{\dagger} \alpha^i \psi_{+}\right) - \frac 1 2
J_{pzm}^i\left( \frac{1}{\Delta_\perp} \ast J_{pzm}^i\right) - \frac 1 2
\left(\frac{1}{\Delta_{\perp}} \ast \partial_i
J_{pzm}^{i}\right)^2 \nonumber\\ 
& - & J^{-}_{pzm} \frac{1}{\Delta_\perp}\ast J_{pzm}^{+} -  \frac 1 2 \left( \Lambda_{pzm} \right)^2 \; , \nonumber 
\end{eqnarray}
where 
\begin{eqnarray}
V^{Coul}_i & = & {A}^{Tr}_i + q_{i}\ ,\\
V^{Coul}_{-} & = & \partial_{-} \Phi + q_{-} -
\frac{1}{\Delta_\perp}\ast J_{nm}^{+} \ ,\\
V^{Coul}_{+} & = & \partial_{+} \Phi -
\partial_{+}\frac{1}{\Delta_\perp} \ast  \Lambda_{pzm}\ .
\end{eqnarray}

\subsection{Fermion field sector}

Before writing down the Lagrangian for the fermion field we have decided to
introduce the {\it dressed} fermion field
\begin{equation}
\widetilde{\psi}_{\pm} (\vec{x}) = {\psi}_\pm(\vec{x})
\exp\left\{- i  e \frac{1}{\Delta_\perp} \ast \Lambda_{pzm}(x_\perp)\right\}
\end{equation}
and this leads to 
\begin{eqnarray}
{\cal L}^{fer}_{Coul} & = & i \sqrt{2} \widetilde{\psi}_{+}^{\dagger}
\partial_{+} \widetilde\psi_{+} + i \sqrt{2} \widetilde{\psi}_{-}^{\dagger}
\partial_{-} \widetilde\psi_{-} + i \widetilde{\psi}_{-}^{\dagger} \alpha^i
\partial_{i} \widetilde\psi_{+} + i \widetilde{\psi}_{+}^{\dagger} \alpha^i
\partial_{i} \widetilde\psi_{-}\nonumber\\ 
&- & m \widetilde{\psi}_{+}^{\dagger} \gamma^0 \widetilde\psi_{-} - m
\widetilde{\psi}_{-}^{\dagger} \gamma^0 \widetilde\psi_{+}-  e \sqrt{2}
\widetilde{\psi}_{-}^{\dagger} \widetilde\psi_{-}{V}_{-}^{Coul}-  e \sqrt{2}
\widetilde{\psi}_{+}^{\dagger}
\widetilde{\psi}_{+}\widetilde{V}_{+}^\prime \nonumber\\
& - &e{V}_i^{Coul}\left(\widetilde{\psi}_{+}^{\dagger} \alpha^i 
\widetilde\psi_{-}+ \widetilde{\psi}_{-}^{\dagger} \alpha^i
\widetilde\psi_{+}\right)- \frac 1 2 \widetilde{J}^i_{pzm} 
\left(\frac{1}{\partial_{\perp}} \ast \widetilde{J}^i_{pzm}
\right) \;.\nonumber \\ 
& - & \frac 1 2 \left(\Lambda_{pzm} -
\frac{1}{\Delta_\perp} \ast \partial_i 
\widetilde{J}_{pzm}^{i}\right)^2 - \frac 1 2 \left( \Delta_{\perp} \Phi -
\frac{1}{\partial_-} \ast \widetilde{J}^{+} \right)^2\ ,
    \label{fvCoulfermionlagr1}
\end{eqnarray}
where
\begin{equation}
\widetilde{V}_{+}^\prime = \partial_{+} \Phi.
\end{equation}
Now we have the same form of constraints for dependent
fermions as in the case of LF Weyl gauge in Section
\ref{chapfvWeyl} 
\begin{eqnarray}
\hspace{-20pt}\left( i \sqrt{2} \partial_{-} - e \sqrt{2}
V_{-}^{Coul} \right) \widetilde\psi_{-} & = & - i \alpha^i
\partial_i \widetilde\psi_{+} + M \gamma^0 \widetilde\psi_{+} +
\ e \alpha^i \left( V_i^{Coul} - 
\frac{1}{\Delta_\perp} \ast \widetilde{J}^{i}_{pzm} \right)
\widetilde\psi_{+} \ , \label{fvCouleqpsim}\\ 
0 & = & \int \ d^3 {x}\ (\widetilde\psi^{\dagger}_{+} \alpha^i
\widetilde{\psi}_{-} + \widetilde\psi^{\dagger}_{-} \alpha^i
\widetilde\psi_{+})(x) \ , 
\end{eqnarray}
thus we can adopt those results directly and only the equation
for $\Lambda_{pzm}$ is new  
\begin{equation}
\Delta_\perp \Lambda_{pzm} = - \partial_i \widetilde{J}^i_{pzm}
\ .
\end{equation}
We easily obtain 
\begin{equation}
\frac{1}{\Delta_\perp} \widetilde{J}^i_{pzm} - q_{i} = 
 - \frac{e}{2L} {\cal G}_{\perp}[\widetilde{\cal M}^2] \ast
\widetilde\Gamma^i\label{fvCoulsolji}
\end{equation}
and further the effective Lagrangian density for fermions
takes the following form:
\begin{equation}
{\cal L}_{eff}^{fer} = i \sqrt{2}\widetilde\psi^{\dagger}_{+}
\partial_{+} \widetilde\psi_{+} - e \sqrt{2}
\widetilde{\psi_{+}}^{\dagger} \widetilde{\psi}_{+}
\widetilde{V}_{+}^\prime - \frac{1}{\sqrt{2}}
\widetilde\xi^{\dagger} \left(\frac{1}{i \partial_{-} - e
V_{-}^{Coul}} \ast \widetilde\xi \right) - \frac 1 2 
\frac{e^2}{4L^2} \widetilde\Gamma^i \left({\cal G}_{(\perp)}
[\widetilde{\cal M}^2_{Coul}] \ast \widetilde\Gamma^i \right)
\label{fvCoulefflagrfer1}\ , 
\end{equation}
where
\begin{eqnarray}
\widetilde\xi(x) & = & \left[ M \gamma^0 - i\alpha^i \partial_i +
e \alpha^i {V}_i^{Coul}(x) \right] \widetilde\psi_{+}(x) \ , 
\label{fvCouldefxi} \\ 
\widetilde\Gamma^i(x_\perp) & = & \frac{1}{\sqrt{2}} \int dx^{-}
\widetilde \psi_{+}^{\dagger}(x) \alpha^i \left(
\frac{1}{i \partial_{-} - e V_{-}^{Coul}} \ast
\widetilde \xi \right) (\vec{x}) \nonumber\\ 
& + &  \frac{1}{\sqrt{2}} \int dx^{-}
\widetilde \xi^{\dagger}(x) \alpha^i \left(\frac{1}{i \partial_{-}
- e V_{-}^{Coul}} \ast \widetilde \psi_{+} \right) (\vec{x}) \
, \\
\widetilde{\cal M}^2_{Coul}(x_\perp) & = & {\sqrt{2}} \int dx^{-}
\widetilde \psi_{+}^{\dagger}(x) \left(\frac{1}{i \partial_{-} - e
V_{-}^{Coul}} \ast \widetilde \psi_{+} \right) (\vec{x})\ .
\label{fvCoulgamM} 
\end{eqnarray}
The equation for the subsidiary dependent field $\Lambda_{pzm}$
\begin{equation}
\Delta_\perp \Lambda_{pzm} = - \frac{e}{2L} \partial_i {\cal
G}_{(\perp)}[\widetilde{\cal M}^2_{Coul}]\ast \widetilde\Gamma^i
\end{equation}
can be uniquely solved but we really do not need this solution.
Since $\Lambda_{pzm}$ has disappeared from the effective
Lagrangian (\ref{fvCoulefflagrfer1}), we can omit it in
further steps. \\ 
Our analysis of constraints for the complete QED in the
Coulomb gauge leads to the final effective Lagrangian
density 
\begin{eqnarray}
{\cal L}_{eff}^{Coul} & = & \partial_{+} {A}^{Tr}_{i}\partial_{-}
{A}^{Tr}_{i} - \frac 1 2 \left( \partial_i {A}_j^{Tr}\right)^2
+ \partial_{+} \partial_i{\Phi} \partial_{-} \partial_i \Phi -
\frac 1 2 \left( \Delta_{\perp} \Phi - e \sqrt{2} 
\frac{1}{\partial_{-}} \ast  (\widetilde{\psi}^{\dagger}_{+}
\widetilde{\psi}_{+}) \right)^2 \nonumber\\ 
& + & \frac 1 2 \left( \partial_{+} q_{-}\right)^2 
+ i \sqrt{2} \psi^{\dagger}_{+} \partial_{+}
\psi_{+} - e \sqrt{2} \psi^{\dagger}_{+}
\psi_{+} \partial_{+}\Phi  \nonumber \\
& + & \frac{1}{\sqrt{2}} \left(\xi^{\dagger} + e
\psi^{\dagger}_{+} \alpha^i \partial_i\phi\right) \frac{1}{i
\partial_{-} - e V_{-}^{Coul}} \ast \left(\xi + e \psi_{+} \partial_i
\phi\right)   \nonumber\\
& - & \frac 1 2 \frac{e^2}{4L^2} \left(\Gamma^i + e {\cal
M}^2_{Coul} \partial_i \phi\right) {\cal G}_{(\perp)} [{\cal
M}^2] \ast \left(\Gamma^i + e {\cal M}^2_{Coul} \partial_i
\phi\right)) \ . \label{fvCoultotefflagr}
\end{eqnarray}

\subsection{Canonical quantization}

From the classical effective Lagrangian
(\ref{fvCoultotefflagr}) we can infer the structure of quantum
(anti)commutation relations for all independent fields. Besides
the canonical relations
\begin{eqnarray}
2\partial_{-}^x\left [A_i^{Tr}(\vec{x}), A_{j}A_i^{Tr}
(\vec{y}) \right ] & = & - i \delta_{ij}A_i^{Tr} \delta_n^{(3)}
(\vec{x} - \vec{y}) \ , \label{fvCoulcomAiAj}\\ 
2\partial^x_{-} \left [\Phi(\vec{x}), \Phi (\vec{y}) \right ] &
= & i \Delta_\perp \delta_n^{(3)} (\vec{x} -
\vec{y}) \ , \label{fvCoulcomPhiPhi}\\ 
\left \{ {\psi'}_{+}^{\dagger}(\vec{x}),{{\psi'}_{+}}(\vec{y})
\right \} & = & \frac{1}{\sqrt{2}} \ \Lambda^{(+)}
\delta^{(3)}_a(\vec{x}- \vec{y})\ , \\ 
\left[ p^{-}, {q}_{-} \right ] & = & - i \ ,
\label{fvCoulcompmqm}
\end{eqnarray}
we find also two noncanonical commutators between $\Phi$ and
the fermion fields 
\begin{eqnarray}
2 \partial_{-}^x\left [{\Phi}(x^{+}_0, \bar{x}),
\widetilde{\psi}_{+}(x^{+}_0, \bar{y}) \right ] & = &  e
\delta^3_p (\bar{x} - \bar{y}) \ , \label{fvCoulcomPhipsip}\\
2 \partial_{-}^x\left [{\Phi}(x^{+}_0, \bar{x}),
\widetilde{\psi}_{+}^{\dag} (x^{+}_0, \bar{y}) \right ] & = & -
e\delta^3_p (\bar{x} - \bar{y}) \ , \label{fvCoulcomPhipsidagp}
\end{eqnarray}
which come from the term $\psi^{\dag}_{+} \psi_{+}
\partial_{+} \Phi$. The quantum Hamiltonian is assumed in the
form: 
\begin{eqnarray} 
H^{Coul}_{quan} & =& \int d^3{x}\left[ \frac 1 2 \left(
\partial_i{A}_j^{Tr} \right)^2 + \frac 1 2 \left(
\Delta_{\perp} \Phi - e \sqrt{2} \frac{1}{\partial_{-}} 
\ast ( \widetilde{\psi}^{\dagger}_{+}
\widetilde{\psi}_{+}) \right)^2 \right] \nonumber\\ 
& + &\frac{1}{16 L L^2_\perp} (p^{-})^2 + \frac{1}{\sqrt{2}}
\int d^3{x} {\xi}_{Coul}^{\dagger}(\vec{x}) \left(\frac{1}{i
\partial_{-} - eV_{-}^{Coul}} \ast \xi_{Coul}\right) (\vec{x})
\nonumber \\
& + &\frac 1 2 \frac{e^2}{2L} \int d^2x_\perp {\Gamma}^i_{Coul} (x_\perp)
\left( {\cal G}_{(\perp)} [{\cal M}^2_{Coul}] \ast {\Gamma}^i
_{Coul} \right)(x_\perp) \label{fvCoulquanHam}
\end{eqnarray}
tacitly defining proper ordering of non-commuting propagators.
Thus we see that the transverse Coulomb gauge condition has the
same number of independent excitations as the physical
subsystems defined either in the LF Weyl or in the Lorentz
covariant gauges. Here one encounters a strange phenomenon
that one boson field does not commute with fermion fields and
this is connected with the rather extravagant choice of gauge
which imposes the constraint on transverse gauge fields which
are known to describe two physical photon excitations at LF.
This non-commutativity is the lowest price we can pay for such
careless choice of gauge fixing condition and it can be solved
if we introduce another dressed {\it physical } fermion fields
\begin{eqnarray}
\psi_{phys} (\vec{x}) & = &\exp{i e
\Phi(\vec{x})} \widetilde{\psi}_{+}(\vec{x}) \,  \\
\psi_{phys}^{\dag} (\vec{x}) & = &\widetilde{\psi}_{+}^{\dag}
(\vec{x}) \exp{-i e \Phi(\vec{x})} \ , 
\end{eqnarray}
which already commute with $\Phi$ fields. Now the quantum
Hamiltonian (\ref{fvCoulquanHam}) takes the form 
\begin{eqnarray} 
H^{phys}_{quan} & =& \int d^3{x}\left[ \frac 1 2 \left(
\partial_i{A}_j^{Tr} \right)^2 + \frac 1 2 \left(
\Delta_{\perp} \Phi - e \frac{1}{\partial_{-}} 
\ast \sqrt{2} {\psi}^{\dagger}_{phys}
{\psi}_{phys} \right)^2 \right] \nonumber\\ 
& + &\frac{1}{16 L L^2_\perp} (p^{-})^2 + \frac{1}{\sqrt{2}}
\int d^3{x} {\xi}_{phys}^{\dagger}(\vec{x}) \left(\frac{1}{i
\partial_{-} - eV_{-}^{phys}} \ast \xi_{phys}\right) (\vec{x})
\nonumber \\
& + &\frac 1 2 \frac{e^2}{2L} \int d^2x_\perp {\Gamma}^i_{phys}
(x_\perp) \left( {\cal G}_{(\perp)} [{\cal M}^2_{phys}] \ast
{\Gamma}^i_{phys} \right)(x_\perp) \ , \label{fvCoulphysquanHam} 
\end{eqnarray}
where
\begin{eqnarray}
V^{phys}_i(x) & = & {A}^{Tr}_i(x) - \partial_i \Phi(x) +
q_{i}(x^{+}) \ , \\
V^{phys}_{-}(x) & = & q_{-}(x^{+}) - \frac{e\sqrt{2}}{2L}
\frac{1}{\Delta_\perp} \ast \int dx^{-}
\psi_{phys}^{\dagger}(x) \psi_{phys} ({x})\ , \\
\xi_{phys}(x) & = & \left[ M \gamma^0 - i\alpha^i \partial_i
+ e \alpha^i {V}_i^{phys}(x) \right] \psi_{phys}(x)\ , 
\label{fvphysdefxi} \\ 
\Gamma^i_{phys}(x_\perp) & = & \frac{1}{\sqrt{2}} \int dx^{-}
\psi_{phys}^{\dagger}(x) \alpha^i \left(
\frac{1}{i \partial_{-} - e V_{-}^{phys}} \ast
\xi_{phys} \right) ({x}) \nonumber\\ 
& + &  \frac{1}{\sqrt{2}} \int dx^{-}
\xi_{phys}^{\dagger}(x) \alpha^i \left(\frac{1}{i \partial_{-}
- e V_{-}^{phys}} \ast \psi_{phys} \right) ({x})\ , \\
{\cal M}^2_{phys} (x_\perp) & = & {\sqrt{2}} \int dx^{-}
\psi_{phys}^{\dagger}(x) \left(\frac{1}{i \partial_{-} - e
V_{-}^{phys}}  \ast  \psi_{phys} \right) ({x}) \ ,\label{fvphysgamM}  
\end{eqnarray}
and agrees with the former expressions for the physical
Hamiltonians in the LF Weyl gauge (\ref{fvWeylfinalHamil}) and
the Lorentz covariant gauge (\ref{fvLorfizHam}).

\subsection{Physical gauge conditions}

The physical field variables that we have just found after
redefinitions of the  fields can be obtained directly from the
following gauge conditions: 
\begin{eqnarray}
A_{-} & = & 0 \ , \\
\partial_{+} a_{-} & = & \partial_\perp a_\perp\ , \\
q_{+} & = & 0\ .
\end{eqnarray}
Only the sector of normal modes has not been analysed by us so
far \cite{LCgaugeQED}, therefore below we sketch some crucial
points. From the Lagrangian density 
\begin{eqnarray}
{\cal L}^{nm} & = & \left(\partial_{+} {A}_i - \partial_i
{A}_{+}\right)\partial_{-} {A}_i  +  \frac 1 2 \left(
\partial_{-}{A}_{+}\right)^2 - \frac 1 4 \left( \partial_i
{A}_j - \partial_j {A}_i\right)^2 + {A}_{+} J_{nm}^{+} + {A}_i
J_{nm}^i \ , 
\end{eqnarray}
where the gauge fixing condition $A_{-} = 0$ has been
explicitly implemented, we derive the Euler-Lagrange equations
\begin{eqnarray}
\partial_{-}\left(\partial_{-} A_{+} - \partial_i A_i
\right) & = &  J_{nm}^{+} \ , \label{fvphyseqAplus}\\ 
 \left( 2 \partial_{+} \partial_{-} - \Delta_\perp \right)
{A}_i & = & \partial_i\left(\partial_{-} A_{+} - \partial_j A_j
\right) - J^i_{nm}\ . \label{fvphyseqAi} 
\end{eqnarray}
Evidently, $A_{+}$ is not a dynamical variable and its equation
of motion (\ref{fvphyseqAplus}) can be solved as
\begin{eqnarray}
A_{+} &=& \frac{1}{2\partial_{-}} \ast \left( \partial_i
A_i  + \frac{1}{2\partial_{-}} \ast J^{+}_{nm} \right).
\end{eqnarray}
Only one Dirac bracket is nonzero
\begin{eqnarray}
2\partial_{-}^x \left \{{A}_{i}(\vec{x}), {A}_j (\vec{y})
\right \}_{DB} & = & - \delta_{ij}
\delta^3(\vec{x}-\vec{y})\label{fvphysDbraAi} 
\end{eqnarray}
and the Dirac Hamiltonian is
\begin{equation}
{\cal H}_D = \frac 1 2 (\partial_i A_i +
\frac{1}{\partial_{-}} \ast J^{+}_{nm})^2 + \frac 1 4 \left( 
\partial_i {A}_j - \partial_j {A}_i\right)^2 - A_i J^i_{nm} \ .
\label{fvphysDHamnm} 
\end{equation}
These results can be incorporated into the whole procedure of
equivalent Lagrangians if the above results are transformed into 
the effective Lagrangian density for the normal mode part 
\begin{equation}
{\cal L}_{eff}^{nm} = \partial_{+} A_i \partial_{-}A_i - \frac 1
2 (\partial_i A_i +  \frac{1}{\partial_{-}} \ast J_{nm}^{+})^2 - \frac 1 4
\left( \partial_i {A}_j - \partial_j {A}_i\right)^2 + A_i J^i_{nm}\ .
\end{equation}
Then following the analysis from Subsection
\ref{fvcorpropzersec} and (with trivial modifications) from
Subsection \ref{fvcovfersect},  one finds the commutator
relations for physical fields (\ref{fvCoulcomAiAj} -
\ref{fvCoulcompmqm}) and the physical Hamiltonian
(\ref{fvCoulphysquanHam}). 

\newpage

\part{Conclusions and perspectives}

In this paper we have presented a novel method of canonical
quantisation for  constrained systems when applying it to the analysis 
of QED at LF. Separating different sectors of fields, we have
been able to find effective description of them avoiding an
explicit implementation of Dirac's procedure for the whole
system of entangled constraints. Different gauge conditions
proved to be legitimate choices for the quantum gauge field
system interacting with  fermions.\\
Feynman rules for perturbative calculations have been found and
they have common properties. Generally, the canonical
propagators for gauge fields have additional noncovariant terms
which behave ~non-causally i.e. in the Fourier representation
they are given solely by the CPV poles. These terms are
cancelled by the 
direct interaction of currents which canonically appear in the
interaction Hamiltonians. The fermion parts of interaction
Hamiltonians are factorized and the non-covariant factors cancel
with the additional non-covariant term in the fermion field
propagator at LF. In this manner there are two equivalent
settings of Feynman perturbative rules, the first (canonical)
rules can contain non-causal and ~non-covariant terms, the second
(effective) rules are causal and covariant.\\
The LC-gauge has been investigated  as the limit of the general axial
gauge and the flow covariant gauge. The ML-prescription for the
LC-gauge propagator has been derived only in the second choice
and these results hold for the interactions with fermions.\\
In the DLCQ method three choices of gauge ~conditions have been
canonically analysed. In the cases of the LF Weyl gauge, the
Lorentz covariant gauge and the transverse Coulomb gauge, the
same physical Hamiltonian has been found though via different
methods of Gauss' law implementation, the Lautrup-Nakanishi
condition and the redefinition of fermion fields, respectively.
Finally, the set of {\it physical gauge conditions}, which
~straightforwardly leads to the above physical Hamiltonian, has
been proposed.\\
All the above results have been derived for fermion currents. The
problems which can appear for other charged matter fields have
been discussed in the simplest model of 1+1 dimensional LF-Weyl
QED with scalar fields. A reasonable analysis of
this model has been formulated starting with the nonlocal
redefinition of one scalar field at LF. For these new fields,
the canonical analysis and structure of perturbative
calculations have been studied. Both methods and results were
similar to those for fermion fields.\\
We have found that the presence of field derivatives in the
matter currents (as in the scalar case) can be neutralized at
LF by a suitable ~nonlocal redefinition of these fields. We
expect that this is a general property of the LF formulation and
can also be applied for more physically relevant models.
Therefore we plan to analyse, first, the QED for charged vector
fields and then the non-Abelian interactions. Especially the latter
case will ultimately settle the status of the LC-gauge at LF,
first of all the question of proper prescription for ~spurious
poles within the canonical LF formulation.\\

\vspace*{.5cm}

\noindent{\large \bf Acknowledgements}

\vspace{.5cm}

\noindent I thank Professor Dominik Rogula for his encouragement and
many invaluable remarks during my research on this project and
Dr Stanis\l aw D. G\l azek for many discussions on different
aspects of light-front physics. I am also indebted to so many
people who have helped me to solve all problems that I have
encountered while preparing this thesis. \\
I express my gratitude towards Dr Alex Kalloniatis and Dr Rik
Naus for their collaboration in our joint paper which was the
starting point of my present research.\\

\appendix

\noindent{\Large \bf Appendices}
	
\addcontentsline{toc}{part}{Appendices}

\renewcommand{\theequation}{\Alph{section}.\arabic{equation}}

\section{LF Notation}\label{Appnotation}\label{DMnotacja}

\setcounter{equation}{0}

\subsection{Coordinates}\label{LFcoorappend}

In 3+1 dimensions we define longitudinal coordinates $x^\pm
= \frac{x^{0} \pm x^1}{\sqrt{2}}$ and take $x^{+}$ as the
dynamical evolution parameter. We denote transverse components
$x_\perp = (x^2, x^3)$ by Latin indices (i, j, \ldots).
~Similarly we define components of any 4-vector and a scalar
product of two 4-vectors decomposes as $A \cdot B = A^{+} B^{-}
+ A^{-} B^{+} - A^i B^i$. The metric has ~non-vanishing components
$g_{+-} = 1 , g_{ij} = - \delta_{ij}$. Partial derivatives are
defined as $\partial_{\pm} = \partial / \partial x^{\pm}, \
\partial_i = \partial / \partial x^i$. Tensor components are
defined analogously e.g. $T^{\pm \mu} = \frac{1}{\sqrt{2}}
(T^{0\mu} \pm T^{1\mu})$ and summation over repeated indices is
understood. \\ 
Also we introduce a vector notation for components $\vec{x} =
(x^{-}, x_\perp)$ which lie on a light-front surface $x^{+} =
const.$ and for momenta associated with them $\vec{k} = (k_{-},
k_\perp)$. Scalar product of such 3-vectors decomposes as
$\vec{k} \cdot \vec{x} = k_{-} x^{-} - k_i x_i$. \\
In the D+1 dimensions the notation generalizes trivially with
the number of transverse directions changing from 2 to d.\\

\subsection{Dirac matrices}\label{Diracmatappend}

The Dirac matrices $\gamma^\mu$ satisfy anticommutation
relation 
\begin{equation}
\gamma^\mu \gamma^\nu + \gamma^\nu \gamma^\mu = 2 g^{\mu \nu}
\ ,
\end{equation}
where their components are defined analogously to coordinates
e.g. $\gamma^\pm = \frac{\gamma^{0} \pm \gamma^1}{\sqrt{2}}$.
Thus $\gamma^{\pm}$ are nilpotent matrices $\left(\gamma^{\pm}
\right)^2 = 0$.\\
For the projection operators 
\begin{equation}
\Lambda_{\pm} = \frac{1}{\sqrt{2}} \gamma^0 \gamma^{\pm} =
\frac{1}{2} \gamma^\mp \gamma^{\pm}\ ,  
\end{equation}
 we have useful relations
\begin{eqnarray}
\Lambda_{\pm} \Lambda_{\pm} & = & \Lambda_{\pm}\ , \\
\Lambda_{+} \Lambda_{-} & = & \Lambda_{-} \Lambda_{+} =  0\ , \\
\Lambda_{+} + \Lambda_{-} & = & 1\ , \\
\gamma^{\pm} \Lambda_{\mp} & = & \Lambda_{\pm} \gamma^{\mp} = 0\ ,\\
\gamma^{\pm} \Lambda_{\pm} & = & \Lambda_{\pm} \gamma^{\pm} \
,\\ 
\gamma^{0} \Lambda_{\pm} & = & \Lambda_{\mp} \gamma^{0}\ ,\\
\gamma^{i} \Lambda_{\pm} & = & \Lambda_{\pm} \gamma^{i}.
\end{eqnarray}
In many places we also use the standard notation
\cite{BjorkenDrell1964} 
\begin{eqnarray}
\gamma^0 & = & \beta\ ,\\
\gamma^0 \gamma^i& = & \alpha^i.
\end{eqnarray} 

\section{Green Functions}

\setcounter{equation}{0}

\subsection{Feynman propagator functions}\label{dodfenGrefun}

\noindent In 1+1 dimensions we define
noncovariant Feynman Green functions $E^1_F(x)$ and
$E^2_F(x)$ 
\begin{eqnarray}
 E^1_F(x) & \stackrel{df}{=} & i\int_0^{\infty} \frac{dk_{-}}{2 \pi}
 \ \left[\Theta(x^{+})e^{-i {k_{-}}x^{-}} -
\Theta(-x^{+})e^{i {k_{-}}{x}^{-}}\right] \nonumber\\
&  = & \frac{1}{2 \pi} \left( \frac{\Theta(x^{+})}{x^{-} -i
\epsilon} + \frac{\Theta(-x^{+})}{x^{-} +i\epsilon} \right)\ ,\\
 E^2_F(x) & \stackrel{df}{=} & -
\frac{x^+}{2 \pi} \left( \frac{\Theta(x^{+})}{x^{-} -i\epsilon}+
\frac{\Theta(-x^{+})}{x^{-} +i\epsilon} \right)\ ,
\end{eqnarray}
which can be also represented by the 2-dimensional Fourier integrals 
\begin{eqnarray}
E^1_F(x) 
& = & - \int_{-\infty}^{\infty} \frac{d^2{k}}{(2 \pi)^2} 
e^{-i {k}\cdot {x}} \frac{1}{k_{+} + i \epsilon \ {\rm
sgn}(k_{-}) }\label{kspace11E1}\ ,\\
E^2_F(x) & = & - i \int_{-\infty}^{\infty} \frac{d^2{k}}{(2 \pi)^2}
e^{-i {k}\cdot x} \frac{1}{[k_{+} + i \epsilon \ {\rm
sgn}(k_{-})]^2 } \label{kspace11E2}.
\end{eqnarray}
\noindent In D+1 dimensions we define covariant Feynman Green
function as:
\begin{eqnarray}
D_F^{d+2}(x) & \stackrel{df}{=} & \int_{-\infty}^{\infty}
\frac{d^dk_\perp}{(2 \pi)^d} \int_{0}^{\infty}
\frac{dk_{-}}{2\pi \ 2k_{-}} \ \left[\Theta(x^{+})e^{-i
{k}\cdot {x}} + \Theta(-x^{+})e^{+i {k}\cdot
{x}}\right]_{k_{+} = \frac{k_\perp^2}{2k_{-}}}\nonumber\\
& = & i \int \frac{d^{d+2}{k}}{(2 \pi)^{d+2}} \frac{e^{-i {k}\cdot
({x}-{y})}}{k^2 + i \epsilon}. \label{kspaceDF}
\end{eqnarray}
The Feynman propagator function for the massive fields in 3+1
dimensions are given by 
\begin{eqnarray}
\Delta_F(x, M^2) & = & i \int \frac{d^4k}{(2 \pi)^4} \frac{e^{-
ik\cdot x}}{2k_{+}k_{-} - k^2_\perp - M^2 + i \epsilon}\ ,
\end{eqnarray}
thus the functions for massless fields can be defined as the
limits 
\begin{eqnarray}
{D}_F(x) & = & \lim_{m^2 \rightarrow 0} \Delta_F(x,M^2) = i \int
\frac{d^4k}{(2 \pi)^4} \frac{e^{- ik\cdot x}}{2k_{+}k_{-} -
k^2_\perp + i \epsilon}\ ,\\ 
{E}_F(x) & = & - \lim_{m^2 \rightarrow 0} \frac{\partial}{\partial
m^2} \Delta_F(x, m^2) = i  \int \frac{d^4k}{(2 \pi)^4} \frac{e^{-
ik\cdot x}}{(2k_{+}k_{-} - k^2_\perp + i \epsilon)^2}.
\end{eqnarray}
In the main text we use the following property
\begin{eqnarray}
\left[ 2 \partial_{-}^x D_F^{d+2}(x) + E_F^{1}({x_L})\delta^d(x_\perp)
\right] & = & (- i) \int_{-\infty}^{\infty}
\frac{d^dk_\perp}{(2 \pi)^d} \int_{0}^{\infty}
\frac{dk_{-}}{2\pi } \ \left[\Theta(x^{+})\left( e^{-i
\frac{k_\perp^2}{2k_{-}}x^{+}} -1\right) e^{-i \vec{k}\cdot
\vec{x}} \right. \nonumber\\ 
&+ &\left. \Theta(-x^{+}) \left(e^{i
\frac{k_\perp^2}{2k_{-}}x^{+}} -1\right) e^{-i \vec{k}\cdot 
\vec{x}}\right] = - \Delta_\perp \int_0^{x^{+}} d\xi
D^{d+2}_F(\xi, \vec{x})\nonumber \\
& = & - \ \Delta_\perp \int \frac{d^{d+2}{k}}{(2 \pi)^{d+2}}
\frac{e^{-i {k}\cdot ({x}-{y})}}{k^2 + i \epsilon}
\frac{1}{k_{+} + i \epsilon \ {\rm
sgn}(k_{-}) }, \label{wlasDeltaFEF}
\end{eqnarray}
and a similar property holds also for the massive case 
\begin{eqnarray}
\left[ 2 \partial_{-}^x \Delta_F(x, m^2) + E_F^{1}({x_L})\delta^2(x_\perp)
\right] =  - \ (\Delta_\perp- m^2) \int \frac{d^{4}{k}}{(2 \pi)^{4}}
\frac{e^{-i {k}\cdot ({x}-{y})}}{k^2  - m^2 + i \epsilon}
\frac{1}{k_{+} + i \epsilon \ {\rm
sgn}(k_{-}) }.
\end{eqnarray}
The noncovariant Feynman propagator functions in 3+1 dimensions
are given by
\begin{eqnarray}
G^1_{\alpha F}(x) & = & i\int \frac{d^4k}{(2 \pi)^4} \frac{e^{-
ik\cdot x}}{(1 + \alpha)k_{+}k_{-} - \alpha k^2_\perp + i
\epsilon}\ ,\\ 
G^2_{\alpha F}(x) & = & i\int \frac{d^4k}{(2 \pi)^4} \frac{e^{-
ik\cdot x}}{[(1 + \alpha)k_{+}k_{-} - \alpha k^2_\perp + i
\epsilon]^2}.
\end{eqnarray}

\subsection{Integral operators in $x^{-}$ direction}
\label{dodscalfun} 

We use the basic inversion of $\partial_{-}$ defined by the
Fourier integral 
\begin{equation}
(\partial_{-} )^{-1}(x^{-}- y^{-}) = i \int_{-\infty}^{\infty}
dk \exp{-i k_{-} (x^{-}- y^{-})} \ {\rm \large CPV }
\frac{1}{k_{-}} = \frac 1 2 {\rm sgn}(x^{-} - y^{-}).
\end{equation}
However, for the fermion and scalar fields it is more
convenient to use another Green function $(i
\partial_{-})^{-1}(x^{-}- y^{-})= -i(\partial_{-} )^{-1}(x^{-}-
y^{-}) = - i \frac 1 2 {\rm sgn}(x^{-} - y^{-})$, which under
the complex conjugation behaves like a Hermitian matrix 
\begin{equation}
\left[ (i \partial_{-})^{-1}(x^{-}- y^{-})\right]^* = (i
\partial_{-})^{-1}(y^{-}- x^{-}). 
\end{equation}
Therefore, the Green function for the covariant derivative $(i
\partial_{-} - eA_{-})^{-1}$ can be given in the terms of
integral operators 
\begin{eqnarray}
(i \partial_{-} - e A_{-})^{-1}[x^{-}, y^{-}] & =
& \int dz^{-} (i\partial_{-})^{-1}(x^{-}-z^{-}) {\cal W}_{-1}
[z^{-}, y^{-}; \widehat{a}] \\
& = & \int dz^{-} {\cal W}_{-1} [x^{-}, z^{-}; \widehat{a}^{\dag}]
(i\partial_{-})^{-1}(z^{-}-y^{-}) \ ,
\end{eqnarray}
where 
\begin{eqnarray}
{\cal W}_{-1}[x^{-}, y^{-}; \widehat{a}] & = & \delta(x^{-} -
y^{-}) + \sum_{k=1}^{\infty} (\widehat{a})^k [x^{-}, y^{-}],\\
\widehat{a}[x^{-}, y^{-}] & = & e A_{-}(x^{-}) (i
\partial_{-})^{-1}(x^{-}-y^{-}),\\
(\widehat{a})^n[x^{-}, y^{-}] & = & \int dz^{-}
(\widehat{a})^{n-1}[x^{-}, z^{-}] \widehat{a}[z^{-}, y^{-}] =
\int dz^{-} \widehat{a}[x^{-}, z^{-}] (\widehat{a})^{n-1}
[z^{-}, y^{-}],\\
\widehat{a}^{\dag}[x^{-}, y^{-}] & = & e  (i
\partial_{-})^{-1}(x^{-}-y^{-})A_{-}(y^{-}), \\
(\widehat{a}^{\dag})^n[x^{-}, y^{-}] & = & \int dz^{-}
(\widehat{a}^{\dag})^{n-1}[x^{-}, z^{-}]
\widehat{a}^{\dag}[z^{-}, y^{-}] = \int dz^{-}
\widehat{a}^{\dag}[x^{-}, z^{-}] (\widehat{a}^{\dag})^{n-1}
[z^{-}, y^{-}]. \nonumber\\
\end{eqnarray}
One can easily check that under the complex conjugation these
expressions behave as follows:
\begin{eqnarray}
\left((\widehat{a})^n[x^{-}, y^{-}]\right)^* & = & 
(\widehat{a}^{\dag})^n[y^{-}, x^{-}] \, \\
\left((\widehat{a}^{\dag})^n[x^{-}, y^{-}]\right)^* & = & 
(\widehat{a})^n[y^{-}, x^{-}] \, \\
\left({\cal W}_{-1}[x^{-}, y^{-}; \widehat{a}]\right)^* & = & 
{\cal W}_{-1}[y^{-}, x^{-}; \widehat{a}^{\dag}] \, \\
\left({\cal W}_{-1}[x^{-}, y^{-}; \widehat{a}^{\dag}]\right)^* & = & 
{\cal W}_{-1}[y^{-}, x^{-}; \widehat{a}] \, \\
\left((i \partial_{-} - e A_{-})^{-1}[x^{-}, y^{-}]\right)^* & = & 
(i \partial_{-} - e A_{-})^{-1}[y^{-}, x^{-}] \ . 
\end{eqnarray}
If all two-argument expressions are treated as generalized
(infinite dimensional) matrices and their convolutions are denoted by
asterisks then one can introduce the self-explanatory matrix notation 
\begin{equation}
(i \partial_{-} - e A_{-})^{-1} =  (i\partial_{-})^{-1} \ast
{\cal W}_{-1} [\widehat{a}] =  {\cal W}_{-1} [\widehat{a}^{\dag}] \ast
(i\partial_{-})^{-1}  \label{appmatnotdefcovgrefun}\ ,
\end{equation}
where 
\begin{eqnarray}
{\cal W}_{-1}[\widehat{a}] & = & 1 + \sum_{k=1}^{\infty}
(\widehat{a})^k \ ,\\ 
\widehat{a} & = & e A_{-} (i
\partial_{-})^{-1}\ ,\\
(\widehat{a})^n & = & (\widehat{a})^{n-1} \ast
\widehat{a} =  \widehat{a} \ast (\widehat{a})^{n-1} \ ,\\
\widehat{a}^{\dag} & = & e  (i
\partial_{-})^{-1} A_{-} \ , \\
(\widehat{a}^{\dag})^n & = & (\widehat{a}^{\dag})^{n-1} \ast
\widehat{a}^{\dag} = \widehat{a}^{\dag} \ast
(\widehat{a}^{\dag})^{n-1}.  
\end{eqnarray}
Also one can check the following property
\begin{equation}
i \partial_{-} {\cal W}_{-1}[\widehat{a}^{\dag}] \ast = i
\partial_{-} + e A_{-} {\cal W}_{-1}[\widehat{a}^{\dag}] \ast \ ,
\end{equation}
so the integral operator ${\cal W}_{-1}[\widehat{a}^{\dag}]$
effectively transforms the covariant derivative into the
partial derivative
\begin{equation}
(i \partial_{-} - eA_{-}) {\cal W}_{-1}[\widehat{a}^{\dag}]
\ast = i \partial_{-}.
\end{equation}
Another simple calculation shows that
\begin{equation}
(i \partial_{-} - eA_{-}) (i\partial_{-})^{-1} \ast {\cal
W}_{-1}[\widehat{a}] = {\cal W}_{-1}[\widehat{a}] - \widehat{a}
\ast {\cal W}_{-1}[\widehat{a}] = 1 \,
\end{equation}
and this constitutes the proof that the integral operator defined
by (\ref{appmatnotdefcovgrefun}) is the Green function for the
covariant derivative operator $(i \partial_{-} - eA_{-})$. All
above results can be easily transformed to the quantum theory,
where the real-valued field $A_{-}(x^{-})$ is substituted by the
Hermitian operator and the complex conjugation is replaced by
the Hermitian conjugation.  \\
All charged fields can be easily incorporated into the matrix
notation, where and the scalar field $\phi$ and fermion field $\psi$ 
are treated as the one-column matrices while their Hermitian
conjugated counterparts $\phi^{\dag}$ and $\psi^{\dag}$ as the
one-row matrices, respectively.\\

\subsection{Inverse Laplace operators}\label{InvLaploperdod}

The Green function for the Laplace operator is well
defined for $d > 2$ dimensions 
\begin{equation}
[\Delta^d_\perp]^{-1}(x_\perp) = - \int \frac{d^dk_\perp}{(2\pi)^d}
\frac{e^{i k_\perp \cdot x_\perp}}{k_\perp^2} = -
\frac{1}{2^{2-d//2}}\frac{1}{(2 \pi)^{d/2}}
\frac{1}{(x_\perp)^{d/2-1}} \Gamma(d/2 -1)
\end{equation}
and for $d=2$ it is singular. From \cite{Balasinetal1992} we
know that there are also other regularizations, strictly in $d
= 2$ dimensions. In the paper, we use 
the massive regularization - when the pole at $k^2_\perp=0$
is shifted by the mass parameter $m^2$ 
\begin{equation}
[\Delta_\perp]^{-1}(x_\perp)  \rightarrow
[\Delta_\perp - m^2]^{-1}(x_\perp) = - \int \frac{d^2k_\perp}{(2\pi)^2}
\frac{e^{i k_\perp \cdot x_\perp}}{k_\perp^2+m^2} = - \frac{1}{2
\pi} K_{0}(m\sqrt{x_\perp^2}),
\end{equation}	
which naturally appears for the massive electrodynamics.\\
Another regularized Green function appears in Section \ref{sectiongenaxigau}
\begin{equation}
\frac{1}{\Delta_\alpha^{-}}(\vec{x}) = \int
\frac{d^2k_\perp}{(2\pi)^2} \frac{dk^{-}}{2\pi}
\frac{e^{-i \vec{k} \cdot \vec{x}}}{k_\perp^2 - 2 \alpha k_{-}^2}
\end{equation}
and for $\alpha < 0$ there is no ambiguity in the integrand.

\subsection{Finite volume Green functions}

One can safely invert differential operators
such as $\partial_-$ and $\partial_\perp^2$ in terms of
well-defined Green's functions, taking care of the
respective mode sector in which the operator acts.
With the covariant derivative $D_\mu = \partial_\mu + i eV_\mu$, 
we define 
the operator-valued Green's function to $(i D_-)$:
\begin{equation}
(i D_-^x) \frac{1}{i \partial_{-} - e V_{-}}[{\vec x},{\vec y}] \equiv
\delta^{(3)}({\vec x} - {\vec y}) - [Sub.]
\;,
\end{equation}
where $[Sub.]$ denotes possible subtractions corresponding to zero
eigenvalues of the operator in question.
A nonperturbative construction for this particular
Green's function is presented in Appendix A of \cite{PrzNausKall}.
The Green's function
${\cal G}_{(\perp)}[x_\perp , y_\perp; {\cal O}]$ is defined by
the relation
\begin{equation}
[ \Delta_\perp^x - \frac{e^2}{2L} {\cal O}(x_\perp) ]
{\cal G}_{(\perp)}[x_\perp , y_\perp; {\cal O}]
\equiv
 \delta^{(2)} (x_\perp - y_\perp ) \; ,
\label{eq:Gperp}
\end{equation}
where $\Delta_\perp \equiv \partial_\perp^2$,
and now ${\cal O}$ is some field operator of mass dimension one.
Eq.(\ref{eq:Gperp}) can be elucidated order by order in
perturbation theory for which one uses the
basic inversion of $\Delta_\perp$.
A nonperturbative definition is however nontrivial
and therefore the above operation is, at best, merely
formal and its concrete implementation remains an open problem.\\
The convolutions of the above Green's functions are also denoted by
asterisks.

\end{document}